\documentclass[12pt]{article}

\usepackage{youngtab}

\usepackage{color}
\usepackage[dvips]{graphicx}
\usepackage{graphicx}
\usepackage{amsmath,amssymb}
\usepackage{feynmp}
\usepackage{epsfig}

\def\Z{\mathbb{Z}}

\def\R{\mathbb{R}}

\input{colordvi.tex}

\newcommand{\sfrac}[2]{\left(\frac{#1}{#2}\right)}

\newcommand{\bra}[1]{ \langle {#1} | }
\newcommand{\ket}[1]{ | {#1} \rangle }

\def\diag{\mathop{\rm diag}\nolimits}
\def\tr{\mathop{\rm tr}}

\def\SO{\mathop{\rm SO}}

\def\SU{\mathop{\rm SU}}
\def\U{\mathop{\rm U}}

\begin{document}

\begin{titlepage}
   
\begin{flushright}
UT-12-42 \\
IPMU12-0225
\end{flushright}
    
\vskip 1cm
\begin{center}
   
{\large \bf Skewness Dependence of GPD / DVCS, \\
    Conformal OPE and AdS/CFT Correspondence I: \\
  --- Wavefunction of Regge Trajectory --- }
 
\vskip 1.2cm
    
 Ryoichi Nishio$^{1,2}$ and Taizan Watari$^2$
     
\vskip 0.4cm
  
 {\it
   $^1$Department of Physics, University of Tokyo, Tokyo 113-0033, Japan  
    \\[2mm]
    
   $^2$Institute for the Physics and Mathematics of the Universe, University of Tokyo, Kashiwano-ha 5-1-5, 277-8583, Japan
    }
\vskip 1.5cm
    
\abstract{
Traditional idea of Pomeron/Reggeon description for hadron scattering 
is now being given theoretical foundation in gravity dual descriptions, 
where Pomeron corresponds to exchange of spin-$j\in 2\Z$ states in the 
graviton trajectory.
Deeply virtual compton scattering (DVCS) 
is essentially a 2 to 2 scattering process
of a hadron and a photon, and hence one should be able to 
study non-perturbative aspects (GPD) of this process 
by the ``Pomeron'' process in gravity dual.
We find, however, that even the most developed formulation of gravity 
dual Pomeron (Brower--Polchinski--Strassler--Tan (BPST) 2006) is not able to
capture skewness dependence of GPD properly. 
Conformal operator product expansion allows us to determine 
DVCS amplitude in terms of matrix elements of primary operators, 
which should then be given by wavefunctions on warped spacetime.
We determined all the necessary wavefunctions on AdS$_5$ as 
an expression holomorphic in $j$, which 
will then be used (in our forthcoming publication) to determine GPD 
through inverse Mellin transformation. 
This approach will extend the formulation of BPST.
} 
    
\end{center}
\end{titlepage}



\section{Introduction}
\label{sec:intro}

In this article, we extend AdS/CFT technique to study a class 
of hadron scattering processes at high energy. Although we cannot 
expect gravitational ``dual'' descriptions to be perfectly equivalent 
to the QCD for high-energy scattering at the quantitative level, 
yet we still hope to learn, at the qualitative level, 
non-perturbative information in hadron scattering that is not 
available in perturbative or lattice QCD calculation \cite{PS-01-PRL, 
PS-02-DIS, BPST-06}.  

AdS/CFT technique can be used to study not just amplitudes 
of hadron scattering as a whole, but also to extract information 
of partons within hadrons \cite{PS-02-DIS}. 
Parton distribution functions are defined by using hadron matrix 
elements of parton-bilinear operators in QCD, and gravity dual descriptions 
can be used to determine matrix elements of the gauge singlet operators. 
The PDF extracted in this way satisfies DGLAP ($q^2$-evolution) and 
BFKL ($\ln (1/x)$-evolution) equations \cite{NW-first}.\footnote{
Just like in perturbative QCD, the essence is to follow the kinematical 
variable dependence of where in the complex angular momentum plane 
the matrix element has a saddle point.}
Thus, the parton information studied in this way may be used to 
understand non-perturbative issues with partons in a hadron 
in the real world at qualitative level. 

In a series of articles consisting of part I (this one) and 
part II \cite{NW-more}, 
we study 2-body--2-body scattering between a hadron and 
a photon (that is possibly virtual) in gravitational dual descriptions; 
$\gamma^* (q_1) + h (p_1) \rightarrow \gamma^{(*)} (q_2) + h(p_2)$. 
A special case of this scattering---the forward scattering with 
$q_1 = q_2$ and $p_1 = p_2$---has been studied extensively in the
literature (e.g., \cite{PS-02-DIS, Hatta-07, CC-DIS, 
BDST-10, NW-first}) for study of DIS and PDF, and some references 
also cover the case with $(q_1-q_2)^2 \neq 0$ (e.g. \cite{BDST-10, NW-first}). 
This series of articles extend the analysis so that all kinds of skewed cases 
are covered. In hadron physics, therefore, the kinematics needed for 
deeply virtual Compton scattering, hard exclusive vector meson
production and time-like Compton scattering processes \cite{GPD-review}
is covered in this analysis. With the full skewness dependence included 
in this analysis, it is also possible to use the result of this study 
to bridge a gap between data in such scattering processes at non-zero 
skewness and the transverse profile of partons in a hadron, 
which is encoded by the generalized parton distribution functions 
at zero skewness \cite{GPD-transv-profile}. 

For this study, we have to extend the formalism developed in 
\cite{PS-02-DIS, BPST-06} (see also 
\cite{Hatta-07, BDST-10, NW-first}) in a number of points.
First, hadron matrix elements of total derivative operators are irrelevant 
for the $h$--$\gamma^{(*)}$ scattering with zero skewness (like DIS), 
but they do contribute to the skewed scattering amplitude. 
This contribution needs to be implemented in the language of gravity dual. 
Secondly, Pomeron/Reggeon propagators have been treated as if it were for a 
scalar field in \cite{PS-02-DIS, BPST-06, NW-first}, but 
they correspond to exchange of stringy states with non-zero (arbitrary high)
spins; for the study of scattering with non-zero skewness, the 
polarization of higher spin state propagator should also be treated
with care. Finally, this also means that infinitely many gauge degrees 
of freedom in string theory (which extends the general coordinate 
transformation of the graviton) need to be dealt with properly. 

This article is organized as follows. 
In this article (part I), we begin in section
\ref{ssec:review-conf-OPE-GPD} with a review of parametrization of 
GPD in terms of conformal OPE, because the expansion in a series of 
conformal primary operators becomes the key concept in using AdS/CFT
correspondence (cf \cite{CC-DIS}).  
After plainly stating what is need to be done by using gravity dual 
in section \ref{ssec:idea-simply}, we proceed to explain our basic
gravity dual setting and idea of how to construct a scattering
amplitude of our interest by using string field theory in sections 
\ref{sec:set-up} and \ref{sec:cubic-sft}. 
Section \ref{sec:mode-fcn} shows the results of computing 
wavefunctions of spin-$j$ fields on AdS$_5$, while more detailed 
account of derivation of the wavefunctions are given in the appendix. 
Classification of eigenmodes is given in section \ref{ssec:eigen-q=0}, 
and wavefunctions are presented as analytic functions of the 
complex spin (angular momentum) variable $j$ in section
\ref{ssec:eigen-q=not0} for eigenmodes that turn out to be relevant for 
the ``twist-2'' operators in \cite{NW-more}.
Those wavefunctions are organized into irreducible representations 
of conformal algebra in section \ref{ssec:repr-dilatation}; 
the representation for spin-$j$ primary fields contain more 
eigenmode components than the those treated by the Pomeron exchange 
amplitude in the formalism of \cite{BPST-06}. 
These wavefunctions (and propagators) is used in part II \cite{NW-more}
in organizing scattering amplitude on AdS$_5$. Most of the physics 
results on GPD/DVCS are deferred to \cite{NW-more}, and part I (this
article) focuses on technical development that may also be of value 
from perspectives of formal theory.

\section{Our Approach: Conformal OPE and Gravity Dual}

\subsection{Review: Conformal OPE of DVCS Amplitude}
\label{ssec:review-conf-OPE-GPD}

\subsubsection{Notation and Conventions}

\begin{figure}[htbp]
\begin{center}
\begin{tabular}{cc}
  \includegraphics[scale=0.4]{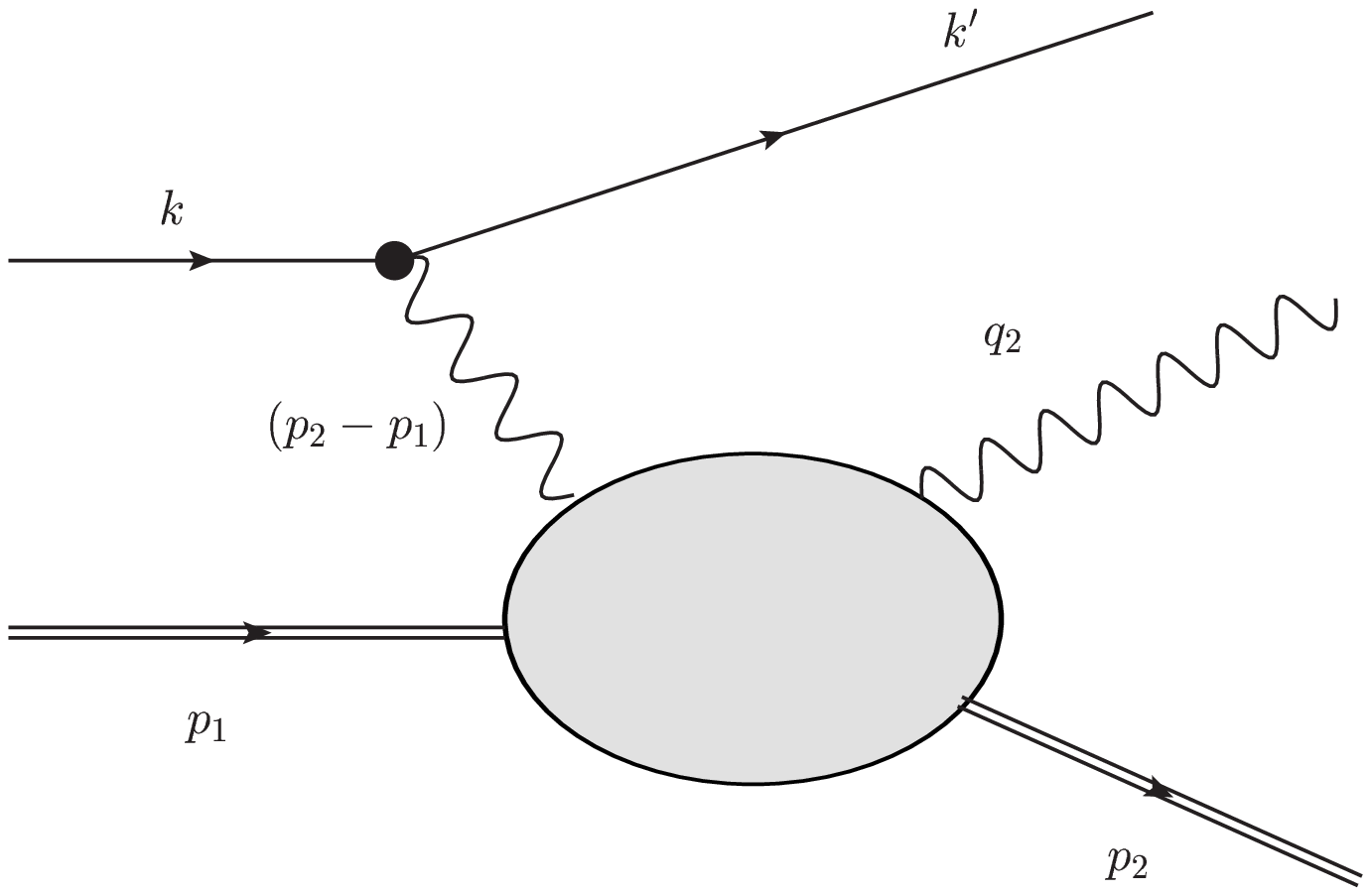} &
  \includegraphics[scale=0.4]{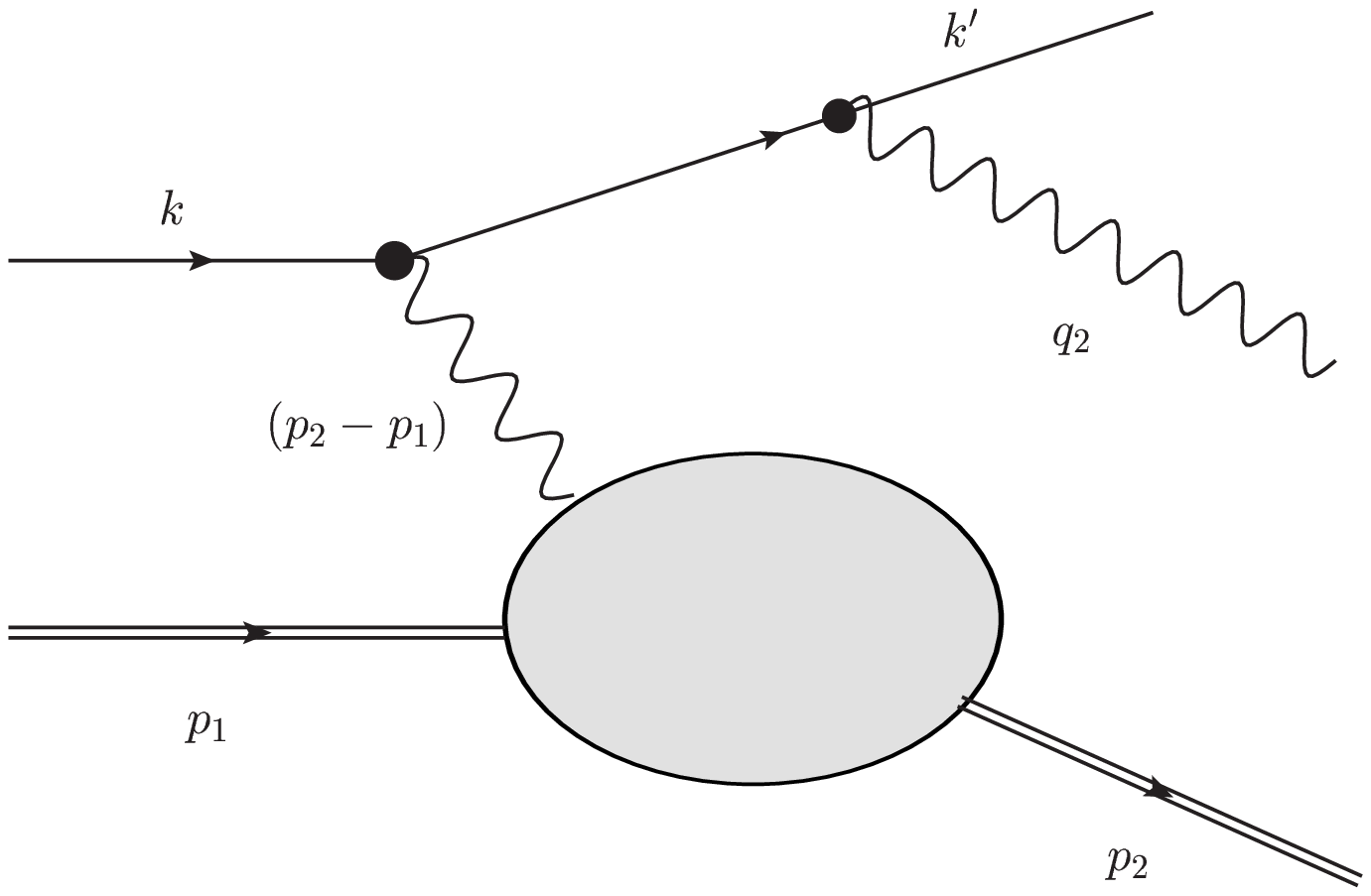} \\
 (a) DVCS-Compton & (b) DVCS-BH \\ 
\hline
  \includegraphics[scale=0.4]{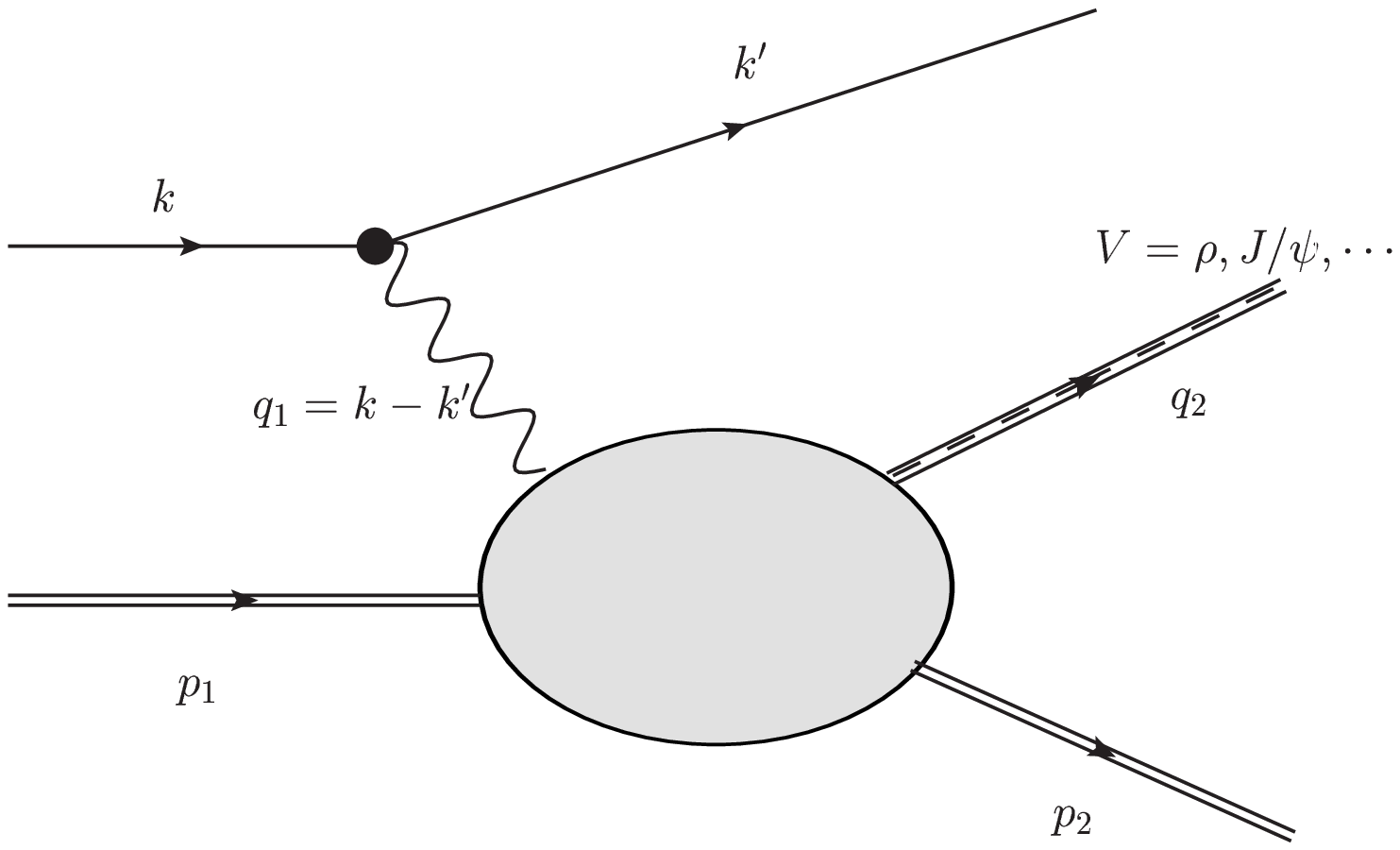} &
  \includegraphics[scale=0.4]{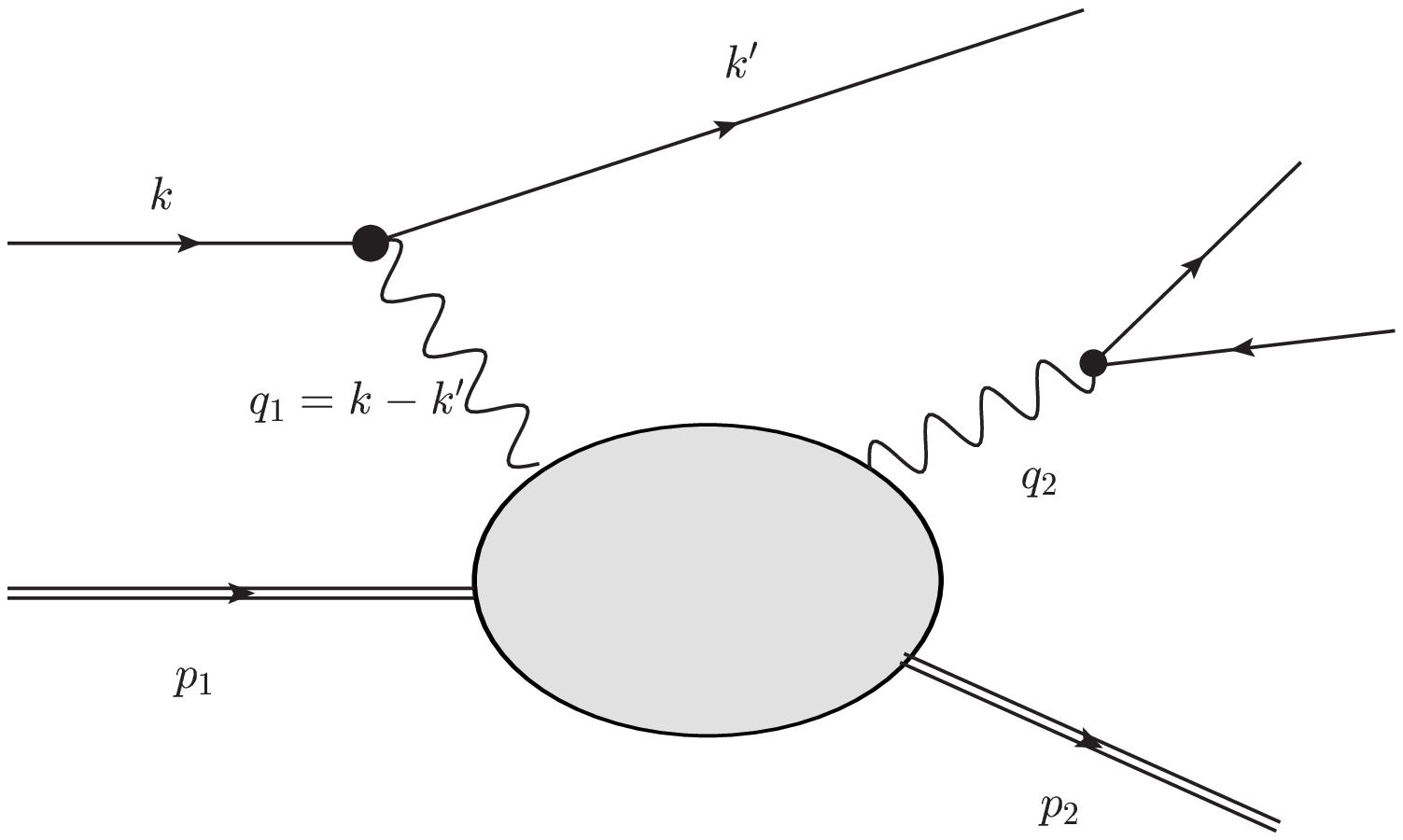}  \\
(c) VMP & (d) TCS 
\end{tabular}
  \caption{\label{fig:DVCS-TCS} 
(a, b) are diagrams contributing to the deeply virtual Compton
 scattering process, 
(c) is the exclusive vector meson production process, and 
finally (d) is the time-like Compton scattering process. 
}
\end{center}
\end{figure}

\begin{figure}[tbp]
 \begin{center}
  \includegraphics[scale=0.8]{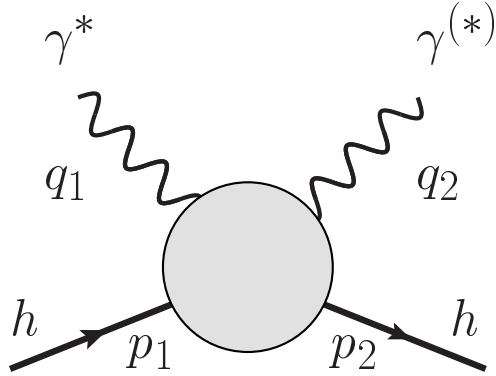} 
  \caption{\label{fig:DVCS-DDVCS} photon--hadron 2-body to 2-body
  scattering amplitude}
 \end{center}
\end{figure}

Deeply virtual Compton scattering $e+h \rightarrow e+h+\gamma$ (DVCS), 
hard exclusive vector meson production $e+h \rightarrow e+h+V$ (VMP) and 
time-like Compton scattering processes $e+h \rightarrow e+h+e^+e^-$ (TLC)
are shown in Figure \ref{fig:DVCS-TCS} (a, b), (c) and (d), respectively. 
As a part of all these processes, the photon--hadron 2-body to 2-body
scattering amplitude 
\begin{equation}
 {\cal M}(\gamma^\ast h \rightarrow \gamma^{(\ast)}h) = (\epsilon_1^\mu) T_{\mu\nu} (\epsilon_2^\nu)^\ast.
\end{equation}
is involved.\footnote{The two contributions from the 
$\gamma+h \rightarrow\gamma+h$ Compton scattering (a) and 
Bethe--Heitler process (b) can be separated experimentally 
by exploiting kinematical dependence and polarization. 
It thus makes sense to focus only on the amplitude (a).} 
This 2-body to 2-body scattering with this exclusive choice of the 
final states (\ref{fig:DVCS-DDVCS}) is truly non-perturbative
information, and this is the subject of this article. 
Because the ``final state'' photon is required to be on-shell 
$q_2^2 = 0$ in DVCS and time-like\footnote{We use the $(-,+++)$ metric 
throughout this paper. } $q_2^2 < 0$ in VMP and TLC, 
we are interested in developing a theoretical framework to calculate 
this non-perturbative amplitude in the case $q_2^2$ is different from 
space-like $q_1^2 > 0$.

Just like in QCD / hadron literature, we used the following 
notation for Lorentz invariant kinematical variables:
\begin{align}
\label{kinematics1}
  p^\mu &= \frac{1}{2}(p_1^\mu+p_2^\mu),& q^\mu &= \frac{1}{2}(q_1^\mu+q_2^\mu),& \Delta^\mu &=p_2^\mu -p_1^\mu = q_1^\mu - q_2^\mu,
\end{align}
\begin{align}
\label{kinematics2}
  x &= \frac{-q^2}{2p\cdot q},& \eta &= \frac{-\Delta \cdot q}{2 p\cdot
 q}, &s&= W^2 = -(p+q)^2, & t&=-\Delta^2.
\end{align}
$\eta$ is called skewness; in the scattering process of our interest, 
$q_1^2 = q^2 + \Delta^2/4 + q \cdot \Delta$ and 
$q_2^2 = q^2 + \Delta^2/4 - q \cdot \Delta$ are not the same, and 
hence the skewness does not vanish.
We will focus on high-energy scattering; for typical energy scale 
of hadron masses / confinement scale $\Lambda$, we assume that 
\begin{equation}
 \Lambda^2 \ll q_1^2, \; \; W^2, \quad {\rm while} \quad 
 |t| \lesssim {\cal O}(\Lambda).
\label{eq:Bjorken-2}
\end{equation}

The photon--hadron scattering amplitude (Figure~\ref{fig:DVCS-DDVCS}) 
in the real-world QCD (where all charged partons are fermions), the 
Compton tensor is given by the hadron matrix element with insertion 
of two QED currents $J^\mu$,
\begin{equation}
T^{\mu\nu} = i
 \int d^4 x e^{-iqx} 
   \langle {h(p_2)}|T \{ J^\nu(x/2) J^\mu(-x/2) \} |{h(p_1)}\rangle.
\label{eq:Compton tensor = <p|JJ|p>}
\end{equation}
For simplicity, we assume that the target hadron is a scalar, and 
further pay attention only to the structure function $V_1$ appearing 
in the gauge-invariant decomposition\footnote{
Here, we introduced a convenient notation 
\begin{align}
  P[q]_{\mu\nu}=\left[\eta_{\mu\nu}-\frac{q_\mu q_\nu}{q^2}\right].
\end{align}
} of the Compton tensor:
\begin{align}
   T^{\mu\nu}=&V_1 P[q_1]^{\mu\rho} P[q_2]^{\nu}_{\rho}
           +V_2 (p\cdot P[q_1])^\mu (p\cdot P[q_2])^\nu
           +V_3 (q_2\cdot P[q_1])^\mu (q_1\cdot P[q_2])^\nu
\notag \\ \label{eq:structure functions of Compton tensor}
           &+V_4 (p\cdot P[q_1])^\mu (q_1\cdot P[q_2])^\nu
           +V_5 (q_2\cdot P[q_1])^\mu (p\cdot P[q_2])^\nu
           +A \epsilon^{\mu\nu\rho\sigma}q_{1\rho}q_{2\sigma}.
\end{align}
Those structure functions, $V_{1,2,3,4,5}(x,\eta,t,q^2)$, should be 
expressed in terms of the kinematical variables $x, \eta$ and $t$, and 
our primary purpose of this article is to study how the structure 
functions depend on the skewness $\eta$.

\subsubsection{Light-cone Operator Product Expansion}

The light-cone operator product expansion (OPE)
can be applied to the product of currents 
$T\left\{ J^\nu  J^\mu \right\}$, before evaluating it as a hadron 
matrix element. Let the expansion be 
\begin{align}
 i  \int d^4 x e^{-iqx} T \{ J^\nu(x/2) J^\mu(-x/2) \}
= \sum_I {\cal C}_{I\: \rho_1 \dots \rho_{j_I}}^{\mu\nu}(q) 
    {\cal O}_I^{\rho_1\dots \rho_{j_I}}(0;q^2)
\label{eq:OPE of two currents}
\end{align}
for some basis of local operators 
${\cal O}_{I}^{\rho_1 \cdots \rho_{j_I}}$ renormalized at $\mu^2 = q^2$.
$C^{\mu\nu}_{I\; \rho_1 \cdots \rho_{j_I}}$ is the corresponding Wilson 
coefficients renormalized at $\mu^2 = q^2$. If we were to evaluate these 
local operators on the right-hand side with the same states as bra and
ket, $\bra{h(p_2)}$ and $\ket{h(p_1)}$ with $p_2^\mu=p_1^\mu$, then 
the Compton tensor and its structure functions do not receive non-zero 
contributions from local operators that are given by total derivative of 
some other local operators. In the case of our interest, however, 
such operators do contribute. 

Let us take a series of operators in QCD that are called twist-2
operators in the weak coupling limit. The twist-2 operators in the 
flavor non-singlet sector are labeled by two integers, $j, l$,
\begin{equation}
 {\cal O}^\alpha_{j,l}:=
   \left[(-i)^{j+l} \partial^{\mu_{j+1}} \cdots \partial^{\mu_{j+l}} 
    \bar{\Psi}_a \gamma^{\mu_1}
           \left(\overleftrightarrow{D}\right)^{\mu_2} \cdots 
           \left(\overleftrightarrow{D}\right)^{\mu_j} \; \lambda^\alpha_{ab} 
    \Psi_b\right]_{{\rm t.s.t.l.}}(0;q^2),
\label{eq:fl-nonsgl-twist2-op}
\end{equation}
with a $N_F \times N_F$ flavor matrix $(\lambda^\alpha)_{ab}$. Similarly 
there are two series of twist-2 operators with the label $j, l$, given by 
quark bilinear and gluon bilinear. Here, these operators are made 
totally symmetric and traceless (t.s.t.l) in the $j+l$ Lorentz indices in order 
to make it transform in an irreducible representation of the Lorentz
group $\SO(3,1)$. 
$\overleftrightarrow{D} := \overrightarrow{D}-\overleftarrow{D}$.

Suppose that the hadron matrix element of the operator ${\cal
O}^\alpha_{j,l}$ is given by 
\begin{equation}
 \bra{h(p_2)} {\cal O}^\alpha_{j,l} \ket{h(p_1)} = 
   \sum_{k=0}^j \left[ \Delta^{\mu_1} \cdots \Delta^{\mu_{k+l}}
       p^{\mu_{k+l+1}} \cdots p^{\mu_{j+l}}\right]_{{\rm t.s.t.l.}} 
  A_{j,k}(t; q^2) (-2)^{j-k};
\end{equation}
the reduced matrix element $A^\alpha_{j,k}(t)$ is non-perturbative
information and cannot be determined by perturbative QCD.
If we pay attention only to Wilson coefficients 
$C^{\mu\nu}_{j,l,\alpha; \mu_1 \cdots \mu_{j+l}}$ that are proportional
to $\eta^{\mu\nu}$, and denote them by
%
%
\begin{equation}
 \eta^{\mu\nu} {\cal C}^\alpha_{j,l} 
   \frac{q_{\rho_1} \cdots q_{\rho_{j+l}} }{(q^2)^{j+l}}, 
\end{equation}
then the twist-2 flavor non-singlet contribution to the structure
function $V_1$ becomes
\begin{equation}
 V_1 \simeq \sum_{j,l} {\cal C}^\alpha_{j,l} \frac{1}{x^{j+l}} 
   \sum_{k=1}^{j}A^\alpha_{j,k}(t;q^2) \eta^{k+l} =: 
  \sum_j {\cal C}^\alpha_{j} \sfrac{\eta}{x} \frac{1}{x^j}  
    A^\alpha_{j}(\eta, t; q^2),
\end{equation}
where 
${\cal C}^\alpha_j(\eta/x) := \sum_{l=0}^{\infty}  
{\cal C}^\alpha_{j,l}(\eta/x)^l$, and 
$A^\alpha_j(\eta,t) := \sum_{k=0}^j \eta^k A^\alpha_{j,k}(t)$.
If the structure function $V_1$ receives only from even $j \in \Z$, 
then this $j$-summation is rewritten as 
\begin{equation}
 V_1(x,\eta,t;q^2) \simeq - \int \frac{dj}{4i}
    \frac{1+e^{-\pi i j}}{\sin(\pi j)} {\cal C}^\alpha_j \sfrac{\eta}{x}
    \frac{1}{x^j} A^\alpha_j (\eta,t;q^2)
\end{equation}
in the form of inverse Mellin transformation; here, 
${\cal C}^\alpha_j(\eta,x)$ and $A^\alpha_j(\eta,t;q^2)$ are now 
meant to be holomorphic functions on $j$ (possibly with some poles and
cuts) that coincide with the original ones at $j \in 2 \Z$.
Precisely the same story holds also for flavor-singlet sector.

Because the structure function is given by the inverse Mellin transform 
of a product of the signature factor $[1+e^{-\pi i j}]/\sin (\pi j)$, 
Wilson coefficients ${\cal C}^\alpha_j$ and hadron matrix elements 
$A^\alpha_j$, it can be regarded as a convolution of inverse Mellin
transforms of those three factors. The inverse Mellin transform 
of the signature factor becomes 
\begin{equation}
 \int \frac{dj}{2\pi i} \frac{1}{x^j}
   \frac{\pi}{2} \frac{\mp[1 \pm e^{-\pi i j}]}{\sin (\pi j)} = \frac{-1}{2} 
   \left[ \frac{1}{1-x+i\epsilon} \pm \frac{1}{1+x } \right] ,
\end{equation} 
which corresponds to propagation of the parton in perturbative
calculation \cite{Ji}, and the inverse Mellin transform of the matrix
element is defined as the generalized parton distribution:
\begin{equation}
 H^\alpha(x,\eta,t;\mu^2 = q^2) = \int \frac{dj}{2\pi i} \frac{1}{x^j} 
  A_j(\eta,t;\mu^2 = q^2).
\label{eq:GPD-def-natural}
\end{equation}

Generalized parton distribution (GPD) $H^\alpha(x,\eta,t;\mu^2)$ of a
hadron $h$ is a non-perturbative information, just like the ordinary
PDF, which is obtained by simply setting $\eta=0$ and $t=0$.
For phenomenological fit of experimental data of DVCS and VMP, 
some function form of the GPD needs to be assumed, because of the 
convolution involved in the scattering amplitude.
Setting up a model (and assuming a function form) for the 
non-perturbative information in terms of $A_j(\eta, t;q^2)$ rather than 
the GPD itself $H(x,\eta,t;q^2)$ is called dual 
parametrization \cite{dual-para}, and some phenomenological ans\"{a}tze 
have been proposed.
In this article, we aim at deriving qualitative form of $A_{j}(\eta, t)$ 
by using gravitational dual (that is analytic in $j$), instead of 
assuming the form of $A_j(\eta,t)$.

\subsubsection{Renormalization and OPE in dilatation eigenbasis}

Remembering that the distinction between the $\gamma^*+h\rightarrow \gamma+h$ 
scattering amplitude and GPD originates from the factorization 
into the Wilson coefficients and local operators (and their matrix
elements), one will notice that the GPD defined in this way should 
depend on the choice of the basis of local operators.
Although the choice of operators ${\cal O}^\alpha_{j,l}$ with $j \geq 1$
and $l \geq 0$ in (\ref{eq:fl-nonsgl-twist2-op}) appears to be the most 
natural (and intuitive) one for the twist-2 operators in the flavor non-singlet
sector, there is nothing wrong to take a different linear combinations 
of these operators as a basis, when the corresponding Wilson
coefficients also become linear combinations of what they are for 
${\cal O}^\alpha_{j,l}$. Given the fact that the operators 
${\cal O}^\alpha_{j,l}$ mix with one another under renormalization,  
it is not compulsory for us to stick to the basis ${\cal O}^\alpha_{j,l}$.

Under the perturbation of QCD, the flavor non-singlet twist-2 operators 
are renormalized under 
\begin{align}
  \mu\frac{\partial}{\partial \mu}\:
 [{\cal O}_{j-m,m}(0;\mu^2)] =
 -[\gamma^{(j)}]_{mm'} \:
  [{\cal O}_{j-m',m'}(0;\mu^2)];
\label{eq:q-evolution in nozero MT}
\end{align}
because operators can mix only with those with the same number of
Lorentz indices, the anomalous dimension matrix is block diagonal 
in the basis of ${\cal O}^\alpha_{j,l}$. The $j\times j$ matrix 
for the operators ${\cal O}^\alpha_{j-m,m}$ ($m=0,\cdots,j-1$) is denoted
by $[\gamma^{(j)}]$. This matrix is upper triangular in this basis, 
and the diagonal entries are given by the anomalous dimensions  
of the twist-2 spin-$j$ operators without a total derivative:
%
%
\begin{equation}
 \left[\gamma^{(j)}\right]_{mm} = \gamma(j-m).
\end{equation}
Therefore, the eigenvalues of the anomalous dimension matrix is 
$\left\{ \gamma(j-m) \right\}_{m=0,\cdots,j-1}$ in this diagonal block, 
and the corresponding operator\footnote{
At the leading order in QCD, this operator basis---a basis that
diagonalizes the anomalous dimension matrix---is given by 
$\overline{\cal O}^\alpha_{n,m} = (-i)^{n+m+1} (\partial^{\mu})^m 
[\bar{\Psi}_a \gamma^\mu
C^{3/2}_n(\overleftrightarrow{D}^\mu/\partial^\mu) 
  \lambda^\alpha_{ab} \Psi_b] = 
 (-i \partial)^m \overline{\cal O}_{n,0}$, where $C^{3/2}_n$ is the Gegenbauer
  polynomial.}
 $\overline{{\cal O}}^\alpha_{j-m-1,m}$ 
is a linear combination of operators ${\cal O}_{j-m',m'}$ with 
$m'=m ,\cdots, j-1$ \cite{ER}. The corresponding Wilson coefficient 
$\overline{C}^\alpha_{j-m-1,m}$ for such an operator is a linear combination of 
${\cal C}^\alpha_{j-m',m'}$ with $m'=m,\cdots,0$.
In this operator basis, matrix elements and Wilson coefficients
renormalize multiplicatively, without mixing.\footnote{In reality, 
the anomalous dimension matrix depends on the coupling constant
$\alpha_s$, and $\alpha_s$ changes over the scale. Thus, the
eigenoperator of the renormalization / dilation also changes over the 
scale. In scale invariant theories (and in theories only with slow running 
in $\alpha_s$), however, this multiplicative renormalization is exact or 
a good approximation. (c.f. \cite{BM-conf-viol})} 

In this new basis of local operator, the structure function becomes 
\begin{equation}
 V_1 \simeq \sum_{n,K} \overline{\cal C}^\alpha_{n,K} \frac{1}{x^{n+1+K}} 
   \sum_k \overline{A}^\alpha_{n+1,k}(t;\mu^2) \eta^{K+k} =:
   \sum_n \overline{C}^\alpha_n \sfrac{\eta}{x} \frac{1}{x^{n+1}}
          \overline{A}^\alpha_{n+1}(\eta, t;\mu^2), 
\end{equation}
where 
\begin{equation}
 \overline{C}^\alpha_n = \sum_{K=0}^\infty \overline{C}^\alpha_{n,K}
  \sfrac{\eta}{x}^K, 
\end{equation}
and $\overline{A}^\alpha_{n+1,k}(t;\mu^2)$ is the reduced matrix element of 
the operator\footnote{Just like 
${\cal O}_{j,l} = (-i \partial)^l {\cal O}_{j,0}$, there is a relation 
$\overline{\cal O}_{n,K} = (-i \partial)^K \overline{\cal O}_{n,0}$ in the new
basis. This is why all the hadron matrix elements of 
$\overline{\cal O}_{n,K}$ can be parametrized by $\overline{A}_{n+1,k}$, just
like those of ${\cal O}_{j,l}$ are by $A_{j,k}$.} $\overline{\cal O}^\alpha_{n,0}(0;\mu^2)$.
The structure function is therefore written as yet another inverse
Mellin transform
\begin{equation}
 V_1 \simeq - \int \frac{dj}{4i} \frac{1+e^{-\pi i}}{\sin (\pi j)} 
    \overline{C}^\alpha_{j-1}(\eta/x) \frac{1}{x^j}
    \overline{A}^\alpha_{j}(\eta, t; \mu^2).
\end{equation}
Yet another GPD can also be defined by using
$\overline{A}^\alpha$, instead of $A^\alpha_j(\eta, t;q^2)$.
\begin{equation}
 \overline{H}(x,\eta,t;\mu^2) = \int \frac{dj}{2\pi i} \frac{1}{x^j} 
    \overline{A}^\alpha_{j}(\eta, t; \mu^2).
\end{equation}
When it comes to the description of the 
$\gamma^* + h \rightarrow \gamma+h$ scattering amplitude as a whole, 
it does not matter which operator basis is used. Although we need GPD 
rather than the scattering amplitude in order to talk about the 
distribution of partons in the transverse directions in a hadron, yet 
we only need GPD at $\eta = 0$. Thus, the newly defined GPD 
$\overline{H}$ does just as good a job as $H$ defined 
in (\ref{eq:GPD-def-natural}); they are the same at $\eta = 0$.

Even within the dual parametrization approach, it has been advantageous 
to use this operator basis, because it becomes much easier to implement 
a phenomenological assumption (function form) of 
$\overline{A}^\alpha_j(\eta,, t; \mu^2)$ that is consistent with
renormalization group flow \cite{GPD-review}.


\subsubsection{Conformal OPE}

Although the hadron matrix element is essentially non-perturbative, 
and is not calculable within perturbative QCD, more discussion has been 
made on the Wilson coefficients $\overline{C}^\alpha_{n,K}$. They still have 
to be calculated order by order in perturbation theory, if one is 
interested strictly in the QCD of the real world. If one is interested
in gauge theories that are more or less ``similar'' to QCD, however, 
stronger statements can be made for a system with higher symmetry: conformal
symmetry. One can think of ${\cal N}=4$ super Yang--Mills theory or 
${\cal N} = 1$ supersymmetric $\SU(N) \times \SU(N)$ gauge theory of 
\cite{Klebanov-W} as an example of theories with exact (super)
conformal symmetry. The QED probe in the real world QCD can be replaced 
by gauging global symmetries (such as (a part of) $\SU(4)$ R-symmetry 
of ${\cal N}=4$ super Yang--Mills theory and 
$\SU(2) \times \SU(2) \times \U(1)$ symmetry of \cite{Klebanov-W}).
By applying the conformal symmetry, one can derive stronger statements 
on the Wilson coefficients appearing in OPE of primary operators (like
conserved currents).

Suppose that we are interested in the OPE of two primary operators, 
$A$ and $B$, that are both scalar under $\SO(3,1)$. If we take 
the basis of local operators for the expansion to be primary operators 
$\overline{\cal O}_n$ (with $j_n$ Lorentz indices and $l_n$ scaling dimension) 
and their descendants $\partial^K \overline{\cal O}_n$ (with $j_n + K$ 
Lorentz indices), 
\begin{equation}
 T\left\{ A(x) B(0) \right\} = \sum_n 
   \left(\frac{1}{x^2}\right)^{\frac{1}{2}(l_A + l_B - l_n + j_n)} 
 \sum_{K=0}^{\infty} c_{n,K}
    \frac{ x^{\rho_1} \cdots x^{\rho_{j_n+K}} }{ (x^2)^{j_n+K} } 
    [\partial^K \overline{\cal O}_n(0)]_{\rho_1 \cdots \rho_{j_n+K}}, 
\end{equation}
then the conformal symmetry determines all the coefficients of the 
descendants $c_{n,K}$ ($K \geq 1$) in terms of that of the primary
operator, $c_{n,0} =: c_n$.
Ignoring the mixture of non-traceless contributions, one finds that 
\cite{FGG-confl-HG}
\begin{equation}
 T \left\{ A(x) B(0) \right\} \simeq \sum_n
   \left( \frac{1}{x^2} \right)^{\frac{1}{2}(l_A + l_B - l_n + j_n)}
    \!\!\!\!\!\!\!\!\! \!\!\!\!\!\!\!\!\! \!\!\!\!\!\!\!\!\! 
    x^{\rho_1} \cdots x^{\rho_{j_n}} c_n 
    {}_1 F_1( (l_A-l_B+l_n+j_n)/2 , l_n + j_n; x \cdot \partial) 
  [\overline{\cal O}_n(0)]_{\rho_1 \cdots \rho_{j_n}}.
\end{equation}

Questions of real interest to us is the OPE conserved currents $J^\nu$ 
and $J^\mu$. They are not scalars of $\SO(3,1)$, 
but the same logic as in \cite{FGG-confl-HG} can be used also to 
show that,
in the terms with Wilson coefficients proportional to $\eta^{\mu\nu}$, 
\begin{equation}
 T\left\{ J^\nu(x) J^\mu(0) \right\} \simeq \eta^{\mu\nu} \sum_n
   \left( \frac{1}{x^2} \right)^{3-\frac{\tau_n}{2}}
    \!\!\!\!\!\!\!\!\! \!\!
    x^{\rho_1} \cdots x^{\rho_{j_n}} c_n 
    {}_1 F_1( (l_n+j_n)/2 , l_n + j_n; x \cdot \partial) 
  [\overline{\cal O}_n(0)]_{\rho_1 \cdots \rho_{j_n}} + \cdots,
\end{equation}
where $\tau_n := l_n - j_n$ is the twist, mixture of the non-traceless 
(and hence higher twist) contributions are ignored, and terms with 
Wilson coefficients without $\eta^{\mu\nu}$ are all omitted here. 
The scaling dimension of conserved currents $l_A = l_B = 3$ have been
used. The momentum space version of the OPE is \cite{Mueller-PRD98}
\begin{eqnarray}
 i \int d^4 x \; e^{- iq_2 \cdot x} \;  T\left\{ J^\nu(x) J^\mu(0)
					 \right\}
 & \simeq & \eta^{\mu\nu} \sum_n 
  \frac{(2\pi)^2 \Gamma\left(\frac{l_n+j_n-2}{2} \right)}
       {4^{2-\frac{\tau_n}{2}} \Gamma\left(3-\frac{\tau_n}{2} \right)}
  c_n \; 
  \frac{(-2i)^{j_n} q_2^{\rho_1} \cdots q_2^{\rho_{j_n}}} 
       {(q^2_2)^{\frac{\tau_n}{2}-1} (q^2_2)^{j_n}}
    \nonumber  \\
 && \!\!\!\!\!\!\!\!\!\!\!\!\!\!\!\!\!\!\!\!\!\!\!\!\!\!
 {}_2 F_1 \left( \frac{l_n + j_n}{2}, \frac{l_n+j_n}{2} -1, l_n + j_n; 
    \frac{- 2i q_2 \cdot \partial}{q_2^2} \right) \overline{\cal O}_n(0) + \cdots,
 \label{eq:conf-OPE-2side}
\end{eqnarray}
or equivalently \cite{KK-Mueller-07}, 
\begin{eqnarray}
 i \int d^4(x-y) e^{-i q \cdot (x-y)} \; T\left\{ J^\nu(x) J^{\mu}(y) \right\} 
  & \simeq & \eta^{\mu \nu} \sum_n 
  \frac{(2\pi)^2 \Gamma\left(\frac{l_n+j_n-2}{2} \right)}
       {4^{2-\frac{\tau_n}{2}} \Gamma\left(3-\frac{\tau_n}{2} \right)}
  c_n \; 
  \frac{(-2i)^{j_n} q^{\rho_1} \cdots q^{\rho_{j_n}}} 
       { (q^2)^{\frac{\tau_n}{2}-1}  (q^2)^{j_n}}
  \nonumber \\
 && \!\!\!\!\!\!\!\!\!\!\!\!\!\!\!\!\!\!\!\!\!\!\!\!\!\!\!\!\!\!\!\!\!\!\!\!\!\!\!\!\!\!\!\!\!\!\!\!\!\!\!\!\!\!\!\!\!\!\!\!\!\!\!
 {}_2 F_1 \left( \frac{l_n + j_n-2}{4}, \frac{l_n+j_n}{4}, \frac{l_n + j_n}{2}; 
    \left(\frac{i q \cdot \partial}{q^2}\right)^2 \right)
   \overline{\cal O}_n \left(\frac{x+y}{2} \right)
   + \cdots.
\label{eq:conf-OPE-center}
\end{eqnarray}
Either in the form of (\ref{eq:conf-OPE-2side}) or
(\ref{eq:conf-OPE-center}), the primary operators $\overline{\cal O}_n$
and corresponding coefficients $c_n$ are renormalized multiplicatively. 

\subsection{AdS/CFT Approach}
\label{ssec:idea-simply}

In AdS/CFT correspondence, 
Type IIB string theory on AdS$_5 \times W$ with a 5-dimensional Einstein 
manifold $W$ would correspond to a gauge theory on $\R^{3,1}$ with an exact 
conformal symmetry, which is qualitatively different from the QCD in the 
real world. But the Type IIB string on a geometry that is close to 
AdS$_5 \times W$, but with confining end in the infrared, may be used to 
extract qualitative lesson on strongly coupled gauge theories with confinement.

In a dual pair of a CFT and a string theory on a background AdS$_5 \times W$,
primary operators of the CFT are in one to one correspondence with string 
states on AdS$_5 \times W$, and their correlation functions can be 
calculated by using the wavefunctions of the string states on AdS$_5$.
When the background geometry is changed from AdS$_5 \times W$ to 
some warped geometry that is nearly AdS$_5$ with an end in the infrared, 
then the wavefunctions might be used to calculate matrix elements of 
the corresponding ``primary'' operators in an almost conformal theory. 
The correspondence between the operator and string states can be made 
precise, because they are both classified in terms of representation 
of the conformal algebra, which is shared by both of the dual theories. 

In order to determine GPD $\overline{H}$ in gravitational dual descriptions, 
it is therefore sufficient to determined wavefunctions of string states 
corresponding to the ``primary'' operators of interest. 
Although there are plenty of literature discussing the correspondence between 
the (superconformal) primary operators and string states at the supergravity 
level, it is known that the flavor-singlet twist-2 operators (labeled by 
the spin $j$) correspond to the stringy excitations with arbitrary high spin 
$j$ that are in the same trajectory as graviton \cite{GKP-02, BPST-06}. 
Our task is therefore to determine the wavefunctions of such string states. 
Needless to say, one has to fix all the gauge degrees of freedom associated 
with string component fields (not just the general coordinate invariance 
associated with the graviton) before working out the mode decomposition. 
Furthermore, wavefunctions need to be grouped together properly so that 
they form an irreducible representation of the conformal group, in order 
to establish correspondence with a primary operator of the gauge theory 
side, which also forms an irreducible representation of the conformal 
group along with its descendants.  

It will be clear by the end of this article that all of such 
technical works is necessary and essential for the purpose of 
extracting skewness dependence of GPD. 

The DVCS $\gamma^* + h \rightarrow \gamma^{(*)} + h$ amplitude can be 
calculated in gravitational dual descriptions in two different ways.
One is to use the matrix elements of spin-$j$ primary operators calculated 
in gravitation dual (as we discussed just above) and combine them with 
the Wilson coefficients which are governed by the conformal symmetry 
(see (\ref{eq:conf-OPE-center})). 
The other is to calculate disc/sphere amplitude directly, with the 
vertex operators given (approximately) by using the wavefunctions 
associated with the target hadron (see sections \ref{sec:set-up} 
and \ref{sec:cubic-sft}); in this approach, contributions from operators 
with higher twists are also included. 
In this article, we adopt the latter approach. In the resulting 
expression for the $\gamma^{(*)}+h \rightarrow \gamma+h$ scattering 
amplitude in gravity dual, however, we can easily identify the 
structure of conformal OPE (as we will see in \cite{NW-more}), so that 
we can identify the contributions from the ``twist-2'' operators. 



\section{Gravity Dual Settings}
\label{sec:set-up}

A number of warped solutions to the Type IIB string theory has been 
constructed, and they are believed to be dual to some strongly coupled 
gauge theories. When the geometry is close to AdS$_5 \times W$ with 
some 5-dimensional Einstein manifold $W$, with weak running of the 
AdS radius along the holographic radius, the corresponding gauge 
theory will also have approximate conformal symmetry, and 
the gauge coupling constant runs slowly. If the ``AdS$_5 \times W$'' 
geometry has a smooth end at the infra red as in
\cite{Klebanov-Strassler}, then the dual gauge theory will end up 
with confinement. Gravitational backgrounds in the Type IIB string
theory with the properties we stated above all provide a decent framework 
of studying qualitative aspects of non-perturbative information
associated with gluons/Yang--Mills theory on 3+1 dimensions. 

In studying the $h+\gamma^{*} \rightarrow h + \gamma$ scattering
process in gravitational dual, we need a global symmetry to be gauged 
weakly, just like QED for QCD. In Type IIB D-brane constructions of gauge
theories that have gravity dual, U(1) subgroups of an R-symmetry 
or a flavor symmetry on D7-branes can be used as the models of the 
electromagnetic U(1) symmetry. Therefore, we have in mind 
gravity dual models on a background that is approximately 
``AdS$_5 \times W$'' with a non-trivial isometry group on $W$, or 
with a D7-brane configuration on it. 

Our motivation, however, is not so much in writing down an exact 
mathematical expression based on a particular gravity dual model 
that is equivalent to a particular strongly coupled gauge theory, 
but more in extracting qualitative information of partons in hadrons 
of confining gauge theories in general. It is therefore more suitable 
for this purpose to use a simplified set-up that carries common (and essential)
features of the Type IIB models that we described above. 
Throughout this article, we assume pure AdS$_5 \times W$ metric background, 
\begin{align}
 \label{eq:AdS5timesS5}
  ds^2  =G_{MN}dx^M dx^N&=g_{mn}dx^m dx^n +R^2 (g_{W})_{ab}d\theta^a d\theta^b,\\
 \label{eq:hwb}
  g_{mn}dx^m dx^n&=e^{2A(z)}(\eta_{\mu\nu}dx^\mu dx^\nu+dz^2), \;\;\;
 e^{2A(z)}=\frac{R^2}{z^2}.\;\;\; 
\end{align}
that is, we ignore the running effect, and we do not specify the 
5-dimensional manifold $W$.
The dilaton vev is simply assumed to be constant, $e^\phi = g_s$. 
Confining effect---the infra-red end of this geometry---can be 
introduced, for example, by sharply cutting of the AdS$_5$ space at 
$z = \Lambda^{-1}$ (hard wall models), or by similar alternatives 
(soft wall models). We are not committed to a particular implementation 
of the infra-red cut-off in this article, except in a couple of places 
where we write down some concrete expressions for illuminating purposes.
The energy scale $\Lambda$ associated with (any form of implementation of) 
the infra-red cut-off corresponds to the confining energy scale in the 
dual gauge theories. 
When we consider (simplified version of the) models with D7-branes for 
flavor, we assume that the D7-brane worldvolume wraps on a 3-cycle on $W$, 
and extends all the way down to the infra-red end of the holographic radius 
$z$; i.e., all of $0 \leq z \leq \Lambda^{-1}$. This corresponds to 
assuming massless quarks. In this article, we will not pay attention to 
physics where spontaneous chiral symmetry breaking is essential. 

As we stated earlier, we would like to work out the 
$h + \gamma^{*} \rightarrow h+\gamma$ scattering amplitude by 
using the gravity dual models. This is done by summing up 
sphere / disc amplitudes, along with those with higher genus worldsheets. 
We will restrict our attention to kinematical regions where saturation 
is not important (i.e., large $q^2$ and/or not too small $x$, and 
large $N_c$).  That allows us to focus only on sphere / disc amplitudes, 
with insertion of four vertex operators corresponding to the incoming
and outgoing hadron $h$ and (possibly virtual) photon $\gamma$.

As a string-based model of the target hadron $h$ (that is $\SO(3,1)$ 
scalar), we have in mind either a scalar ``glueball''\footnote{By
``glueball'', we only mean a bound state of fields in super Yang--Mills
theory. } that has non-trivial R-charge, or a scalar meson made of 
matter fields. The former corresponds to a vertex operator 
(in the $(-1,-1)$ picture)
\begin{equation}
 V(p) = : e^{i p_{\mu} \cdot X^{\hat{\mu}} }
    \psi^m \tilde{\psi}^n g_{mn} \Phi(Z; m_n) Y(\Theta):,
\end{equation} 
where $Y(\Theta)$ is a ``spherical harmonics'' on $W$, and 
the latter to 
\begin{equation}
 V(p) = : e^{i p_\mu \cdot X^{\hat{\mu}} }  \psi \Phi(Z; m_n):,
\end{equation}
where $\psi$ corresponds to the D7-brane fluctuations in the transverse
direction. $\Phi(Z)$ is the wavefunction on AdS$_5$, with the argument 
promoted to the field on the world sheet. Vertex operators above 
are approximate expressions in the $(\alpha'/R^2) \sim 1/\sqrt{\lambda}$
expansion in a theory formulated with a non-linear sigma model given by 
(\ref{eq:AdS5timesS5}). If we are to employ the hard-wall implementation 
of the infra-red boundary, with the AdS$_5$ metric without modification, 
then the wavefunction $\Phi(Z;m_n)$ is of the form
\begin{equation}
  \sqrt{c_\phi}\Phi(z; m_n) = 
 \sfrac{2\kappa_5^2}{R^3}^{1/2}\sqrt{2}\Lambda z^2 \frac{J_{\Delta-2}(j_{{\Delta-2},n}\Lambda z)}{J_{\Delta-2}(j_{{\Delta-2},n})}.
\label{eq:hadron-wvfc}
\end{equation}
%

The ``photon'' current in the correlation function/matrix element 
$T^{\nu\mu}$ in the gauge theory description corresponds to insertion of 
vertex operators associated with non-renormalizable wavefunctions,
rather than to the normalizable wavefunctions (\ref{eq:hadron-wvfc}) 
for the target hadron state. If we are to employ an R-symmetry current 
as the string-based model of the QED current, then the corresponding 
closed string vertex operator is 
\begin{equation}
 V(q) = : e^{i q_\mu \cdot X^{\hat{\mu}} } v_a(\Theta) A_m(Z; q)
     (\psi^a \tilde{\psi}^m + \psi^m \tilde{\psi}^a):,
\end{equation}
with some Killing vector $v_a \partial/\partial \theta^a$ on $W$. The
vertex operator in the case of D7-brane U(1) current is 
\begin{equation}
 V(q) = : e^{i q_\mu \cdot X^{\hat{\mu}} } A_m(Z; q) \psi^m :.
\end{equation}
The wavefunction $A_m(Z; q)$ on AdS$_5$ is of the form 
\begin{align}
A_{\mu}(z; q)  = &
  \left[ \delta_\mu^{\; \hat{\kappa}}
       - \frac{q_\mu q^{\hat{\kappa}} }{q^2} \right]
 \epsilon_\kappa(q) (qz) K_1(qz) + &
 q_\mu \frac{q^{\hat{\kappa}} \epsilon_\kappa(q)}{2 q^2}
 (qz)^2 K_2(qz), \\
A_z (z; q)  = & &
 \!\!\! -i \partial_z  \frac{q^{\hat{\kappa}} \epsilon_\kappa(q)}{2 q^2}
 (qz)^2 K_2(qz).
\label{eq:photon-wvfc}
\end{align}
Rationale for our choice of the terms proportional to 
$(q \cdot \epsilon)$ will be explained later on \ref{ssec:vector}, 
but those terms should 
not be relevant in the final result, because of the gauge invariance 
of $T^{\nu\mu}$. When the infra-red boundary is implemented by the hard
wall, $K_1(qz)$ should be replaced by
$K_1(qz)+[K_0(q/\Lambda)/I_0(q/\Lambda)] I_1(qz)$; and 
$K_2(qz)$ by arbitrary linear combination of $K_2(qz)$ and $I_2(qz)$.

It is not as easy to calculate the sphere/disc amplitudes in practice,
however. It has been considered that the parton contributions to 
$\gamma^{*} + h \rightarrow \gamma^{(*)} + h$ scattering is given by 
amplitude with states in the leading trajectory with arbitrary high spin 
being exchanged \cite{BPST-06}. Those fields are not scalar on AdS$_5$
but come with multiple degree of freedom associated with polarizations.
As we will see clearly in \cite{NW-more},\footnote{At least it will 
be evident by the end of section \ref{ssec:repr-dilatation}, that the 
existing treatment of Pomeron wavefunction in the literature---dealing
with Pomeron propagator as if it were a scalar---is not appropriate for
the purpose of obtaining matrix elements of higher spin primary
operators. } 
such polarization of higher
spin fields propagating on AdS$_5$ needs to be treated
properly---including such issues as covariant derivatives and kinetic 
mixing among different polarizations---in gravity dual descriptions, 
in order to be able to discuss skewness dependence of GPD / DVCS amplitude.  
Direct impact of the curved background geometry can be implemented 
through the non-linear sigma model on the world sheet, but one has to 
define the vertex operators as a composite operator properly in such an 
interacting theory. Ramond--Ramond background is an essential ingredient 
in making the warped background metric stable, yet non-zero 
Ramond--Ramond background cannot be implemented in the NSR formalism. 

Instead of world-sheet calculation in the NSR formalism in implementing 
the effect of curved background (\ref{eq:AdS5timesS5}), we use 
string field theory action on flat space in this article, and make it 
covariant. 
Because the gravity dual set-up of our interest is in the Type IIB 
string theory, we are thus supposed to use superstring field theory 
for closed string and open string. In order to avoid technical 
complications associated with the interacting superstring field
theories, however, we employ a sort of toy-model approach by using 
the cubic string field theory for bosonic string theory.

In our toy-mode approach, we deal with the cubic string field theory 
on AdS$_5$ ($\times$ some internal compact manifold),
and ignore instability of the background geometry. 
The probe photon in this toy-model gravity dual set-up will be the 
massless vector state of the bosonic string theory with the
wavefunction (\ref{eq:photon-wvfc}). The target hadron can be any 
scalar states, say, the tachyon, with the 
wavefunction (\ref{eq:hadron-wvfc}). We are to construct a 
toy-model amplitude of the $h+\gamma^{(*)} \rightarrow h+\gamma^{(*)}$ 
scattering, by using the 2-to-2 scattering of the massless photon and 
some scalar in the bosonic string on the AdS$_5$ background. 
Clearly one of the cost of this approach (without technical complexity
of interacting superstring field theory) is that we have to restrict our 
attention to the Reggeon exchange (flavor non-singlet) amplitude.  
The amplitude constructed in this way is certainly not faithful to 
the equations of the Type IIB string theory, either.
Since our motivation is not in constructing yet another 
exact solution to superstring theory, however, we still expect 
that this (flavor non-singlet) toy-mode amplitude in bosonic string 
still maintains some fragrance of hadron scattering amplitude to be 
calculated in superstring theory.

\section{Cubic String Field Theory}
\label{sec:cubic-sft}


Section \ref{ssec:cubic-action} summarizes technical details of 
cubic string field theory that we need in later sections and in 
part II \cite{NW-more}. We then move on in section \ref{ssec:Veneziano}
to explain an idea of how to reproduce disc amplitude only 
from string-field-theory $t$-channel amplitude, using photon--tachyon 
scattering on a flat spacetime background.  
This idea of constructing amplitude is generalized in 
part II \cite{NW-more} for scattering on a warped spacetime, and we will 
see that this construction of the amplitude allows us to 
cast the amplitude almost immediately into the form of conformal 
OPE (\ref{eq:conf-OPE-2side}, \ref{eq:conf-OPE-center}).  

\subsection{Action of the Cubic SFT on a Flat Spacetime}
\label{ssec:cubic-action}

The action of the cubic string field theory (cubic SFT) is given by
\cite{SFT-Witten-cubic}
\begin{eqnarray}
 S  & = &  - \frac{1}{2\alpha'} \int \left(
    \Phi * Q_B \Phi +  \frac{2}{3} g_o \Phi * \Phi * \Phi \right), 
 \label{eq:SFT-action} \\
& = & -\frac{1}{2\alpha'}\left(
  \Phi\cdot Q_B \Phi +\frac{2g_o}{3}\Phi \cdot \Phi * \Phi \right), 
 \label{eq:SFT-action-prev}
\end{eqnarray}
where $g_o$ is a coupling constant of mass dimension $(1-D/2)$, where 
$D=26$ is the spacetime dimensions of the bosonic string theory.
The string field $\Phi$ is, as a ket state, expanded in terms of the 
Fock states as in 
\begin{align}
 \Phi = |\Phi \rangle=&
\phi(x)|\downarrow\rangle 
+(A_M(x)\alpha^M_{-1}+C(x)b_{-1}+\bar C(x)c_{-1})|\downarrow\rangle 
 \notag
 \\
& + 
 \left(
 f_{MN}(x)\frac{1}{\sqrt{2}}\alpha_{-1}^M\alpha_{-1}^N
+ig_{M}(x)\frac{1}{\sqrt{2}}\alpha_{-2}^M
+h(x)b_{-1}c_{-1}+\cdots \right)|\downarrow\rangle, 
\end{align}
with component fields 
$\phi, A_M, C, \bar{C}, f_{MN}, g_M, h, \cdots$; we have already chosen the
Feynman--Siegel gauge here. We will eventually be interested only in 
the states with vanishing ghost number, $N_{\text{gh}} = 0$, because 
states with non-zero ghost number does not appear in the $t$-channel / 
$s$-channel exchange for the disc amplitude. 

The Hilbert space of one string state is spanned by the Fock states 
given (in this gauge) by 
\begin{align}
 \prod_{a=1}^{h_a} \alpha_{-n_a}^{M_a} \prod_{b=1}^{h_b} b_{-l_b}
 \prod_{c=1}^{h_c} c_{-m_c} |\downarrow\rangle, 
\label{eq:Fock-bos-string}
\end{align}
%
with $1\le  n_1\le n_2 \le \dots \le n_{h_a}$,
$1\le  l_1< l_2 < \dots < n_{h_b}$ and $1\le  m_1< m_2 < \dots < m_{h_c}$.
Let us use 
$Y := \left\{ \left\{ n_i \right\}\mbox{'s}, \; \left\{ l_i
\right\}\mbox{'s}, \; \left\{ m_i \right\}'s \right\}$ as the 
label distinguishing different Fock states of string on a flat spacetime.
%
%
Mass of these Fock states are determined by 
\begin{equation}
\alpha' k^2 + (N^{(Y)} - 1)  = 0, \qquad  
N^{(Y)} = \sum_{a=1}^{h_a} n_a + \sum_{b=1}^{h_b} l_b + \sum_{c=1}^{h_c} m_c.
\label{eq:level-bos-string}
\end{equation}
The corresponding component field of those Fock states may be decomposed 
into multiple irreducible representation of the Lorentz group, but at least, 
the rank-$h_a$ totally symmetric traceless tensor representation 
is always contained. 
%
%
%
%
Fock states of particular interest to us are the ones in the leading
trajectory: $Y = \left\{ 1^N, 0, 0 \right\}$, so that 
$N^{(Y)} = h_a$, $h_b = h_c = 0$ and all $n_a$'s are $1$. 
The totally symmetric traceless tensor component field of these
states are denoted by $(N !)^{-1/2} A^{(Y)}_{M_1 \cdots M_{h_a}}$. 

The kinetic term---the first term of (\ref{eq:SFT-action}, 
\ref{eq:SFT-action-prev})---is written down in terms of the component
fields as follows:
\begin{align}
 -\frac{1}{2\alpha'}\Phi\cdot Q_B \Phi&=
\frac{1}{2}\int d^{26}x
 \left[
\phi(x) \left(\partial^2+\frac{1}{\alpha'} \right) \phi(x)+
A_M(x)\partial^2A^M(x)+
\right. \notag
\\
&\left.
f_{MN}(x) \left(\partial^2-\frac{1}{\alpha'}\right)f^{MN}(x)+
g_M(x) \left(\partial^2-\frac{1}{\alpha'}\right) g^M(x)-
h(x) \left( \partial^2-\frac{1}{\alpha'} \right) h(x) + \cdots
\right].
\end{align}
%
%
The totally symmetric tensor component field of the Fock states in the 
leading trajectory $Y = \left\{ 1^N,0,0\right\}$ has a kinetic term
\begin{align}
 \frac{1}{2} \int d^{26}x A^{M_1\dots M_j}\left(\partial^2 -\frac{N-1}{\alpha '} \right) A_{M_1\dots M_j}.
\end{align}
The cubic string field theory action in the Feynman--Siegel gauge has 
two nice properties; first, the kinetic terms of those Fock states do 
not mix in the flat spacetime background, and second, 
the second derivative operators are simply given by d'Alembertian, 
without complicated restrictions or mixing among various polarizations 
in the component fields.

%

The second term of the action (\ref{eq:SFT-action},
\ref{eq:SFT-action-prev}) gives rise to interactions involving three 
component fields. Interactions involving Fock states with small $N$ 
are \cite{SFT-interaction}
\begin{align}
 - \frac{1}{2\alpha'}\frac{2g_o}{3}\Phi \cdot \Phi * \Phi= - 
 \int d^{26}x \frac{g_o \lambda}{3\alpha'}\hat E& \left(
  \tr \left[ \phi^3(x) \right]
  + \sqrt{\frac{8\alpha'}{3}} 
      \tr \left[ (-i A_M) \left(\phi \overleftrightarrow{\partial}^M
                                  \phi  \right)\right]   \right.
  \notag  \\
& \quad \left.  -\frac{8\alpha'}{9\sqrt{2}}
    \tr \left[ f_{MN}\left( \phi \overleftrightarrow{\partial}^M
                   \overleftrightarrow{\partial}^N \phi \right) \right]
-\frac{5}{9\sqrt{2}}\tr \left[f^M_M\phi^2 \right]
\right.
\notag
\\&\quad +
\left.
  \frac{2\sqrt{\alpha'}}{3}\tr \left[ (\partial_Mg^M)\phi^2 \right] 
 -\frac{11}{9} \tr \left[h \phi^2 \right]
\right) + \cdots,
\label{eq:cubicSFT-interaction-A}
\end{align}
where $\lambda=3^{9/2}/2^6$ \cite{SFT-Hata-lec.note}, 
$\overleftrightarrow{\partial^M}=({\overleftarrow{\partial^M}-\overrightarrow{\partial^M}})$,
and
\begin{align}
 \hat E = \exp\left[-2\alpha'\ln\sfrac{2}{3^{3/4}}(\partial_{(1)}^2+\partial_{(2)}^2+\partial_{(3)}^2)\right].
\end{align}
The $\partial^2_{(1,2,3)}$ means derivatives for the 1st, 2nd, and 3rd field\footnote{
Concretely,
\begin{align}
 \hat E A(x)B(x)C(x) =
 \left(\exp\left[-2\alpha'\ln\sfrac{2}{3^{3/4}}\partial^2\right]A(x)\right)
 \left(\exp\left[-2\alpha'\ln\sfrac{2}{3^{3/4}}\partial^2\right]B(x)\right)
 \left(\exp\left[-2\alpha'\ln\sfrac{2}{3^{3/4}}\partial^2\right]C(x)\right)
\notag.
\end{align}
}. 

Interactions involving totally symmetric leading trajectory states are 
also of interest to us. 
The tachyon--tachyon--$Y=\{1^N,0,0\}$ cubic coupling with
$N$-derivatives is given by 
\begin{equation}
 - \frac{g_o \lambda}{\alpha'} \int d^{26}x \; \hat{E} \; 
  {\rm Tr} \left[ A^{(Y)}_{m_1 \cdots m_N}
                   \left( \phi (-i \overleftrightarrow{\partial}^{m_1})
                         \cdots (-i \overleftrightarrow{\partial}^{m_N})      
                          \phi \right) 
      	   \right] 
  \left( \frac{8\alpha'}{27} \right)^{\frac{N}{2}}   
  \frac{1}{\sqrt{N!}}
\label{eq:cubicSFT-interaction-B}
\end{equation}
%
in the interaction part of the action. 
The photon (the level-1 state)--photon--$Y=\{1^N,0,0 \}$ coupling 
in the cubic string field theory includes  
\begin{eqnarray}
&& - \frac{g_o \lambda}{\alpha'} \int d^{26}x \; \hat{E} \; 
  {\rm Tr} \left[ A^{(N)}_{m_1 \cdots m_N}
                  \left( A_l (-i \overleftrightarrow{\partial}^{m_1})
                         \cdots (-i \overleftrightarrow{\partial}^{m_N})
                         A_k \right)
  \left( \frac{8\alpha'}{27} \right)^{\frac{N}{2}}   
      \frac{\eta^{kl} \frac{16}{27}}{\sqrt{N!}}+ \cdots 
  \right],
\label{eq:cubicSFT-interaction-C}
\end{eqnarray}
where we kept only the terms that have $N$-derivatives and are
proportional to $\eta^{kl}$, as they are necessary in deriving 
(\ref{eq:tch-leading-tp}).

%
%

\subsection{Cubic SFT Scattering Amplitude and $t$-Channel Expansion}
\label{ssec:Veneziano}

Before proceeding to study the 
$h + \gamma^{(*)} \rightarrow h + \gamma$ scattering amplitude 
by using the cubic string field theory on the warped spacetime
background, let us remind ourselves how to obtain $t$-channel 
operator product expansion from the the amplitude calculation 
based on string field theory, by using tachyon--photon scattering 
on the flat spacetime as an example.   

\begin{figure}[tbp]
  \begin{center}
\begin{tabular}{cccc}
 \includegraphics[scale=0.35]{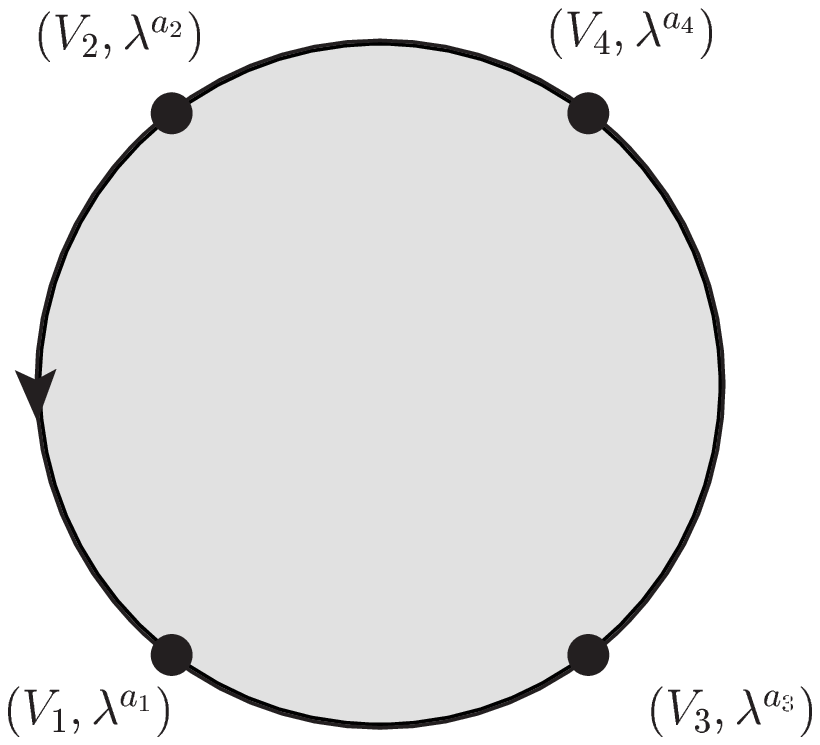} &
 \includegraphics[scale=0.35]{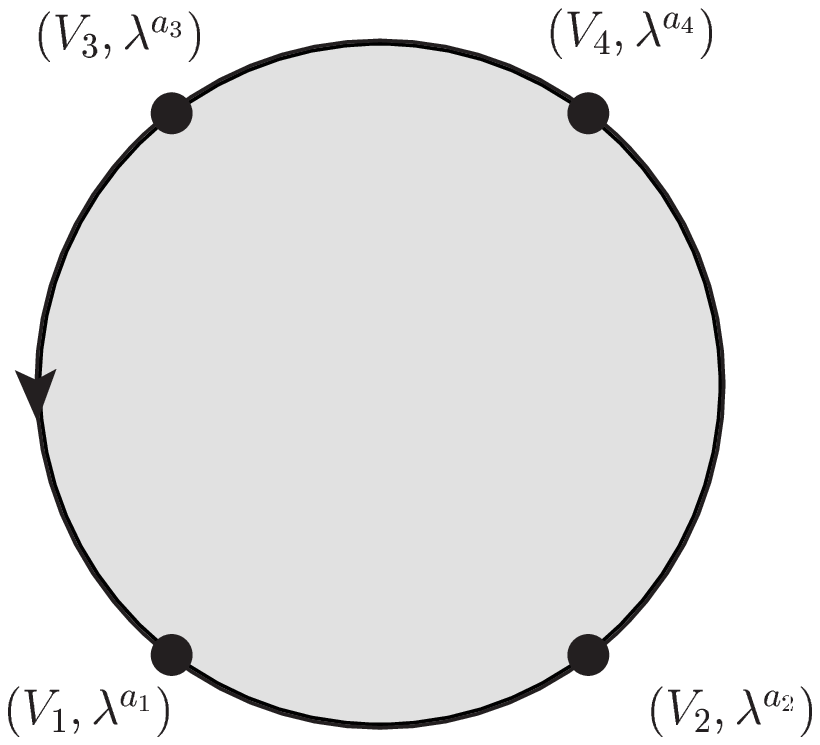} &
 \includegraphics[scale=0.35]{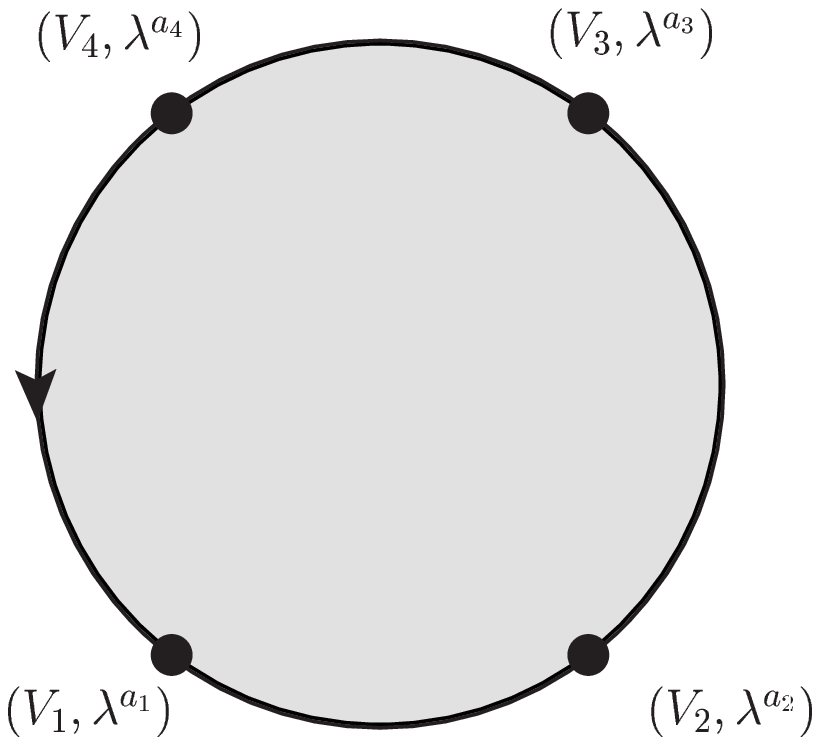} &
 \includegraphics[scale=0.35]{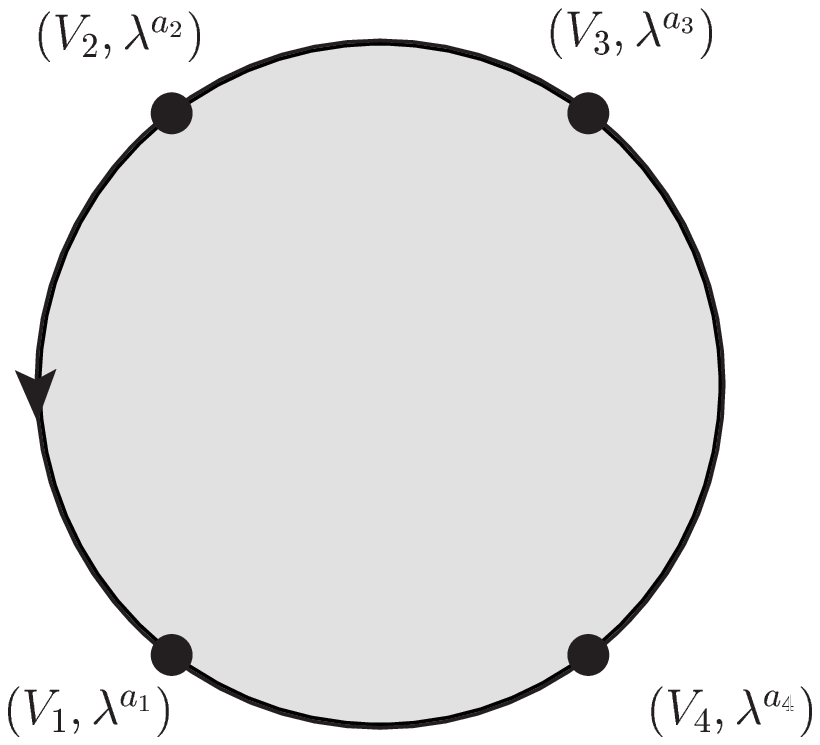}   \\
(a) 1342 & (b) 1243 & (c) 1234 & (d) 1432
\end{tabular}
\caption{Disc amplitudes with two photon vertex operators ($V_1$ and
   $V_2$) and two tachyon vertex operators ($V_3$ and $V_4$) inserted. 
Kinematical amplitudes given by the disc amplitudes above are multiplied 
by the Chan--Paton factors 
$\tr [\lambda^{a_1} \lambda^{a_3} \lambda^{a_4} \lambda^{a_2}]$ in (a), 
$\tr [\lambda^{a_1} \lambda^{a_2} \lambda^{a_4} \lambda^{a_3}]$ in (b), 
$\tr [\lambda^{a_1} \lambda^{a_2} \lambda^{a_3} \lambda^{a_4}]$ in (c) 
and $\tr [\lambda^{a_1} \lambda^{a_4} \lambda^{a_3} \lambda^{a_2}]$ in
(d), respectively. The two disc amplitudes (a, b) become 
${\cal M}_{\rm Ven}(s,t)$, while (c, d) ${\cal M}_{\rm Ven}(u,t)$.}
\label{fig:disc-tp-tp}
  \end{center}
\end{figure}

Let us consider the disc amplitude of tachyon--photon scattering. 
The vertex operators labeled by $i=1,2$,
$V_i = :\epsilon^{i}_\mu \partial X^\mu e^{i k_i \cdot X }:$, are for photon
incoming ($i=1$) and outgoing states ($i=2$), which come with 
Chan--Paton matrices $\lambda^{a_i}$.
Tachyon incoming $(i=3)$ and outgoing $(i=4)$ states correspond to 
vertex operators $V_i = : e^{i k_i \cdot X} :$ with Chan--Paton matrices 
$\lambda^{a_i}$. The photon--tachyon scattering amplitude  
$A + \phi \rightarrow A + \phi$ in bosonic open 
string theory (Veneziano amplitude) is given by\footnote{
Here, $p := (k_3 - k_4)/2$, averaged momentum of tachyon before and after
the scattering, just like in (\ref{kinematics1}). } 
\begin{eqnarray}
{\cal M}_{\rm Ven}(s,t) & = & - \left(\frac{g_o^2}{\alpha'}\right) 
    \frac{\Gamma(-\alpha' t-1) \Gamma(-\alpha' s-1)}
      {\Gamma(-\alpha'(s+t) -1)}
   \epsilon_{\mu}(k_2) \epsilon_{\nu}(k_1)  \nonumber \\
 & & \times \left\{
  \left[ \eta^{\mu\nu} - \frac{k_1^\mu k_2^\nu}{k_1 \cdot k_2} \right]
  (\alpha' s + 1)  \right.  \label{eq:Veneziano-amplitude-tp}\\
 & & \left. \quad + 2\alpha' 
  \left( \left[p^\mu - k_1^\mu\frac{k_2 \cdot p}{k_1 \cdot k_2}\right]
       - \frac{k_2^\mu}{2} \right)
  \left( \left[p^\nu - k_2^\nu\frac{k_1 \cdot p}{k_2 \cdot k_1}\right]
       - \frac{k_1^\nu}{2} \right)(\alpha' t+1) 
  \right\},    \nonumber 
\end{eqnarray}
which is to be multiplied by the Chan--Paton factor 
${\rm Tr}\left[\lambda^{a_2}\lambda^{a_4}\lambda^{a_3}\lambda^{a_1} 
 +\lambda^{a_4}\lambda^{a_2}\lambda^{a_1}\lambda^{a_3}\right]$.
(see Figure \ref{fig:disc-tp-tp}~(a,~b).)
If the Chan--Paton matrices of a pair of incoming and outgoing vertex
operators, $\lambda^{a_1}$ and $\lambda^{a_2}$, commute with each
other,\footnote{Just like in the case both $\lambda^{a_1}$ and
$\lambda^{a_2 }$ are an $N_F \times N_F$ matrix 
$\diag(2/3, -1/3,-1/3)$.} 
then the Chan--Paton factors from the diagrams 
Figure~\ref{fig:disc-tp-tp}~(c,~d) are the same, and the total 
kinematical part of the amplitude for this Chan--Paton factor becomes 
${\cal M}_{\rm Ven}(s,t) + {\cal M}_{\rm Ven}(u,t)$.

Let us stay focused on ${\cal M}_{\rm Ven}(s,t)$ alone for now. 
The amplitude proportional to $\eta^{\mu\nu}$ can be expanded, as 
is well-known, as a sum only of $t$-channel poles:\footnote{It is also
possible to expand this as a sum only of $s$-channel poles; that's the
celebrated $s$-$t$ duality of the Veneziano amplitude. } 
\begin{eqnarray}
 \frac{g_o^2}{\alpha'}
 \frac{\Gamma(-\alpha' t-1) \Gamma(-\alpha' s)}
      {\Gamma(-\alpha'(s+t)-1)} & = &
 \frac{g_o^2}{\alpha'} \int_0^1 dx \;
  x^{-\alpha' t-2}(1-x)^{-\alpha's-1},
 \label{eq:Veneziano-str-fcn-tp}
 \\
 & = &  \frac{g_o^2}{\alpha'}
  \sum_{N=0}^{\infty} \frac{-1}{\alpha' t - (N-1)} 
   \frac{(\alpha's+1)\cdots (\alpha' s+N)}{N!}.
\label{eq:pole-expansion-of-Veneziano-Amplitude-tp}
\end{eqnarray}
%

%

\begin{figure}[tbp]
  \begin{center}
\begin{tabular}{ccc}
 \includegraphics[scale=0.5]{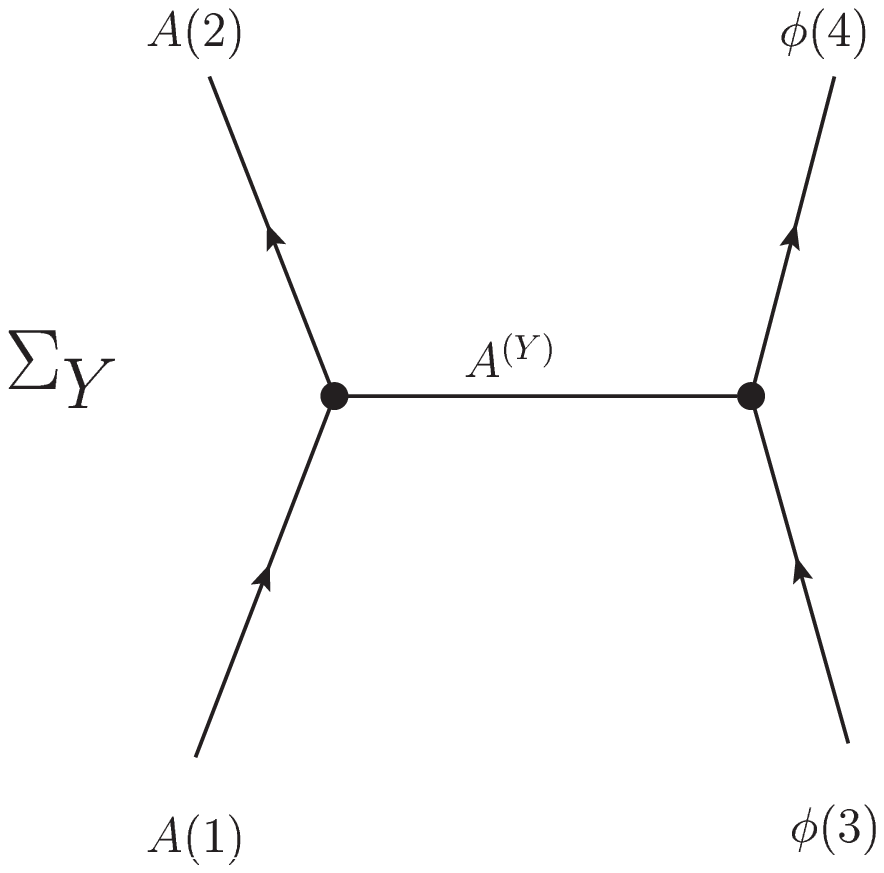} & $\qquad$ &
 \includegraphics[scale=0.5]{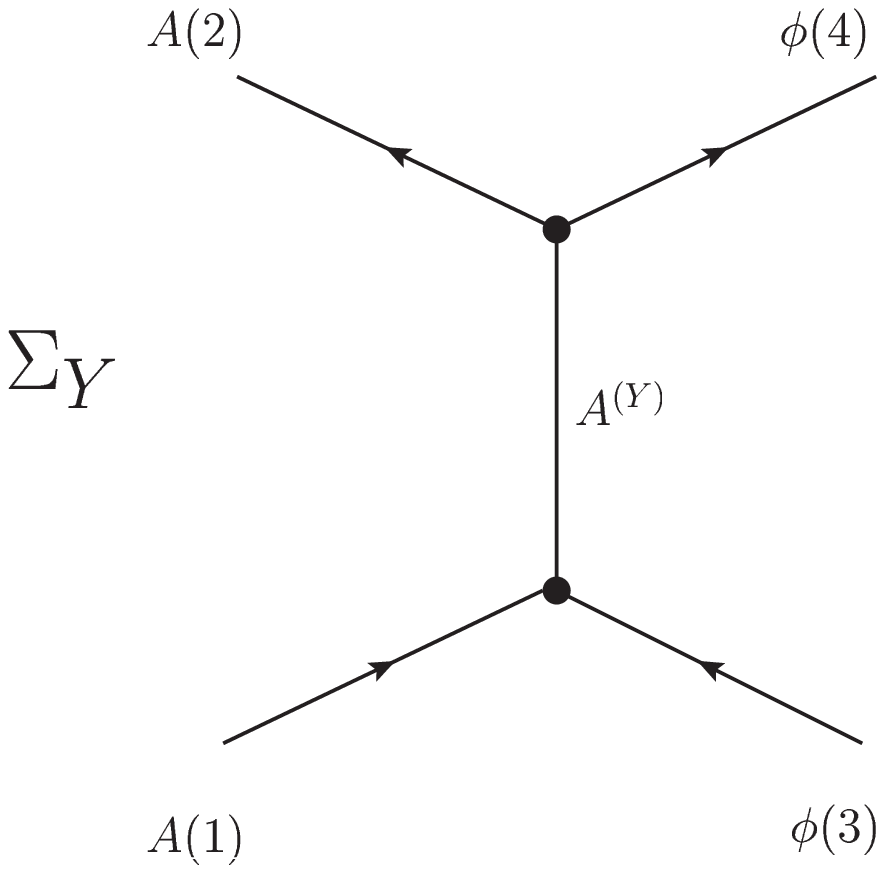} \\
\end{tabular}
\caption{\label{fig:SFT-TandS}Two types of diagrams contribute to the 
photon--tachyon scattering amplitude ${\cal M}_{\rm Ven}(s,t)$ 
in cubic string field theory: 
$t$-channel exchange of one string states labeled by $Y$ (left), 
and $s$-channel exchange (right). }
  \end{center}
\end{figure}

The Veneziano amplitude (\ref{eq:Veneziano-amplitude-tp}) can also be 
obtained in cubic string field theory \cite{SFT-Giddings-Veneziano}.
In the cubic SFT, the scattering amplitude consists of two pieces, 
a collection of $t$-channel exchange diagrams and that of $s$-channel 
diagrams (Figure~\ref{fig:SFT-TandS}).
\begin{align}
 &{\cal M}_\text{Ven}(s,t)= \sum_Y {\cal M}^{(t)}_Y(s,t)
                          + \sum_Y {\cal M}^{(s)}_y(s,t).
\label{eq:Veneziano-in-SFT}
\end{align}
%
%
Infinitely many one string states (\ref{eq:Fock-bos-string}) with zero 
ghost number ($h_b = h_c$)---labeled by $Y$---can be exchanged 
in the $t$-channel or in the $s$-channel, and the corresponding 
contributions are in the form of 
\begin{align}
&{\cal M}^{(t)}_Y= \frac{f^{(t)}_Y(s,t)}{-\alpha't-1+N^{(Y)}}, \quad
{\cal M}^{(s)}_Y= \frac{f^{(s)}_Y(t,s)}{-\alpha's-1+N^{(Y)}},
\label{eq:pole-SandT}
\end{align}
where $f^{(t)}_Y$ and $f^{(s)}_Y$ are regular function at finite $s,t$, 
and $N^{(Y)}$ is the excitation level (\ref{eq:level-bos-string}) of 
a component field $A^{(Y)}$.

Because both the world-sheet calculation (\ref{eq:Veneziano-amplitude-tp}, 
\ref{eq:pole-expansion-of-Veneziano-Amplitude-tp}) and the cubic SFT
calculation (\ref{eq:Veneziano-in-SFT}, \ref{eq:pole-SandT}) are the
same thing, ${\cal M}_{\rm Ven}(s,t)$ in both approaches should be completely 
the same functions of $(s,t)$. 
Therefore, for an arbitrary given value 
of $s$, the residue of all the poles in the complex $t$-plane should be
the same. We also know that the Veneziano amplitude can be expanded
purely in the infinite sum of $t$-channel poles with $t$-independent 
residues. This means that the full Veneziano 
amplitude (\ref{eq:Veneziano-amplitude-tp}) can be reproduced just from 
the $t$-channel cubic SFT amplitude\footnote{The $t$-channel and
$s$-channel amplitudes of the cubic SFT, $\sum_Y {\cal M}_Y^{(t)}$ and 
$\sum_Y {\cal M}_Y^{(s)}$ correspond to the integration over $[0, 1/2]$
and $[1/2, 1]$ in (\ref{eq:Veneziano-str-fcn-tp}), 
respectively \cite{SFT-Giddings-Veneziano}. 
Thus, $\sum_Y {\cal M}_Y^{(s)}$ does not contain a pole in $t$.} 
$\sum_Y {\cal M}_Y^{(t)}(s,t)$, by the following procedure:
\begin{equation}
 \sum_Y \frac{f^{(t)}_Y(s,t)}{-\alpha' t - 1 + N^{(Y)}} \longrightarrow 
 \sum_Y \frac{f^{(t)}_Y(s, (N^{(Y)}-1)/\alpha')}{-\alpha' t-1+N^{(Y)}} = 
  {\cal M}_{\rm Ven}(s,t).
  \label{eq:SFT-Veneziano-prescr}
\end{equation}
%

\begin{figure}[tbp]
  \begin{center}
\begin{tabular}{cccc}
 \includegraphics[scale=0.4]{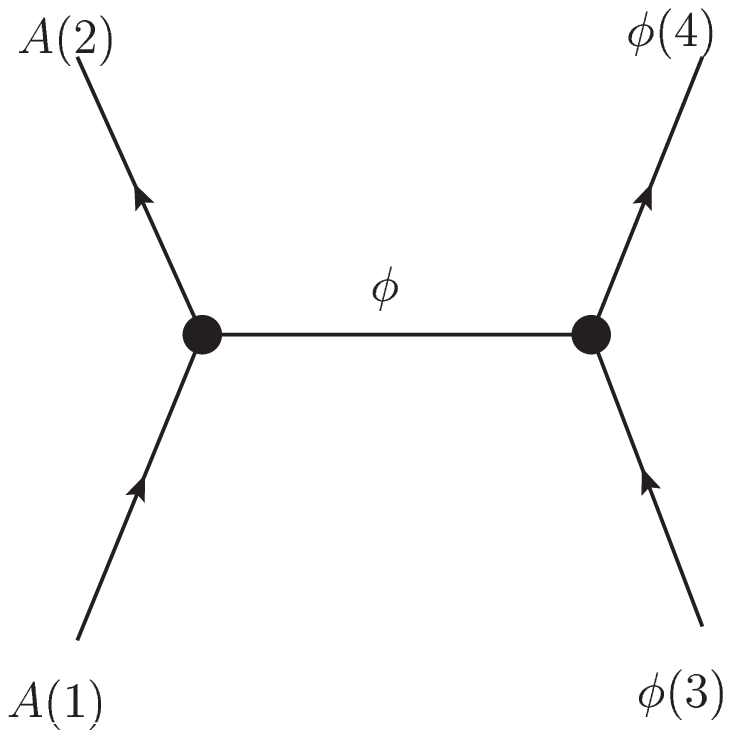} &
 \includegraphics[scale=0.4]{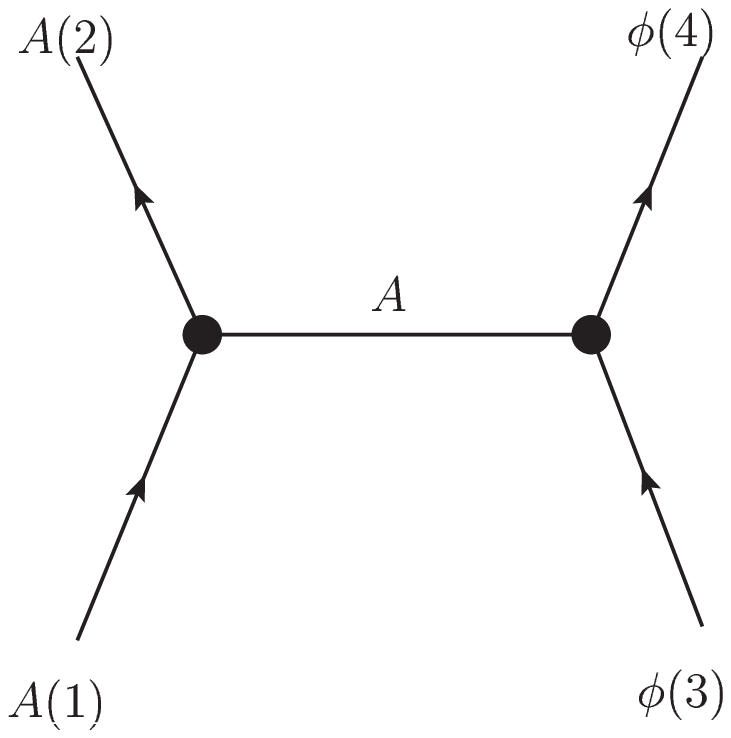} &
 \includegraphics[scale=0.4]{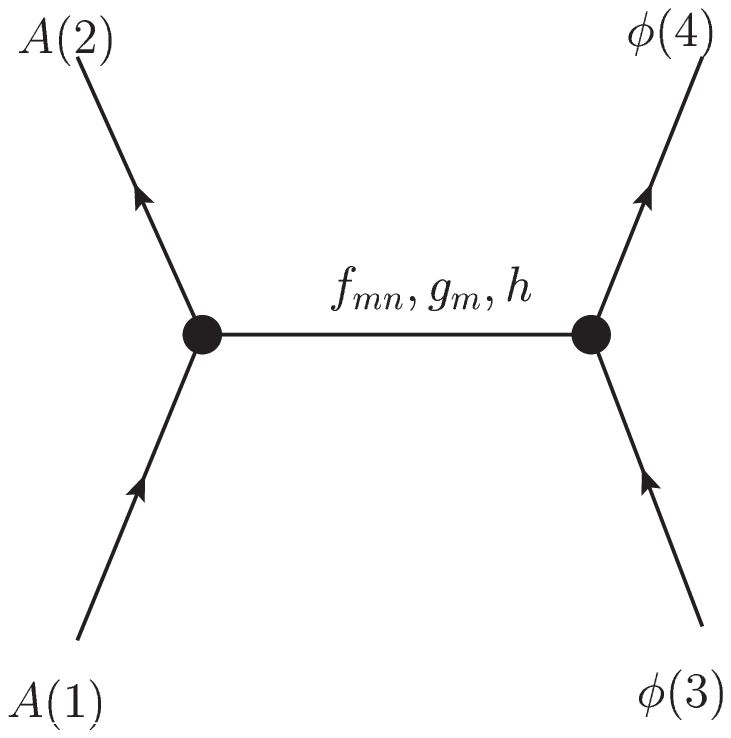} \\
 (a) & (b) & (c) 
\end{tabular}
\caption{\label{fig:SFT-tch-diag} $t$-channel exchange diagrams for 
$A + \phi \rightarrow A + \phi$ scattering in the cubic string field
   theory. The tachyon ($N=0$), photon ($N=1$) and level-2 states are 
exchanged in the diagrams (a), (b) and (c), respectively.}
  \end{center}
\end{figure}

To see that this prescription really works, let us take a look at 
the amplitudes of $t$-channel exchange of one string states with 
small excitation level $N^{(Y)}=0,1,2$. Focusing on the 
amplitude of $A + \phi \rightarrow A + \phi$ proportional to 
$\eta^{\mu\nu}$, we find that the tachyon exchange in the $t$-channel 
(Figure~\ref{fig:SFT-tch-diag} (a)) gives rise to the amplitude 
\cite{SFT-Thorn}
\begin{align}
 {\cal M}_{\phi}^{(t)}(s,t) =
\left(\frac{g_o \lambda}{\alpha'}\right)^2
\left(\frac{2}{3^{3/4}}\right)^{-2\alpha' t -2\alpha' t + 4}
\frac{-1}{t+1/\alpha'} = 
  \frac{g_o^2}{\alpha'}
  \left( \frac{27}{16} \right)^{\alpha' t+1}
  \frac{-1}{\alpha' t+1},
\end{align}
which is obtained simply by using the $\phi$-$\phi$-$\phi$ vertex rule 
(\ref{eq:cubicSFT-interaction-A}) and $A$-$A$-$\phi$ vertex rule 
(\ref{eq:cubicSFT-interaction-B}). 
The prescription (\ref{eq:SFT-Veneziano-prescr}) turns this amplitude 
into 
\begin{align}
 \longrightarrow
 {\cal M}_\phi(s,t)  = \frac{g_o^2}{\alpha'}\frac{-1}{\alpha' t+1},
\end{align}
which reproduces the $N=0$ term of
(\ref{eq:pole-expansion-of-Veneziano-Amplitude-tp}). 

The $t$-channel exchange of level $N^{(Y)} = 1$ excited states can also 
be calculated in the cubic string field theory 
(Figure~\ref{fig:SFT-tch-diag}~(b)). 
The amplitude proportional to $\eta^{\mu\nu}$ is 
\begin{equation}
 {\cal M}^{(t)}_A(s,t)  = 
  \frac{g_o^2}{\alpha'}\left(\frac{27}{16}\right)^{\alpha't}
  \frac{-1}{\alpha't} \; 
  \left[ \frac{\alpha'(s-u)}{2} \right], 
\end{equation}
where $(s-u) = (k_{(1)} - k_{(2)}) \cdot (k_{(4)} - k_{(3)})$. 
Using the relation $\alpha' (s+t+u) = -2$ in the tachyon--photon 
scattering to eliminate $u$ in favor of $s$ and $t$, and following 
the prescription (\ref{eq:SFT-Veneziano-prescr})---which is to 
exploit $\alpha' t = 0$ in the numerator, 
this amplitude is replaced by \cite{SFT-Thorn}
\begin{equation}
\longrightarrow 
{\cal M}_A(s,t) = \frac{g_o^2}{\alpha'} 
   \frac{-(\alpha' s + 1)}{\alpha' t}.
\end{equation}
Once again, this reproduces the level $N=1$ contribution to the 
Veneziano amplitude (\ref{eq:pole-expansion-of-Veneziano-Amplitude-tp}).

Similar calculation for level-2 state exchange can be carried out 
(Figure~\ref{fig:SFT-tch-diag}~(c)). 
Using the vertex rule in (\ref{eq:cubicSFT-interaction-A}) 
for the [level-2]-$\phi$-$\phi$ couplings, and also the 
interactions among [level-2]-$A$-$A$ coupling in the literature, 
the cubic SFT $t$-channel amplitude is given by \cite{SFT-Thorn}
\begin{eqnarray}
 {\cal M}_f^{(t)}(s,t) & = &
  \frac{g_o^2}{\alpha'}\left(\frac{27}{16}\right)^{\alpha't-1}
  \frac{-1}{\alpha't-1} \;
  \left[ \frac{ (\alpha'(s-u))^2 }{8}
       - \frac{5(\alpha' t + 2)}{16 \cdot 2}
       + \frac{ 490 }{ 16^2 \cdot 2} \right],  \\
 {\cal M}_g^{(t)}(s,t) & = &
  \frac{g_o^2}{\alpha'}\left(\frac{27}{16}\right)^{\alpha't-1}
  \frac{-1}{\alpha't-1} \;
  \left[ - \frac{36 \; \alpha' t} {16^2} \right] ,  \\
 {\cal M}_h^{(t)}(s,t) & = &
  \frac{g_o^2}{\alpha'}\left(\frac{27}{16}\right)^{\alpha't-1}
  \frac{-1}{\alpha't-1} \; 
  \left[ - \frac{11^2}{16^2} \right]. 
\end{eqnarray}
After using $\alpha'u = - \alpha'(s+t) - 2$ to eliminate $u$ in favor of
$s$ and $t$, and further following the prescription
(\ref{eq:SFT-Veneziano-prescr}) [$\alpha' t \rightarrow 1$ in the 
numerator], one will see that the level $N^{(Y)} = 2$ amplitude turns into 
\begin{align}
\longrightarrow 
 \left( {\cal M}_f + {\cal M}_g + {\cal M}_h \right)(s,t) =
\frac{g_o^2}{\alpha'}\frac{-1}{\alpha' t-1}\left[
\frac{(\alpha's)^2+3(\alpha' s)+2}{2} \right].
\end{align}
Once again, this is precisely the same as the $N=2$ contribution 
to the Veneziano amplitude (\ref{eq:pole-expansion-of-Veneziano-Amplitude-tp}).

Contributions from the $t$-channel exchange of states in the leading
trajectory can also be examined systematically. 
Using the vertex rule (\ref{eq:cubicSFT-interaction-B}, 
\ref{eq:cubicSFT-interaction-C}) involving the states in the leading
trajectory ($Y=\left\{1^N,0,0 \right\}$), one finds that the amplitude 
proportional to $\eta^{\mu\nu}$ is   
\begin{equation}
 {\cal M}_{\left\{1^N,0,0 \right\} }^{(t)} \simeq \frac{g_o^2}{\alpha'}
   \left(\frac{27}{16}\right)^{\alpha' t -(N-1)}
   \frac{-1}{\alpha' t-(N-1)} 
   \frac{\left(\alpha' (s-u)/2 \right)^N}{N!},
\label{eq:tch-leading-tp}
\end{equation}
where we maintained only the terms with the highest power of either $s$
or $u$.
After using the kinematical relation $\alpha'(s+t+u)+2 = 0$ to eliminate 
$u$ in favor of $s$ and $t$, and following the prescription
(\ref{eq:SFT-Veneziano-prescr}) [$\alpha' t \rightarrow (N-1)$ 
in the numerator], we obtain the large-$\alpha's$ leading power contribution to
the $N$-th term of (\ref{eq:pole-expansion-of-Veneziano-Amplitude-tp}) 
precisely with the correct coefficient. 

We have therefore seen that the prescription (\ref{eq:SFT-Veneziano-prescr})
allows us to use the $t$-channel exchange amplitude in the cubic 
string field theory to construct the full disc scattering amplitude.
In part II \cite{NW-more}, this prescription is extended for the 
disc scattering amplitudes on a spacetime with curved background metric, 
which is the situation of real interest in the context of hadron scattering. 

\section{Mode Decomposition on AdS$_5$}
\label{sec:mode-fcn}

Let us now proceed to work out mode decomposition of 
the totally symmetric (traceless) component field on the warped
spacetime. The correspondence between the primary operators of the 
conformal field theory on the (UV) boundary and wavefunctions on 
AdS$_5$ is made clear. The Pomeron/Reggeon wavefunctions are obtained 
as a {\it holomorphic} function of the spin variable $j$, as we need to 
do so for the further inverse Mellin transformation.
The wavefunction will then be used also to construct 
the scattering amplitude of $h+\gamma^{(*)} \rightarrow h+\gamma$ 
in our future publication \cite{NW-more}.

Let the bilinear (free) part of the (bulk) action of a rank-$j$ tensor 
field on $AdS_5$ to be\footnote{The dimensionless constant $t_y$ 
is something like $1/N_c^2$ for a mode obtained by reduction of closed 
string component fields in higher dimensions. }
\begin{eqnarray}
 S & = & - \frac{1}{2} \frac{t_y}{R^3}\int d^4 x \int dz \sqrt{-g(z)} 
  g^{m_1 n_1} \cdots g^{m_j n_j} \nonumber \\
  & & \qquad \qquad 
  \left[ 
    g^{m_0 n_0} (\nabla_{m_0} A^{(y)}_{m_1 \cdots m_j})
                (\nabla_{n_0} A^{(y)}_{n_1 \cdots n_j}) + 
    \left(\frac{c_y}{R^2} + \frac{N^{(y)}_{\rm eff.}}{\alpha'}\right) 
       A^{(y)}_{m_1 \cdots m_j} A^{(y)}_{n_1 \cdots n_j}
  \right],  
\end{eqnarray}
where we assume that kinetic mixing between different fields is either 
absent or sufficiently small.
Here, the dimensionless parameter $N^{(y)}_{\rm eff.}$ is 
$N^{(Y)}-1$ for an $N^{(Y)} \in {\bf Z}_{\geq 0}$ for bosonic open string 
($j \leq N^{(Y)}$), which would be $4(N^{(Y)}-1)$ for an 
$N^{(Y)} \in {\bf Z}_{\geq 1}$ for closed string ($j \leq 2N^{(Y)}$).
This field is regarded as a reduction of some field with some 
``spherical harmonics'' on the internal manifold,\footnote{The internal 
manifold would be a five-dimensional one, $W$, for
closed string modes in Type IIB, and a three-cycle for open string
states on the flavor D7-branes. For sufficiently small $x$, however, 
amplitudes of exchanging modes with non-trivial ``spherical harmonics'' 
on these internal manifolds are relatively suppressed, and we are not 
so much interested.} and hence $j \leq h_a$ in general. 
Another dimensionless coefficient $c_y$ may contain a contribution from the 
``mass'' associated with the ``spherical harmonics'' over the internal 
manifold, and also include ambiguity (which is presumably of order
unity) associated with making d'Alembertian of the flat metric
background covariant. 

The equation of motion (in the bulk part)\footnote{Conditions need 
to be imposed on the (IR) boundary part of the action as well in the 
hard-wall implementation of the confinement effect, but this issue 
will be discussed elsewhere.} then becomes
\begin{equation}
 g^{m_1 m_2} (\nabla_{m_1} \nabla_{m_2} A^{(y)}_{n_1 \cdots n_j} )
  - \left(\frac{c_y}{R^2}+ \frac{N^{(y)}_{\rm eff.}}{\alpha'}\right)
    A^{(y)}_{n_1 \cdots n_j} = 0. 
\end{equation}
Solutions to this equation of motion can be obtained from 
solutions of the following eigenmode equation\footnote{The 
differential operator $\nabla^2 := g^{mn} \nabla_m \nabla_n$ 
is Hermitian under the measure 
$d^4x  dz \sqrt{-g(z)} g^{m_1n_1} \cdots g^{m_j n_j}$. } 
\begin{equation}
 \nabla^2 A_{m_1 \cdots m_j} =
   - \frac{{\cal E}}{R^2} A_{m_1 \cdots m_j},
  \label{eq:Eigenequation}
\end{equation}
by imposing the on-shell condition 
\begin{equation}
 \frac{({\cal E} +c_y)}{\sqrt{\lambda}}+ N^{(y)}_{\rm eff.} = 0.
\label{eq:mass-shell}
\end{equation}
We will work out the eigenmode decomposition for rank-$j$ tensor 
fields in the following, where we only have to work for separate $j$, 
without referring to the mass parameter.\footnote{There are many states 
with the same value of $j$, but with different $c_y$ and $N^{(y)}_{\rm eff.}$.} 

The eigenmode wavefunctions are used not just for constructing 
solutions to the equation of motions, but also in constructing 
the Reggeon exchange contributions to the 
$h+\gamma^{*} \rightarrow h+\gamma^{(*)}$ scattering amplitude. 
The propagator is proportional to 
\begin{equation}
 \frac{-i}{\frac{ {\cal E} +c_y }{\sqrt{\lambda}}
                  + N^{(y)}_{\rm eff.} - i\epsilon}
  \frac{\alpha' R^3}{t_y}.
\label{eq:propagator-idea}
\end{equation}
%


The mode equation for a rank-$j$ tensor field $A_{m_1 \cdots m_j}$ on 
AdS$_5$ is further decomposed into those of irreducible representations of
$SO(4,1)$. For simplicity of the argument, we only deal with the mode 
equations for the totally symmetric (and traceless) rank$-j$ tensor fields. 
Namely, 
\begin{equation}
 A_{m_1 \cdots m_j} = A_{m_{\sigma(1)} \cdots m_{\sigma(j)}} \qquad
  \qquad  {\rm for~}{}^{\forall} \sigma \in \mathfrak{S}_j.
\end{equation}
We call them spin-$j$ fields. 

The eigenmode equation (\ref{eq:Eigenequation}) for a totally symmetric 
spin $j$ field can be decomposed into $j+1$ pieces, labeled by
$k=0, \cdots, j$:
\begin{eqnarray}
&& \left( (R^2 \Delta_j) 
- \left[(2k+1)j-2k^2+3k\right] \right) A_{z^k \mu_1 \cdots \mu_{j-k} } 
   \nonumber  \\ 
&& + 2zk \partial^{\hat{\rho}} A_{z^{k-1}\rho\mu_1\cdots \mu_{j-k}}
+ k(k-1) A^{\hat{\rho}}_{z^{k-2}\rho \mu_1 \cdots \mu_{j-k} } 
   \nonumber \\
&& - 2z (D[A_{z^{k+1} \cdots}])_{\mu_1 \cdots \mu_{j-k}}
+ (E[A_{z^{k+2} \cdots}])_{\mu_1 \cdots \mu_{j-k}} = -{\cal E} A_{z^k \mu_1
\cdots \mu_{j-k}}.  
\label{eq:Eigenequation-4+1}
\end{eqnarray}
Here, 
\begin{equation}
A_{z^k \mu_1 \cdots \mu_{j-k}} := A_{\underbrace{z \cdots z}_k \mu_1
 \cdots \mu_{j-k}},   
\end{equation}
and can be regarded as a rank-$(j-k)$ totally symmetric tensor of 
$\SO(3,1)$ Lorentz group.\footnote{
The $\SO(3,1)$ indices with $\hat{}$ in the superscript, 
such as ${}^{\hat{\rho}}$ in $\partial^{\hat{\rho}}$, 
are raised by the 4D Minkowski metric $\eta^{\rho\sigma}$ from 
a subscript ${}_\sigma$, not by the 5D warped metric $g^{mn}$.
} $D[ a ]$ and $E[ a ]$ are operations 
creating totally symmetric rank-$(r+1)$ and rank-$(r+2)$ tensors of 
$\SO(3,1)$, respectively, from a totally symmetric rank-$r$ tensor 
of $\SO(3,1)$, $a$; 
\begin{eqnarray}
 \left(D[ a ]\right)_{\mu_1 \cdots \mu_{r+1}} & := & \sum_{i=1}^{r+1}
   \partial_{\mu_i} a_{\mu_1 \cdots \check{\mu_i} \cdots
   \mu_{r+1}}, 
\label{eq:def-D} \\ 
 \left(E[ a ]\right)_{\mu_1 \cdots \mu_{r+2}} & := & 
  2 \sum_{p < q} \eta_{\mu_p \mu_q}
  a_{\mu_1 \cdots \check{\mu_p} \cdots \check{\mu_q} \cdots
  \mu_{r+2}}.
\label{eq:def-E}
\end{eqnarray} 
The differential operator $\Delta_j$ in the first term is defined, 
as in \cite{BPST-06}, by 
\begin{eqnarray}
 R^2 \Delta_j & := & R^2 z^{-j} \left[ 
    \left(\frac{z}{R}\right)^5 \partial_z 
        \left[ \left(\frac{R}{z}\right)^3 \partial_z \right] 
                           \right] z^j
  + R^2 \left(\frac{z}{R}\right)^2 \partial^2, \nonumber \\  
  & = & z^2 \partial_z^2 + (2j-3)z\partial_z +j(j-4) + z^2 \partial^2.
\end{eqnarray}
The eigenmode equation (\ref{eq:Eigenequation},
\ref{eq:Eigenequation-4+1}) is a generalization of the 
``Schr\"{o}dinger equation'' of \cite{BPST-06} determining the Pomeron 
wavefunction. As we will see, the single-component Pomeron wavefunction 
discussed in \cite{BPST-06} etc. corresponds 
to (\ref{eq:Pomeron-wvfc-1compnt})---that of $(n,l,m) = (0,0,0)$ 
eigenmode in our language, and the Schr\"{o}dinger equation to
(\ref{eq:EGeq-q=not0-n=0}, \ref{eq:EGeq-q=0-n=0-app}); 
there are other eigenmodes, whose
wavefunctions are to be determined in the following.  

In the following sections \ref{ssec:eigen-q=0}--\ref{ssec:eigen-q=not0}, 
we simply state the results of the eigenmode decomposition of
(\ref{eq:Eigenequation}, \ref{eq:Eigenequation-4+1}) for spin-$j$
fields. More detailed account is given in the 
appendix \ref{sec:appendix-mode-fcn}.

\subsection{Eigenvalues and Eigenmodes for $\Delta^\mu = 0$}
\label{ssec:eigen-q=0}

Because of the 3+1-dimensional translational symmetry in $\nabla^2$, 
solutions to the eigenmode equations can be classified by the
eigenvalues of the generators of translation, $(-i\partial^\mu)$.
Until the end of section \ref{ssec:eigen-q=not0}, we will focus on eigenmodes 
in the form of 
\begin{equation}
 A_{m_1 \cdots m_j}(x,z) = e^{i \Delta \cdot x} A_{m_1 \cdots m_j}(z;\Delta),
\end{equation}
and study the eigenmode equation (\ref{eq:Eigenequation}) separately 
for different eigenvalues $\Delta^\mu$. 

The eigenmode equation for $\Delta^\mu=0$ and that for 
$\Delta^\mu \neq 0$ are qualitatively different, and need separate 
study. The eigenmodes for $\Delta^\mu \neq 0$ will be presented in 
Section \ref{ssec:eigen-q=not0} 
(and appendix \ref{ssec:appendix-eigen-q=not0}); we begin 
in section \ref{ssec:eigen-q=0} 
(and appendix (\ref{ssec:appendix-eigen-q=0})) with the eigenmode equations 
for $\Delta^\mu = 0$, which is also regarded as an approximation of 
the eigenmode equation for $\Delta^\mu \neq 0$ in the asymptotic UV 
boundary region (at least $\Delta z \ll 1$, and maybe $z$ is as small as $R$).

For now, we relax the traceless condition on the spin-$j$ field $A_{m_1
\cdots m_j}$ ($m_i = 0,1,\cdots,3,z$), and we just assume that the
rank-$j$ tensor field $A_{m_1 \cdots m_j}$ is totally
symmetric.\footnote{This only makes the following presentation more 
far reaching; in the end, it is quite easy to identify which eigenmodes 
fall into the traceless part within $A_{m_1 \cdots m_j}$. 
See (\ref{eq:egval-l=0}--\ref{eq:coeff-l=0-odd}) at the end of 
section \ref{ssec:eigen-q=0}.}
Consider the following decomposition of the space of $z$-dependent 
field configuration $A_{m_1 \cdots m_j}(z; \Delta = 0)$:
\begin{equation}
 A_{z^k \mu_1 \cdots \mu_{j-k}}(z; \Delta^\mu = 0) =
 \sum_{N=0}^{[(j-k)/2]} \left(E^N[ a^{(k,N)}] \right)_{\mu_1 \cdots \mu_{j-k}};
\label{eq:block-dcmp-q=0}
\end{equation}
here, $\left(a^{(k,N)}(z; \Delta^\mu = 0)\right)_{\mu_1 \cdots
\mu_{j-k-2N}}$ is a rank-$(j-k-2N)$ totally symmetric tensor of
$\SO(3,1)$, and satisfies the 4D-traceless condition,  
\begin{equation}
 \eta^{\mu_1 \mu_2} a^{(k,N)}_{\mu_1 \cdots \mu_{j-k-2N}} = 0.
\label{eq:cond-4D-traceless}
\end{equation}
Thus, the field configuration can be described by 
$a^{(k,N)}$'s with $0\leq k \leq j$, $0\leq N \leq [(j-k)/2]$.  
These components form groups labeled by $n = 0, \cdots, j$, 
where the $n$-th group consists of $a^{(k,N)}$'s with 
$k+2N = n$; they are all rank-$(j-n)$ totally symmetric tensors of 
$\SO(3,1)$; let us call the subspace spanned by the components 
in this $n$-th group as the $n$-th subspace. 
The eigenmode equation for $\Delta^\mu = 0$ becomes block diagonal 
under the decomposition into the subspaces labeled by $n = 0, \cdots, j$. 
(See (\ref{eq:Eigen-eq-q=0-app}) in the appendix.) 
Therefore, the eigenmode equation for $\Delta^\mu = 0$ can be 
studied separately for the individual $n$-th diagonal block.

The $n$-th diagonal block contains $[n/2] + 1$ components, and hence 
there are $[n/2]+1$ eigenmodes. Let ${\cal E}_{n,l}$ ($l=0, \cdots,
[n/2]$) be the eigenvalues in the $n$-th diagonal block. 
The corresponding eigenmode wavefunction is of the form 
\begin{equation}
 \left(a^{(k,N)}(z; \Delta^\mu = 0)\right)_{\mu_1 \cdots\mu_{j-n}} 
= c_{k,l,n} \; \left(\epsilon^{(n,l)} \right)_{\mu_1 \cdots \mu_{j-n}} 
            \; z^{2-j-i\nu},
\end{equation}
where $\epsilon^{(n,l)}$ is a $z$-independent $k$-independent 
rank-$(j-n)$ tensor of $\SO(3,1)$. $c_{k,l,n} \in \R$. 
In the eigenmode equation for $\Delta^\mu = 0$, the eigenmode
wavefunctions are all in a simple power of $z$, and the power is
parametrized by $i\nu$ ($\nu \in \R$). 
The eigenvalues ${\cal E}_{n,l}$ are functions of $\nu$; once the
mass-shell condition (\ref{eq:mass-shell}) is imposed, the eigenmodes 
turn into solutions of the equation of motion, and $i\nu$ is determined 
by the mass parameter. 

The eigenmodes with smaller $(n,l)$ are as follows:
\begin{eqnarray}
 {\cal E}_{0,0} = (j + 4 + \nu^2), & \qquad & 
a^{(0,0)}(z)_{\mu_1 \cdots \mu_{j}} = 
 \epsilon^{(0,0)}_{\mu_1 \cdots \mu_{j}} \; z^{2-j-i\nu}, \\
{\cal E}_{1,0} = (3j + 5 + \nu^2), & \qquad & 
a^{(1,0)}(z)_{\mu_1 \cdots \mu_{j-1}} = 
 \epsilon^{(1,0)}_{\mu_1 \cdots \mu_{j-1}} z^{2-j-i\nu}, \\
 {\cal E}_{2,0} =(5j+4+\nu^2), & \qquad &
\left( \begin{array}{c}
 a^{(0,1)}(z)_{\mu_1 \cdots \mu_{j-2}} \\
 a^{(2,0)}(z)_{\mu_1 \cdots \mu_{j-2}}
       \end{array}\right) =
 \left( \begin{array}{c}
  1 \\ -4j
	\end{array}\right) \; 
 \epsilon^{(2,0)}_{\mu_1 \cdots \mu_{j-2}} \; z^{2-j-i\nu}, \\
 {\cal E}_{2,1} = (j+2+ \nu^2), & \qquad & 
\left( \begin{array}{c}
 a^{(0,1)}(z)_{\mu_1 \cdots \mu_{j-2}} \\
 a^{(2,0)}(z)_{\mu_1 \cdots \mu_{j-2}}
       \end{array}\right) =
 \left( \begin{array}{c}
  1 \\ 2
	\end{array}\right) \; 
 \epsilon^{(2,1)}_{\mu_1 \cdots \mu_{j-2}} \; z^{2-j-i\nu}.
\end{eqnarray}
Empirically, the $j$-dependence of the eigenvalues in the $n$-th 
diagonal block appear to be 
${\cal E}_{n,l} = ((2n+1-4l) j + \nu^2 + O(1))$ ($l = 0,\cdots,[n/2]$), 
[see (\ref{eq:bgn-of-empirical-egval}--\ref{eq:end-of-empirical-egval}) 
in the appendix for more samples of the eigenvalues]
and we promote this $j$-dependence to a rule of the labeling of 
the eigenmodes with $l$.

The eigenmode with $l=0$ is found in any one of the $n$-th diagonal
block ($n = 0, \cdots, j$). Its eigenvalue is 
\begin{equation}
  {\cal E}_{n,0} = (2n+1)j + 2n - n^2 + 4+\nu^2, 
\label{eq:egval-l=0}
\end{equation}
and 
\begin{eqnarray}
 c_{2\bar{k},0,2\bar{n}} = (-)^{\bar{k}} 4^{\bar{k}}
   \frac{ \bar{n}! }{ (\bar{n}-\bar{k})! }
   \frac{ (j-\bar{n}+1)! }{ (j-\bar{n}-\bar{k}+1)! }, & \qquad & 
   (n=2\bar{n}, \bar{k}=0,\cdots,\bar{n}),
  \label{eq:coeff-l=0-even} \\
 c_{2\bar{k}+1,0,2\bar{n}+1} = (-)^{\bar{k}} 4^{\bar{k}}
   \frac{ \bar{n}! }{ (\bar{n}-\bar{k})! }
   \frac{ (j-\bar{n})! }{ (j-\bar{n}-\bar{k})! }, & \qquad &
   (n = 2\bar{n}+1, \bar{k}=0, \cdots, \bar{n}). 
 \label{eq:coeff-l=0-odd}
\end{eqnarray}
This $(n,l) = (n,0)$ eigenmodes are characterized 
by the 5D-traceless condition
\begin{equation}
 g^{m_1 m_2} A_{m_1 \cdots m_j} = 0. \nonumber 
\end{equation}
Thus, the eigenmodes within the 5D-traceless (and totally
symmetric) component---spin-$j$ field---for $\Delta^\mu=0$ 
are labeled simply by $n = 0, \cdots, j$.

\subsection{Mode Decomposition for non-zero $\Delta_\mu$}
\label{ssec:eigen-q=not0}

\subsubsection{Diagonal Block Decomposition for the $\Delta^\mu \neq 0$ Case}

The eigenmode equation 
(\ref{eq:Eigenequation}, \ref{eq:Eigenequation-4+1}) 
is much more complicated in the case of $\Delta^\mu \neq 0$, 
because of the 2nd and 4th terms in (\ref{eq:Eigenequation-4+1}). 
The eigenmode equation is still made block diagonal for an appropriate 
decomposition of the space of field $A_{m_1\cdots m_j}(z; \Delta^\mu)$.

Consider a decomposition 
\begin{equation}
 A_{z^k \mu_1 \cdots \mu_{j-k}}(z; \Delta^\mu)  = 
 \sum_{s=0}^{j-k} \sum_{t=0}^{[s/2]} 
   \left( \tilde{E}^N D^{s-2N} [ a^{(k,s,N)} ] \right)_{\mu_1 \cdots \mu_{j-k}},
\label{eq:block-dcmp-q=not0-tilde}  
\end{equation}
where a new operation $a \mapsto \tilde{E}[a]$ on a symmetric 
$\SO(3,1)$ tensor $a$, 
\begin{equation}
  \left(\tilde{E}[ a ]\right)_{\mu_1 \cdots \mu_{r+2}}  :=  
  2 \sum_{p < q}
   \left( \eta_{\mu_p \mu_q} -
          \frac{ \partial_{\mu_p} \partial_{\mu_q} }{ \partial^2 }
  \right)
  a_{\mu_1 \cdots \check{\mu_p} \cdots \check{\mu_q} \cdots
  \mu_{r+2}},
\label{eq:def-E-tilde}
\end{equation}
is used. $a^{(k,s,N)}$ are totally symmetric 4D-traceless 
(i.e. (\ref{eq:cond-4D-traceless})) rank-$(j-k-s)$ tensor fields 
of $\SO(3,1)$ that satisfies an additional condition, 
the 4D-transverse condition:
\begin{equation}
 \partial^{\hat{\rho}}
  \left( a^{(k,s,N)} \right)_{\rho \mu_2 \cdots \mu_{j-k-s} }
 = i \Delta^{\hat{\rho}} 
  \left( a^{(k,s,N)} \right)_{\rho \mu_2 \cdots \mu_{j-k-s} } = 0.
\label{eq:cond-4D-transverse}
\end{equation} 
The space of field configuration $A_{m_1 \cdots m_j}(z; \Delta^\mu)$
is now decomposed into $a^{(k,s,N)}$'s with $0 \leq k \leq j$, 
$0 \leq s \leq j-k$, $0 \leq N \leq [s/2]$; these components form 
groups labeled by $m = 0, \cdots, j$, 
where the $m$-th group consists of $a^{(k,s,N)}$'s with 
$k+s = m$; they are all rank-$(j-m)$ totally symmetric 4D-traceless and 
4D-transverse tensors of $\SO(3,1)$; let us call the subspace spanned by 
the components in this $m$-th group as the $m$-th subspace. 
The eigenmode equation for $\Delta^\mu \neq 0$ becomes block diagonal 
under the decomposition into the subspaces labeled by $m = 0, \cdots, j$. 
The eigenmode equation for the $m$-th sector is given by 
(\ref{eq:EGeq-q=not0-matrix-m}) in the 
appendix \ref{ssec:appendix-eigen-q=not0}.
The $m$-th subspace should have 
\begin{equation}
  \sum_{s=0}^m \left([s/2] + 1 \right) 
\end{equation}
eigenmodes.

Eigenvalues ${\cal E}$ are determined in terms of the characteristic 
exponent in the expansion of the solution in power series of $z$; 
let the first term in the expansion be $z^{2-j-i\nu}$; the eigenvalues 
are functions of $\nu$ then.  Because the indicial equation at 
the regular singular point $z \simeq 0$ allows us to determine 
the eigenvalues in terms of $\nu$, the eigenvalues in the case of 
$\Delta^\mu \neq 0$ cannot be different from the ones 
we have already known in the $\Delta^\mu = 0$ case. 
In the $m$-th diagonal block, the eigenvalues consist of 
${\cal E}_{n,l}$ with $0 \leq n \leq m$, $0 \leq l \leq [n/2]$.

To summarize, the eigenmodes in the totally symmetric rank-$j$ tensor 
field of $\SO(4,1)$ are labeled by $(n,l,m)$ and $\Delta^\mu$ and
$\nu$. Their eigenvalues ${\cal E}_{n,l}$ depend only on $n$ and $l$
(with $0 \leq n \leq j$ and $0 \leq l \leq [n/2]$) and $\nu$. 
Corresponding eigenmodes are denoted by 
\begin{eqnarray}
 A(x,z)^{n,l,m; \Delta, \nu}_{z^k \mu_1 \cdots \mu_{j-k}} & = &
  \nonumber \\
 e^{i \Delta \cdot x} 
   A^{n,l,m}_{z^k \mu_1 \cdots \mu_{j-k}}(z; \Delta^\mu, \nu) & = &
 e^{i \Delta \cdot x} \sum_{N=0}^{[s/2]} 
    \tilde{E}^{N} D^{s-2N}[\epsilon^{(n,l,m)}] 
    \frac{b^{(j-m)}_{s,N}}{\Delta^{s-2N}}
  \Psi^{(j);s,N}_{i\nu;n,l,m}(-\Delta^2,z).
\label{eq:Pomeron-wvfc-general-expans}
\end{eqnarray}
$\epsilon^{(n,l,m)}$ is a ($z$-independent) totally symmetric 
4D-traceless 4D-transverse rank-$(j-m)$ tensor of $\SO(3,1)$, and 
all the $s$'s appearing in the expression above are understood as
$s=m-k$. $b^{(r)}_{s,N}$ is a
constant whose definition is given in (\ref{eq:def-of-b}) in the appendix.

\subsubsection{Single Component Pomeron Wavefunction}

The Pomeron wavefunction that has been discussed in the literature
(e.g. \cite{BPST-06}) does not look as awful 
as (\ref{eq:Pomeron-wvfc-general-expans}). To our knowledge, 
the Pomeron wavefunction in the literature in the context of hadron
high-energy scattering has been a single component one, $\Psi_{i\nu}(t,z)$. 
How is $A^{n,l,m}_{m_1 \cdots m_j}(z; \Delta^\mu, \nu)$ related to
$\Psi_{i\nu}(-\Delta^2; z)$?

In the block diagonal decomposition of the eigenmode equation, 
there is only one subspace where the diagonal block is $1 \times 1$.
That is the $m=0$ subspace, which consists only of $a^{(0,0,0)}$.
The eigenmode equation is 
\begin{equation}
 \left[\Delta_j - \frac{j}{R^2} \right] a^{(0,0,0)}(z; \Delta^\mu) 
 = - \frac{{\cal E}}{R^2} a^{(0,0,0)}(z; \Delta^\mu).
\label{eq:EGeq-q=not0-n=0}
\end{equation}
This equation, as well as (\ref{eq:EGeq-q=0-n=0-app}) in the 
$\Delta^\mu = 0$ case, corresponds to the ``Schr\"{o}dinger equation'' 
in \cite{BPST-06} determining the Pomeron wavefunction. 
It should be noted, however, that we consider that $\nabla^2$ is 
the operator relevant to the eigenmode decomposition\footnote{Thus, 
the propagator (\ref{eq:propagator-idea}) uses the eigenvalue of 
$\nabla^2$, rather than that of $\Delta_j$. The eigenvalue ${\cal E}$ 
of $\nabla^2$ in the $m=0$-th subspace is $(j+4+\nu^2)$ 
(\ref{eq:egval-nlm=000}), instead of $(4+\nu^2)$.} 
rather than $\Delta_j$; furthermore the operator 
$\nabla^2$ and $\Delta_j$ has a simple relation 
$\nabla^2 = \Delta_j - j/R^2$ only on this $m=0$-th subspace 
of a totally symmetric rank-$j$ tensor field of $\SO(4,1)$.

The eigenvalue is 
\begin{equation}
 {\cal E}_{0,0} = (j+4+\nu^2),  
\label{eq:egval-nlm=000}
\end{equation}
when we define the first term in the power series expansion of $z$ to be 
$z^{2-j-i\nu}$. The eigenmode wavefunction is 
\begin{eqnarray}
 a^{(0,0,0)}(z; \Delta^\mu)_{\mu_1 \cdots \mu_j} & = &  
 \epsilon^{(0,0,0)}_{\mu_1 \cdots \mu_j} \;
 \Psi^{(j)}_{i\nu}(- \Delta^2, z), \\
\Psi^{(j)}_{i\nu}(-\Delta^2,z) & := & 
  \frac{2}{\pi}\sqrt{\frac{\nu \sinh (\pi \nu)}{2R}} \; 
  e^{(j-2)A} K_{i\nu}(\Delta z).
\label{eq:Pomeron-wvfc-1compnt}
\end{eqnarray}
The normalization factor is determined \cite{BPST-06}\footnote{See 
also \cite{NW-first}. The Pomeron wavefunction in \cite{NW-first} was 
\begin{eqnarray}
  \Psi^{(j)}_{i\nu}(- \Delta^2, z) & = & e^{(j-2)A}
  \frac{2}{\pi} \sqrt{ \frac{\nu \sinh(\pi\nu)}{2R} } 
  \sqrt{ \frac{ I_{i\nu}}{I_{-i\nu}} }
  \left[ K_{i\nu}(\Delta z)
     - \frac{K_{i\nu}}{I_{i\nu}} I_{i\nu}(\Delta z)
  \right], \nonumber 
\end{eqnarray}
where $K_{\mu} := K_{\mu}(\Delta/\Lambda)$ and 
$I_\mu := I_\mu(\Delta/\Lambda)$, when the argument is not 
specified explicitly. 
This wavefunction becomes (\ref{eq:Pomeron-wvfc-1compnt})
in the main text 
in the limit of $\Lambda \rightarrow 0$, while keeping $z$ and 
$\Delta^\mu$ fixed. 
}
so that it satisfies the normalization condition\footnote{
The normalization condition is generalized 
to (\ref{eq:wvfc-normalization-cond}) later on.}
\begin{equation}
 \int d^4 x \int dz \sqrt{-g(z)} e^{-2jA} \; [e^{i \Delta \cdot x}
  \Psi^{(j)}_{i\nu}(-\Delta^2,z)] \; [\Psi^{(j)}_{i\nu'}(-\Delta^{'2}, z)
  e^{-i \Delta' \cdot x}]
  = (2\pi)^4 \delta^4(\Delta-\Delta') \; \delta(\nu-\nu').
\end{equation}
The single component Pomeron/Reggeon wavefunction 
$\Psi^{(j)}_{i\nu}(-\Delta^2, z)$ is now understood as 
$\Psi^{(j);0,0}_{i\nu; 0,0,0}(-\Delta^2,z)$.

\subsubsection{5D-Traceless 5D-Transverse Modes}

The eigenmode equation (\ref{eq:Eigenequation}) for a totally symmetric 
rank-$j$ tensor field of $\SO(4,1)$ should be closed within its 
5D-traceless component. The subspace of 5D-traceless component is 
characterized by the 5D-traceless condition
\label{page:5D-trl-trv}
\begin{equation}
 g^{m_1 m_2} A_{m_1 \cdots m_j} (z; \Delta^\mu) = 0.
 \label{eq:cond-5D-traceless}
\end{equation}
The fact that the Hermitian operator $\nabla^2$ maps this subspace to
itself implies that the eigenmode equation of $\nabla^2$ is block
diagonal, when the space of (not-necessarily 5D-traceless) 
$A_{m_1 \cdots m_j}$ is decomposed into the sum of the 5D-traceless
subspace and its orthogonal complement. Collection of the eigenmodes 
with $l=0$ correspond to the subspace of 5D-traceless field configuration.

Similarly, one can think of a subspace of field configuration satisfying 
both the 5D-traceless condition (\ref{eq:cond-5D-traceless}) and 
the 5D-transverse condition
\begin{equation}
 g^{n m_1} \nabla_n A_{m_1 m_2 \cdots m_j} = 0.
\label{eq:cond-5D-transverse}
\end{equation}
Obviously this is a subspace of the subspace of 5D-traceless modes we 
discussed above. Since the Hermitian operator $\nabla^2$ on AdS$_5$ 
maps this new subspace also to itself, the eigenmode equation of 
$\nabla^2$ should also become block diagonal, when the subspace of 
5D-traceless modes is decomposed into this new subspace its orthogonal 
complement. 

As we will see in the appendix \ref{ssec:appendix-5D-trltrv}, 
there is only one such mode satisfying this set of conditions 
(\ref{eq:cond-5D-traceless}, \ref{eq:cond-5D-transverse}) 
in each one of the $m$-th diagonal block. Thus, the combination 
of the 5D-traceless and 5D-transverse conditions allow us to 
determine an eigenmode completely. This mode turns out to be 
$(n,l,m) = (0,0,m)$ (for $0 \leq m \leq j$). Put differently, 
the eigenmodes with the eigenvalue 
${\cal E}_{n,l} = {\cal E}_{0,0} = (j+4+\nu^2)$ are charachterized 
by the traceless and transverse conditions on AdS$_5$.

The eigenmode wavefunctions of the 5D-traceless transverse modes 
$(n,l,m) = (0,0,m)$ are (see the appendix \ref{ssec:appendix-5D-trltrv}) 
\begin{equation}
  \Psi^{(j);s,N}_{i\nu;0,0,m}(-\Delta^2,z) =
    \sum_{a=0}^N (-)^a {}_N C_a 
      \left(\frac{z^3 \partial_z z^{-3}}{\Delta}\right)^{s-2a} 
        \left[(z \Delta)^m \Psi^{(j);0,0}_{i\nu;0,0,0}(-\Delta^2, z)
	\right] \times N_{j,m}.
\label{eq:00m-mode-fcn}
\end{equation}
$N_{j,m}$ is a dimensionless normalization constant. We choose it to
be\footnote{Note that $N_{j,m} = 1$, if $m=0$.} 
\begin{equation}
 N_{j,m}^{-2} = {}_jC_m
  \frac{\Gamma(j+1-i\nu)}{\Gamma(j+1-m-i\nu)}
  \frac{\Gamma(j+1+i\nu)}{\Gamma(j+1-m+i\nu)} 
  \frac{\Gamma(3/2+j-m)}{2^m\Gamma(3/2+j)}
  \frac{\Gamma(2+2j)}{\Gamma(2+2j-m)}, 
\end{equation}
so that the eigenmode wavefunctions are normalized as in 
\begin{eqnarray}
&& \int d^4 x \int_0 dz \sqrt{-g(z)} g^{m_1 n_1} \cdots g^{m_j n_j} \; 
   A^{n,l,m; \Delta, \nu}_{m_1 \cdots m_j}(x,z) \; 
   A^{n',l',m'; \Delta',\nu'}_{n_1 \cdots n_j}(x,z) \nonumber \\ 
&=& (2\pi)^4 \delta^4(\Delta+\Delta^{'}) \; \delta(\nu - \nu') \; 
    \delta_{n,n'} \delta_{l,l'} \delta_{m,m'} \;
    \left[ \epsilon^{(n,l,m)}(\Delta) \right] \cdot
    \left[ \epsilon^{(n',l',m')}(\Delta') \right].
\label{eq:wvfc-normalization-cond}
\end{eqnarray}
Here, $[\epsilon^{(n,l,m)}] \cdot [\epsilon^{'(n,l,m)}] :=
 \epsilon^{(n,l,m)}_{\mu_1 \cdots \mu_{j-m}}
 \epsilon^{'(n,l,m)}_{\nu_1 \cdots \nu_{j-m}}
 \eta^{\hat{\mu}_1 \hat{\nu}_1 } \cdots 
 \eta^{\hat{\mu}_{j-m} \hat{\nu}_{j-m} }$.

\subsubsection{Propagator}

The propagator of the totally symmetric rank-$j$ tensor field 
[resp. spin-$j$ field] on AdS$_5$ is given by summing up propagators 
of the $(n,l,m)$ modes [resp. $(n,l,m)$ modes with $l=0$]. 
In order to write down the propagator of a given $(n,l,m)$ eigenmode, 
it is convenient to introduce the following notation:
\begin{equation}
 A^{n,l,m; \Delta,\nu}_{m_1 \cdots m_j}(x,z) = 
 \left[A^{n,l,m; \Delta,\nu}_{m_1 \cdots m_j}(x,z) 
 \right]^{\hat{\kappa}_1 \cdots \hat{\kappa}_{j-m}} 
    \epsilon^{(n,l,m)}_{\kappa_1 \cdots \kappa_{j-m}} = 
 e^{i \Delta \cdot x}
 \left[A^{n,l,m}_{m_1 \cdots m_j}(z; \Delta^\mu, \nu) 
 \right]^{\hat{\kappa}_1 \cdots \hat{\kappa}_{j-m}} 
    \epsilon^{(n,l,m)}_{\kappa_1 \cdots \kappa_{j-m}}.
\end{equation}
With this notation, the propagator of the $(n,l,m)$ mode is given by 
\begin{eqnarray}
 G(x,z; x',z')_{m_1 \cdots m_j;n_1 \cdots n_j}^{(n,l,m)} & = & 
 \int \frac{d^4 \Delta}{(2\pi)^4} \int_0^{\infty} d\nu 
\frac{- i 
      P_{ \rho_1 \cdots \rho_{j-m}; \; \sigma_1 \cdots \sigma_{j-m} } }
     {\frac{{\cal E}_{n,l} + c}{\sqrt{\lambda}} + N_{\rm eff.} - i \epsilon}  
    \frac{\alpha' R^3}{C} \nonumber \\ 
&&
   \left[A^{n,l,m; \Delta,\nu}_{m_1 \cdots
    m_j}(x,z)\right]^{\rho_1\cdots \rho_{j-m}}
   \left[A^{n,l,m; -\Delta,\nu}_{n_1 \cdots n_j}(x',z')
   \right]^{\sigma_1 \cdots \sigma_{j-m}}.
\end{eqnarray}
Here, $P_{\rho_1 \cdots \rho_{j-m}; \; \sigma_1 \cdots \sigma_{j-m}}$ 
is a polarization tensor generalizing 
$\eta_{\rho\sigma} - \partial_\rho \partial_\sigma /\partial^2$.
When an orthogonal basis 
$\epsilon_a(q) \cdot \epsilon_b(-q) = \delta_{a,b} D_a$ of rank-$r$ 
4D-traceless 4D-transverse tensors is given, 
\begin{equation}
 P_{\mu_1 \cdots \mu_r; \; \nu_1 \cdots \nu_r} := \sum_a \frac{1}{D_a} 
  \epsilon(q)_{a; \; \mu_1 \cdots \mu_r}
  \epsilon(-q)_{a; \; \nu_1 \cdots \nu_r}.
\end{equation}
An alternative characterization of this 
$P_{\mu_1 \cdots \mu_r; \nu_1 \cdots \nu_r}$ is given by a 
combination of the two following conditions: one is  
\begin{equation}
 P_{\mu_1 \cdots \mu_r; \; \nu_1 \cdots \nu_r} 
  \epsilon_a^{\nu_1 \cdots \nu_r}  =  
   \epsilon_{a;\mu_1 \cdots \mu_r},  
\end{equation}
and the other is that  $P_{\mu_1 \cdots \mu_r; \; \nu_1 \cdots \nu_r}$ 
be also a totally symmetric 4D-transverse 4D-traceless tensor 
with respect to $(\mu_1 \cdots \mu_r)$ for any choice of 
$(\nu_1 \cdots \nu_r)$. 


\subsection{Representation in the Dilatation Eigenbasis}
\label{ssec:repr-dilatation}

It is an essential process in the application of AdS/CFT correspondence 
to classify solutions to the equation of motions on the gravity dual 
background (AdS$_5$) into irreducible representations of the conformal 
group $\SO(4,2)$ (or possibly its supersymmetric extension). 
In the CFT description, primary operators are in one to one
correspondence with (highest weight) irreducible representations of 
the conformal group, and it is believed that one can establish an 
one-to-one correspondce between i) a primary operator in the CFT
description and ii) a group of solutions to the equation of motion forming 
an irreducible representation in the gravity dual description. 
Once this correspondence is given, then hadron matrix elements of 
the primary operators in a (nearly conformal) field theory can be
calculated by using the corresponding solutions to the equation 
of motions (wavefunctions) on AdS$_5$.
Note that the hadron matrix elements of primary operators are all 
that remains unknown in the formulation of conformal operator product
expansion (\ref{eq:conf-OPE-center}). 

Let $P_\mu$, $K_\mu$, $L_{\mu\nu}$ and $D$ denote the generators 
of the unitary operators of the conformal group transformation on the
Hilbert space. They satisfy the following commutation relations:
\begin{eqnarray}
& [ D, P_\mu ] = i P_\mu,
& [ P_\rho, L_{\mu\nu} ] = i ( \eta_{\rho\mu} P_\nu- \eta_{\rho\nu}P_\mu
), \\
& [ D, K_{\mu} ] = -i K_{\mu}, 
& [K_\rho, L_{\mu\nu}] = i (\eta_{\rho\mu} K_\nu- \eta_{\rho\nu} K_\mu), 
\end{eqnarray}
\begin{align}
 & [ P_\mu, K_\nu ] = -2i ( \eta_{\mu\nu} D + L_{\mu\nu} ), \\ 
 & [L_{\mu\nu}, L_{\rho\sigma}] = i (
    \eta_{\nu\rho} L_{\mu \sigma} - \eta_{\nu \sigma} L_{\mu \rho}
    - \eta_{\mu\rho} L_{\nu \sigma} + \eta_{\mu \sigma} L_{\nu \rho} ).
\end{align}

When such a conformal symmetry exists in a conformal field thoery 
on 3+1 dimensions, those generators have a representation as
differential operators on fields on $\R^{3,1}$; those differential 
operators are denoted by ${\cal P}_\mu$, ${\cal K}_\mu$, 
${\cal L}_{\mu\nu}$ and ${\cal D}$. The generators and the differential
operators on a CFT are in the following relation:
\begin{equation}
 [ {\cal O}(x), P_\mu] = {\cal P}_\mu {\cal O}(x), \quad 
 [ {\cal O}(x), K_{\mu} ] = {\cal K}_\mu {\cal O}(x), \quad 
 [ {\cal O}(x), D ] = {\cal D} {\cal O}(x), \cdots, 
\end{equation}
and those differential operators acts on primary operators as follows:
\begin{eqnarray}
 {\cal D} \overline{\cal O}_n(x)  & = & -i (x \cdot \partial + l_n)
      \overline{\cal O}_n(x), \\
 {\cal L}_{\mu\nu} \overline{\cal O}_n(x) & = &
    \left(i(x_\mu \partial_\nu - x_\nu \partial_\mu) + [S_{\mu\nu}]
    \right) \overline{\cal O}_n(x) \\
 {\cal P}_\mu \overline{\cal O}_n(x)& = &
       -i \partial_\mu \overline{\cal O}_n(x), \\
 {\cal K}_\mu \overline{\cal O}_n(x)& = &
    \left( -i(2x_\mu x\cdot \partial - x^2 \partial_\mu )
           -i 2l_n x_\mu
           - x^\kappa [S_{\mu\nu}]
   \right) \overline{\cal O}_n(x),
\end{eqnarray}
where $[S_{\mu\nu}]$ is a finite dimensional representation of
$\SO(3,1)$ generators satisfying the same commutation relation as 
$L_{\mu\nu}$'s. Thus, for a primary operator $\overline{\cal O}_n(x)$, 
$\overline{\cal O}_n(x = 0)$ plays the role of the highest weight state 
\begin{equation}
 [ \overline{\cal O}_n(0), K_\mu ] = 0, \qquad 
 [ \overline{\cal O}_n(0), D] = -i l_n \overline{\cal O}_n(0); 
\end{equation}
all other states in the highest weight state
representation---decendants---are generated by applying 
$[ \bullet, P_\mu]$ multiple times; the whole representation, therefore, 
is spanned by a collection of 
\begin{equation}
 \left\{ \overline{\cal O}_n(0), \; \partial_\mu \overline{\cal O}_n(0), \; 
         \partial_\mu \partial_\nu \overline{\cal O}_n(0), \; \cdots \right\};
\end{equation}
it is also equivalent to $\overline{\cal O}(x = x_0)$ with arbitrary 
$x_0 \in \R^{3,1}$.

In the preceding sections, we have worked on solutions to the eigenmode 
equation on AdS$_5$; once the mass-shell condition (\ref{eq:mass-shell}) 
is imposed, they become solutions to the equation of motion. 
They are obtained as an eigenmode of the spacetime translation in 3+1 
dimensions, $(-i \partial^\mu) = \Delta^\mu$. Under the conformal group 
$\SO(4,2)$, which contains Lorentz $\SO(3,1)$ symmetry, however, 
an irreducible representation has to include solutions 
with all kinds of eiganvalues $\Delta^\mu$. 

In the case of a scalar field on AdS$_5$, one can think of the following 
linear combination $G(x,z; x_0; R_0)$ (for some $R_0 \ll \Delta^{-1}$):
\begin{equation}
 G(x,z; x_0) =
 \frac{i}{\pi^2}\frac{\Gamma(l_n)}{\Gamma(l_n-2)}
     R_0^{l_n-4}
     \left(\frac{z}{z^2 + (x-x_0)^2}\right)^{l_n} =
 \int \frac{d^4 \Delta}{(2\pi)^4} e^{i \Delta \cdot (x-x_0)}
     \frac{(\Delta z)^2 K_{l_n-2}(\Delta z) }{ (\Delta R_0)^2 K_{l_n-2}(\Delta R_0) }.
\end{equation}
The factor $[e^{i \Delta \cdot x} (\Delta z)^2 K_{l_n-2}(\Delta z)]$ 
in the integrand on the right hand side is a solution to the equation 
of motion of a scalar field on AdS$_5$ whose mass-square 
$M^2_{\rm eff.}$ is given by 
$l_n -2 = i \nu = \sqrt{4 + M^2_{\rm eff.} R^2}$. The coefficient 
of the linear combination, 
$e^{-i \Delta \cdot x_0} [(\Delta R_0)^2 K_{l_n-2}(\Delta R_0)]$, is 
chosen so that the integrand behaves as 
\begin{equation}
 e^{i \Delta \cdot (x-x_0)} \left(\frac{z}{R_0}\right)^{4-l_n} 
\end{equation}
at $0 \leq z \ll \Delta^{-1}$. The space of solutions to the equation of
motion parametrized by $x_0 \in \R^{3,1}$ is spanned by derivatives of 
$G(x,z;x_0)$ with respect to $x^\mu_0$ at $x^\mu_0 = 0$. 
It is easy to see that this basis 
\begin{equation}
 \left\{ G(x,z;0), \; \partial^{(x_0)}_\mu G(x,z; 0), \;
         \partial^{(x_0)}_\mu \partial^{(x_0)}_\nu G(x,z;0), 
    \; \cdots \right\}
\end{equation}
is an eigenbasis under the action of dilatation, 
${\cal D} := i (z \partial_z + x \cdot \partial)$, and their 
weights are $-i l_n$, $-i (l_n + 1)$, $-i (l_n + 2), \cdots$,
respectively. Correspondence between scalar field wavefunctions on 
AdS$_5$ and scalar primary operators of the dual CFT is established 
in this way \cite{AdSCFT-review}. 

Let us now generalize the discussion above slightly, to construct 
an analogue of $G(x,z; x_0)$ for a spin-$j$ field $A_{m_1 \cdots m_j}$ 
on AdS$_5$, from which the dilatation eigenbasis is constructed. 
To this end, note that all the $(0,0,m)$-modes ($m = 0, \cdots, j$) 
have the leading $z^{2-j-i\nu}$ in the power series expansion only 
in the $A_{z^0 \mu_1 \cdots \mu_j}$ component, not in any other 
$A_{z^k \mu_1 \cdots \mu_{j-k}}$ components\footnote{
Use (\ref{eq:00m-mode-fcn}).} with $k > 0$. It is possible 
to choose $\epsilon^{(0,0,m)}(\Delta^\mu)$ properly so that  
\begin{equation}
 \sum_{m=0}^j \left[A^{0,0,m;\Delta,\nu}_{\mu_1 \cdots \mu_j} (x,z) 
 \right]^{\hat{\kappa}_1 \cdots \hat{\kappa}_{j-m}}
 \epsilon^{(0,0,m)}_{\kappa_1 \cdots \kappa_{j-m}}
 e^{-i \Delta \cdot x_0}  \simeq 
 e^{i \Delta \cdot (x-x_0)} \left(\frac{z}{R_0}\right)^{2-j-i\nu}
   \epsilon_{\mu_1 \cdots \mu_j}
\end{equation}
in the region near the UV boundary $z \ll \Delta^{-1}$, where 
$\epsilon_{\mu_1 \cdots \mu_j}$ is a $\Delta^\mu$-independent 
4D-traceless totally symmetric rank-$j$ tensor of $\SO(3,1)$. 
The condition is 
\begin{eqnarray}
  \epsilon_{\mu_1 \cdots \mu_j}  & = &  
   \left(\frac{R_0}{R}\right)^{2-j} \!\!\!\!\! K_{i\nu}(\Delta R_0)
   \frac{2}{\pi}\sqrt{ \frac{\nu \sinh (\pi \nu)}{2R} } \nonumber \\
&& \sum_{m=0}^j \frac{ N_{j,m} \Gamma(m-j-i\nu)}{\Gamma(-j-i\nu)}
\sum_{N=0}^{[m/2]} \frac{b^{(j-m)}_{m,N}}{\Delta^{m-2N}} 
  \left(\tilde{E}^N D^{m-2N}[\epsilon^{(0,0,m)}]\right)_{\mu_1 \cdots \mu_j} 
\end{eqnarray}
It is possible to invert this relation by 
using (\ref{eq:extract-SO(3,1)-tensor}) and write down 
$\epsilon^{(0,0,m)}(\Delta^\mu)$ in terms of 
$\epsilon_{\mu_1 \cdots \mu_j}$, though we will not present the result here. 
What really matters to us is that 
$\epsilon^{(0,0,m)}(\lambda \Delta) = 
 \epsilon^{(0,0,m)}(\Delta) \lambda^{i\nu}$.
With $\epsilon^{(0,0,m)}$'s satisfying the condition above, one can see
that the following linear combination of solutions to the equation of motion, 
\begin{equation}
 G_{m_1 \cdots m_j}(x,z; x_0) := \int \frac{d^4 \Delta}{(2\pi)^4} 
  \sum_{m=0}^j
    \left[ A^{0,0,m;\Delta,\nu}_{m_1 \cdots m_j}(x,z)
    \right]^{\hat{\kappa}_1 \cdots \hat{\kappa}_{j-m}}
    \epsilon^{(0,0,m)}(\Delta)_{\kappa_1 \cdots \kappa_{j-m}}
   e^{-i \Delta \cdot x_0}, 
\end{equation}
has a property 
\begin{equation}
 G_{m_1 \cdots m_j}( \lambda x, \lambda z; \lambda x_0) = 
\lambda^{-(2+j+i\nu)} G_{m_1 \cdots m_j}(x,z; x_0).
\end{equation}
$i\nu$ is determined by the mass parameter on AdS$_5$, once the 
mass shell condition (\ref{eq:mass-shell}) is imposed.
Therefore, $G_{m_1 \cdots m_j}(x,z;0)$ is an eigenstate of dilatation, 
and so are the derivatives of $G_{m_1 \cdots m_j}(x,z;x_0)$ with respect
to $x_0^\mu$ at $x_0^\mu = 0$. All the derivatives combined form a 
dilatation eigenbasis in the space of solutions to the equation of
motion of a spin-$j$ field. 

It is now clear that the eigenmodes with $(n,l,m) = (0,0,m)$ 
($0 \leq m \leq j$) and arbitrary $\Delta^\mu$ as a whole---modes that 
satisfy the 5D-traceless and 5D-transverse conditions 
(\ref{eq:cond-5D-traceless}, \ref{eq:cond-5D-transverse})---forms 
an irreducible representation of the conformal group. 
If one is interested purely in the matrix element of a spin-$j$ 
primary operator $\overline{\cal O}_n(x_0 = 0)$ of an approximately 
conformal gauge theory, then the matrix element can be calculated 
by using the wavefunction $G_{m_1 \cdots m_j}(x,z; 0)$.
Note that the $m=0$ mode alone, where the Pomeron/Reggeon wavefunction 
has a single component as in \cite{BPST-06}, cannot reproduce all the 
matrix element associated with matrix elements of spin-$j$ primary 
operators.


\appendix 

\section{More on the Mode Decomposition on AdS$_5$}
\label{sec:appendix-mode-fcn}

For convenience, let us copy here the eigenmode equation
(\ref{eq:Eigenequation}) for a totally symmetric rank-$j$ tensor field 
on AdS$_5$; the equation consists of the following equations labeled by 
$k=0, \cdots, j$:
\begin{eqnarray}
&& \left( (R^2 \Delta_j) 
- \left[(2k+1)j-2k^2+3k\right]\right) A_{z^k \mu_1 \cdots \mu_{j-k} } 
   \nonumber  \\ 
&& + 2zk \partial^{\hat{\rho}} A_{z^{k-1}\rho\mu_1\cdots \mu_{j-k}}
+ k(k-1) A^{\hat{\rho}}_{z^{k-2}\rho \mu_1 \cdots \mu_{j-k} } 
   \nonumber \\
&& - 2z (D[A_{z^{k+1} \cdots}])_{\mu_1 \cdots \mu_{j-k}}
+ (E[A_{z^{k+2} \cdots}])_{\mu_1 \cdots \mu_{j-k}} = -{\cal E} A_{z^k \mu_1
\cdots \mu_{j-k}}.  
 \label{eq:Eigenequation-4+1-app}
\end{eqnarray}
%

\subsection{Eigenvalues and Eigenmodes for $\Delta^\mu = 0$}
\label{ssec:appendix-eigen-q=0}

{\bf block diagonal decomposition}

In the main text, we considered a decomposition of the rank-$j$ totally
symmetric tensor field with $(-i \Delta_\mu) = \Delta_\mu = 0$ in the
form of 
\begin{equation}
 A_{z^k \mu_1 \cdots \mu_{j-k}}(z; \Delta^\mu = 0) =
 \sum_{N=0}^{[(j-k)/2]} \left(E^N[ a^{(k,N)}] \right)_{\mu_1 \cdots \mu_{j-k}};
 \nonumber 
\end{equation}
where $a^{(k,N)}$'s are $z$-dependent rank-$(j-k-2N)$ totally symmetric 
tensor field of $\SO(3,1)$, satisfying the 4D-traceless condition 
(\ref{eq:cond-4D-traceless}).  
This is indeed a decomposition, in that all the degrees of freedom in 
$A_{z^k \mu_1 \cdots \mu_j-k}(z; \Delta^\mu = 0)$ are described by 
$a^{(k,N)}(z)_{\mu_1 \cdots \mu_{j-k-2N}}$ with $0 \leq N \leq [(j-k)/2]$ 
without redundancy. To see this, 
one only needs to note that there is a relation\footnote{
This relation can be verified recursively in $a$.} that, for 
a totally symmetric 4D-traceless rank-$r$ $\SO(3,1)$-tensor $a$, 
\begin{equation}
 \eta^{\hat{\rho} \hat{\sigma}} 
 E^N[a]_{\rho \sigma \mu_1 \cdots \mu_{r+2N-2}} = 
4N(r+N+1) E^{N-1}[a]_{\mu_1 \cdots \mu_{r+2N-2}}. 
\label{eq:recursion-E-by-eta}
\end{equation}
Using this relation, $a^{(k,N)}_{\mu_1 \cdots \mu_{j-k-2N}}$ can be 
retrieved from $A_{z^k\mu_1 \cdots \mu_{j-k}}$, starting from ones with 
larger $N$ to ones with smaller $N$.

Let us now see that the eigenmode equation 
(\ref{eq:Eigenequation-4+1}=\ref{eq:Eigenequation-4+1-app}) can be made 
block diagonal by using this decomposition.
The eigenmode equation (\ref{eq:Eigenequation-4+1-app}) with the label $k$ 
for $\Delta^\mu = 0$ can be rewritten by using this 
relation (\ref{eq:recursion-E-by-eta}) as follows:
\begin{eqnarray}
\sum_N &&  \left(R^2 \Delta_j - \left[(2k+1)j-2k^2+3k\right] + {\cal E} \right) 
   E^N[a^{(k,N)}]
  \nonumber
   \\ 
&+& k(k-1) \left[4 (N+1) (j-k-N+2) \right] E^N[a^{(k-2,N+1)}]
+ E[E^{N-1}[a^{(k+2, N-1)}]] = 0. \nonumber
\end{eqnarray}
Although this equation has to hold only after the summation in $N$, 
it actually has to be satisfied separately for different $N$'s.
To see this, let us first multiply $\eta^{\hat{\rho}\hat{\sigma}}$
for $[(j-k)/2]$ times and contract indices just like in 
(\ref{eq:recursion-E-by-eta}); we obtain an equation that involves 
only $a^{k,[(j-k)/2]}$, $a^{(k-2,[(j-k)/2]+1)}$ and
$a^{(k+2),[(j-k)/2]-1}$. Next, multiply $\eta^{\hat{\rho}\hat{\sigma}}$
for $[(j-k)/2] - 1$ times, to obtain another eqution involving 
$a^{k,[(j-k)/2]-1}$, $a^{(k-2,[(j-k)/2])}$ and
$a^{(k+2),[(j-k)/2]-2}$. In this way, we obtain 
\begin{eqnarray}
&& \left(R^2 \Delta_j - \left[(2k+1)j-2k^2+3k\right] + {\cal E} \right) 
   a^{(k,N)}    \label{eq:Eigen-eq-q=0-app} \\ 
&+& k(k-1) \left[4 (N+1) (j-k-N+2) \right] a^{(k-2,N+1)}
+ a^{(k+2, N-1)} = 0.   \qquad \qquad ({\rm for~} {}^{\forall} k,N)  \nonumber 
\end{eqnarray}
Fields $a^{(k,N)}$'s with the same $k+2N = n$ form a system of coupled 
equations, but those with different $n = k + 2N$ do not mix.   
Thus, the eigenmode equation for $\Delta^\mu=0$ is decomposed into 
sectors labeled by $n$. The $n$-th sector consists of 
$z$-dependent fields that are all in the rank-$(j-n) = (j-k-2N)$ 
totally symmetric tensor of $\SO(3,1)$.

{\bf classification of eigenmodes for $\Delta^\mu=0$}

Let us now study the eigenmode equations more in detail for the 
separate diagonal blocks we have seen. 
Simultaneous treatmen is possible for all the $n$-th sectors with even 
$n$, and for all the sectors with odd $n$.

Let us first look at the $n$-th sector of the eigenmode problem for 
an $n=2\bar{n} \leq j$.
In the eigenmode equation of $\Delta^\mu=0$, we can assume\footnote{
This is because, in the absence of $z^2 \partial^2$ term in $\Delta_j$, 
it becomes a constant multiplication when it acts on a simple power of
$z$. Upon $z^{2-j-i\nu}$, for example, $R^2 \Delta_j$ returns $-(4+\nu^2)$.} 
the same $z$-dependence for all the fields in this diagonal block:
\begin{equation}
 a^{(k,N)}(z)_{\mu_1\cdots \mu_{j-n}} = 
   \bar{a}^{(k,N)}_{\mu_1 \cdots \mu_{j-n}} z^{2-j-i\nu}, \qquad
   \qquad  k+2N = n
\end{equation}
where $\bar{a}^{(k,N)}$'s are $(x,z)$-independent 4D-traceless 
rank-$(j-n)$ tensor of $\SO(3,1)$. The eigenmode equations with the 
label $(k,N) = (2\bar{k},\bar{n}-\bar{k})$ with 
$\bar{k}=0, \cdots, \bar{n}$ are relevant to the $n = 2\bar{n}$ sector, 
and are now written in a matrix form:
\begin{equation}
 \sum_{\bar{k}' = 0}^{\bar{n}} {\cal D}_{2\bar{k},2\bar{k}'}  \; 
   \bar{a}^{(2\bar{k}',\bar{n}-\bar{k}')} = 
   ((4+\nu^2)-{\cal E}) \; \bar{a}^{(2\bar{k}, \bar{n}-\bar{k})}, 
\label{eq:EGeq-q=0-matrix-n-even}
\end{equation}
where 
\begin{itemize}
 \item diagonal $(k,k')=(2\bar{k}, 2\bar{k})$ entry: 
   ${\cal D}_{2\bar{k},2\bar{k}} = -[(2k+1)j-2k^2+3k]$.
 \item diagonal$^+$; $(k,k') = (2\bar{k}, 2\bar{k}+2)$ entry: 
   ${\cal D}_{2\bar{k}, 2\bar{k}+2} = 1$.
 \item diagonal$^-$; $(k,k') = (2\bar{k}, 2\bar{k}-2)$ entry: 
   ${\cal D}_{2\bar{k},2\bar{k}-2} = k(k-1) \times 
  4(\bar{n}-\bar{k}+1)(j-\bar{n}-\bar{k}+2)$.
\end{itemize}
There must be $(\bar{n}+1)$ independent eigenmodes 
in this $(\bar{n}+1) \times (\bar{n}+1)$ matrix equation. 
Let us denote the collection of eigenvalues in this $n=2\bar{n}$-th 
diagonal block as ${\cal E}_{n,l}$, and 
$l=0,\cdots, \bar{n} = n/2 $ label distinct eigenmodes.  
Corresponding eigenmode wavefunction for the $(n=2\bar{n}, l)$ 
mode is given by 
\begin{equation}
 a^{(k,N)}(z; \Delta^\mu=0) = a^{(2\bar{k},\bar{n}-\bar{k})} 
= c_{2\bar{k},l,n}  \epsilon^{(n,l)} z^{2-j-i\nu},  
\end{equation}
where $\epsilon^{(n,l)}$ is an $(x,z)$-independent 4D-traceless 
totally symmetric rank-$(j-n)=(j-2\bar{n})$ tensor of $\SO(3,1)$, 
and $c_{2\bar{k},l,n}$ are $(x,z)$-independent constants determined 
as the eigenvector corresponding to the eigenvalue ${\cal E}_{n,l}$.

Similarly, in the $n=2\bar{n}+1 \leq j$-th sector of the 
eigenmode problem, with an odd $n$, we can assum a simple power law 
for all the component fields involved in this sector;  
\begin{equation}
 a^{(k,N)}(z)_{\mu_1 \cdots \mu_{j-n}} =
  \bar{a}^{(k,N)}_{\mu_1 \cdots \mu_{j-n}} z^{2-j-i\nu}, \qquad \qquad 
   k+2N = n,
\end{equation}
where $\bar{a}^{(k,N)}$ are $(x,z)$-independent 4D-traceless totally
symmetric tensor of $\SO(3,1)$. The eigenmode equation with the label 
$(k,N) = (2\bar{k}+1,\bar{n}-\bar{k})$ with $\bar{k}=0,\cdots,\bar{n}$ 
are relevant to this sector, and in the matrix form, the eigenmode
equation now looks 
\begin{equation}
 \sum_{\bar{k}' = 0}^{\bar{n}} {\cal D}_{2\bar{k}+1, 2\bar{k}'+1} 
  \bar{a}^{(2\bar{k}'+1,\bar{n}-\bar{k})} = 
  ((4+\nu^2)-{\cal E}) \; \bar{a}^{(2\bar{k}+1, \bar{n}-\bar{k})}, 
\label{eq:EGeq-q=0-matrix-n-odd}
\end{equation}
where 
\begin{itemize}
 \item diagonal $(k,k') = (2\bar{k}+1, 2\bar{k}+1)$ entry:
   ${\cal D}_{2\bar{k}+1,2\bar{k}+1} = -[(2k+1)j+(-2k^2+3k)]$. 
 \item diagonal$^+$ $(k,k') = (2\bar{k}+1, 2\bar{k}+3)$ entry: 
   ${\cal D}_{2\bar{k}+1, 2\bar{k}+3} = 1$.
 \item diagonal$^-$ $(k,k') = (2\bar{k}+1, 2\bar{k}-1)$ entry:
   ${\cal D}_{2\bar{k}+1, 2\bar{k}-1} = k(k-1) \times
    4(\bar{n}-\bar{k}+1)(j - \bar{n} - \bar{k} + 1)$.
\end{itemize}
From here, $\bar{n}+1$ independent modes arise; 
their eigenvalues are denoted by ${\cal E}_{n,l}$, and 
$l = \{ 0, \cdots, \bar{n}\}$ is the label distinguishing different modes.
The eigenmode labeled by $(n=2\bar{n}+1, l)$ has a wavefunction 
\begin{equation}
 a^{(k,N)}(z; \Delta^\mu=0) = 
 a^{(2\bar{k}+1, \bar{n}-\bar{k})} = 
 c_{2\bar{k}+1,l,n} \epsilon^{(n,l)} z^{2-j-i\nu},
\end{equation}
where $\epsilon^{(n,l)}$ is an $(x,z)$-independent 4D-traceless 
rank-$(j-n)$ totally symmetric tensor of $\SO(3,1)$, and 
$c_{2\bar{k}+1,l,n}$ is the eigenvector for the $(n,l)$ eigenmode
determined in the matrix equation above. 

{\bf explicit examples}

Let us take a moment to see how the general theory above works out 
in practice. 

{\bf The easiest of all is the $n = 0$ sector}, which contains 
only one rank-$j$ 4D-traceless field, $a^{(0,0)}$. The eigenmode equation is 
\begin{equation}
 \left[\Delta_j - \frac{\left[(2k+1)j-2k^2+3k\right]_{k=0}}{R^2} \right]
  a^{(0,0)}  = 
 \left[\Delta_j - \frac{j}{R^2} \right]  a^{(0,0)}  = 
 - \frac{ {\cal E}_{0,0}}{R^2} a^{(0,0)}.
\label{eq:EGeq-q=0-n=0-app}
\end{equation}
The eigenmode wavefunction is the form of 
\begin{equation}
 a^{(0,0)}(z)_{\mu_1 \cdots \mu_j} =
  \epsilon^{(0,0)}_{\mu_1 \cdots \mu_j} \; z^{2-j-i\nu}, 
\end{equation}
and the eigenvalue ${\cal E}_{n,l}$ is 
\begin{equation}
 {\cal E}_{0,0} = (j + 4 + \nu^2).
\label{eq:bgn-of-empirical-egval}
\end{equation}

{\bf Also to the $n=1$ sector}, there is only one rank-$(j-1)$
4D-traceless tensor field contributes. That is $a^{(1,0)}$.
The eigenmode equation becomes 
\begin{equation}
 \left[R^2 \Delta_j - [(2k+1)j-2k^2+3k]|_{k=1} \right] \; a^{(1,0)} =
 \left[R^2 \Delta_j - (3j+1) \right] \; a^{(1,0)}
 = - {\cal E}_{1,0} \; a^{(1,0)}.
\end{equation}
The solution is 
\begin{equation}
a^{(1,0)}(z)_{\mu_1 \cdots \mu_{j-1}} = 
 \epsilon^{(1,0)}_{\mu_1 \cdots \mu_{j-1}} z^{2-j-i\nu}, \qquad 
{\cal E}_{1,0} = (3j + 5 + \nu^2). 
\end{equation}

{\bf In the $n=2$ sector}, two rank-$(j-2)$ 4D-traceless fields are
involved. That is $a^{(0,1)}$ and $a^{(2,0)}$. After introducing the 
$z$-dependence $\propto z^{2-j-i\nu}$, the eigenmode equation 
(\ref{eq:EGeq-q=0-matrix-n-even}) in the $n=2$ sector becomes 
\begin{equation}
 \left[ \begin{array}{cc}
  -j & 1 \\ 8j & - (5j-2)
	\end{array}\right]
\left(\begin{array}{c}
 \bar{a}^{(0,1)} \\ \bar{a}^{(2,0)}
      \end{array}\right)
 = ((4+\nu^2)-{\cal E}) 
\left(\begin{array}{c}
 \bar{a}^{(0,1)} \\ \bar{a}^{(2,0)}
      \end{array}\right).
\end{equation}

One of the two eigenmode is 
\begin{equation}
 {\cal E}_{2,0} =(4+5j+\nu^2), \qquad 
\left( \begin{array}{c}
 a^{(0,1)}(z)_{\mu_1 \cdots \mu_{j-2}} \\
 a^{(2,0)}(z)_{\mu_1 \cdots \mu_{j-2}}
       \end{array}\right) =
 \left( \begin{array}{c}
  1 \\ -4j
	\end{array}\right) \; 
 \epsilon^{(2,0)}_{\mu_1 \cdots \mu_{j-2}} \; z^{2-j-i\nu},
\end{equation}
and the other 
\begin{equation}
 {\cal E}_{2,1} = (2+j + \nu^2), \qquad 
\left( \begin{array}{c}
 a^{(0,1)}(z)_{\mu_1 \cdots \mu_{j-2}} \\
 a^{(2,0)}(z)_{\mu_1 \cdots \mu_{j-2}}
       \end{array}\right) =
 \left( \begin{array}{c}
  1 \\ 2
	\end{array}\right) \; 
 \epsilon^{(2,1)}_{\mu_1 \cdots \mu_{j-2}} \; z^{2-j-i\nu}.
\end{equation}

{\bf In the $n=3$ sector}, two rank-$(j-3)$ 4D-traceless tensor fields
are involved: $a^{(1,1)}$ and $a^{(3,0)}$.
The eigenmode equations (\ref{eq:EGeq-q=0-matrix-n-odd}) become
\begin{equation}
 \left[ \begin{array}{cc}
  -(3j+1) & 1 \\ 24 (j-1) & - (7j-9)
	\end{array}\right]
\left(\begin{array}{c}
 \bar{a}^{(1,1)} \\ \bar{a}^{(3,0)}
      \end{array}\right)
 = ((4+\nu^2)-{\cal E}) 
\left(\begin{array}{c}
 \bar{a}^{(1,1)} \\ \bar{a}^{(3,0)}
      \end{array}\right). 
\end{equation}
So, one of the two eigenmodes is
\begin{equation}
{\cal E}_{3,0} = (7j + 1 + \nu^2), \qquad 
 \left( \begin{array}{c}
  a^{(1,1)}(z)_{\mu_1 \cdots \mu_{j-3}} \\
  a^{(3,0)}(z)_{\mu_1 \cdots \mu_{j-3}} 
	\end{array}\right) = 
 \left( \begin{array}{c}
  1 \\ -4(j-1)
	\end{array}\right)
 \epsilon^{(3,0)}_{\mu_1 \cdots \mu_{j-3}} \; z^{2-j-i\nu}, 
\end{equation}
and the other one is 
\begin{equation}
 {\cal E}_{3,1} = (3j -1 + \nu^2), \qquad 
 \left( \begin{array}{c}
  a^{(1,1)}(z)_{\mu_1 \cdots \mu_{j-3}} \\
  a^{(3,0)}(z)_{\mu_1 \cdots \mu_{j-3}} 
	\end{array}\right) = 
 \left( \begin{array}{c}
  1 \\ 6
	\end{array}\right)
 \epsilon^{(3,1)}_{\mu_1 \cdots \mu_{j-3}} \; z^{2-j-i\nu}.
\end{equation}

{\bf Finally, in the $n=4$ sector}, 
the eigenmode equation (\ref{eq:EGeq-q=0-matrix-n-even}) is given by 
\begin{equation}
 \left[ \begin{array}{ccc}
  -j & 1 & 0 \\ 16(j-1) & -(5j-2) & 1 \\ 0 & 48(j-2) & -(9j-20)
	\end{array}\right]
 \left( \begin{array}{c}
  \bar{a}^{(0,2)} \\ \bar{a}^{(2,1)} \\ \bar{a}^{(4,0)}
	\end{array}\right) = 
 ((4+\nu^2)-{\cal E}) 
 \left( \begin{array}{c}
  \bar{a}^{(0,2)} \\ \bar{a}^{(2,1)} \\ \bar{a}^{(4,0)}
	\end{array}\right).
\end{equation}
There are three solutions. First, 
\begin{eqnarray}
 {\cal E}_{4,0} & = & (9j-4+\nu^2),  \\
 (a^{(0,2)}, a^{(2,1)}, a^{(4,0)}) & = & 
 (1, -8(j-1),32(j-1)(j-2)) \; \epsilon^{(4,0)} \; z^{2-j-i\nu}, 
\end{eqnarray}
second, 
\begin{eqnarray}
 {\cal E}_{4,1} & = & (5j-6+\nu^2),  \\
(a^{(0,2)}, a^{(2,1)}, a^{(4,0)}) & = & 
 (1, -(4j-10), -48(j-2)) \; \epsilon^{(4,1)} \; z^{2-j-i\nu}, 
\end{eqnarray}
and finally, 
\begin{eqnarray}
 {\cal E}_{4,2} & = & (j + \nu^2),  \\
(a^{(0,2)}, a^{(2,1)}, a^{(4,0)}) & = & 
 (1, 4, 24) \; \epsilon^{(4,2)} \; z^{2-j-i\nu}.  
\label{eq:end-of-empirical-egval}
\end{eqnarray}

An empirical relation is observed in the $j$-dependence of the 
eigenvalues we have worked out so far. The eigenvalues 
in the $n$-the sector are in the form of 
${\cal E}_{n,l} = \nu^2 + (2n+1-4l)j + {\cal O}(1)$ for 
$0 \leq l \leq [n/2]$.

{\bf 5D-traceless modes: the $l=0$ modes}


Although the precise expressions for the eigenvalues ${\cal E}_{n,l}$ 
and eigenvectors $c_{k,l,n}$ are not given for all the eigenmodes, 
there is a class of eigenmodes whose eigenvalues and eigenvectors 
(wavefunctions) are fully understood. 

As we discussed in p.~\pageref{page:5D-trl-trv}, it is compatible to 
require both a field is an eigenmode and satisfies the 5D-traceless
condition (\ref{eq:cond-5D-traceless}).
In the $n=(k+2N)$-th sector, the 5D-traceless condition becomes 
\begin{eqnarray}
0 & = & 
 \left( E^N[a^{(k,N)}] \right)^{\hat{\rho}}_{\; \rho \mu_3 \cdots \mu_{j-n}}
+ \left( E^{N-1}[a^{(k+2,N-1)}] \right)_{\mu_3 \cdots \mu_{j-n}}, 
 \label{eq:5D-traceless-q=0-explicit} \\
 & = & E^{N-1} \left[ 4N(j-n+N+1) a^{(k,N)} + a^{(k+2,N-1)} \right] 
 \quad 
 \left\{ 
\begin{array}{l}
 k=0,2,\cdots, 2(\bar{n}-1)   \quad ({\rm even~}n), \\
 k=1,3,\cdots, 2\bar{n}-1  \quad ({\rm odd~}n).
\end{array}
 \right. \nonumber
\end{eqnarray}
Thus, the 5D-traceless condition uniquely determines one eigenmode 
in each one of the $n$-th sector. 
\begin{equation}
  {\cal E}_{n,0} = (2n+1)j + 2n - n^2 + 4+\nu^2, 
\end{equation}
and 
\begin{equation}
 c_{2\bar{k},0,2\bar{n}} = (-)^{\bar{k}} 4^{\bar{k}}
   \frac{ \bar{n}! }{ (\bar{n}-\bar{k})! }
   \frac{ (j-\bar{n}+1)! }{ (j-\bar{n}-\bar{k}+1)! }, \quad 
 c_{2\bar{k}+1,0,2\bar{n}+1} = (-)^{\bar{k}} 4^{\bar{k}}
   \frac{ \bar{n}! }{ (\bar{n}-\bar{k})! }
   \frac{ (j-\bar{n})! }{ (j-\bar{n}-\bar{k})! }. 
\end{equation}
%



\subsection{Mode Decomposition for non-zero $\Delta_\mu$}
\label{ssec:appendix-eigen-q=not0}

\subsubsection{Diagonal Block Decomposition for the $\Delta^\mu \neq 0$ Case}

Let us now turn our attention to the eigenmode 
equation (\ref{eq:Eigenequation}, \ref{eq:Eigenequation-4+1}) 
with $\Delta^\mu \neq 0$. Because of the 2nd and 4th terms in 
(\ref{eq:Eigenequation-4+1}), the eigenmode problem becomes much more 
complicated. We begin by finding diagonal block decomposition 
suitable for the case with $\Delta^\mu \neq 0$.

In the main text, we introduced a decomposition of a totally symmetric 
rank-$j$ tensor field $A_{m_1 \cdots m_j}$ of $\SO(4,1)$ into a collection 
of totally symmetric 4D-traceless 4D-transverse tensor fields of
$\SO(3,1)$. Instead of (\ref{eq:block-dcmp-q=0}), a new
decomposition is given by (\ref{eq:block-dcmp-q=not0-tilde}=\ref{eq:block-dcmp-q=not0-tilde-app}):  
\begin{equation}
 A_{z^k \mu_1 \cdots \mu_{j-k}}(z; \Delta^\mu)  = 
 \sum_{s=0}^{j-k} \sum_{t=0}^{[s/2]} 
   \left( \tilde{E}^N D^{s-2N} [ a^{(k,s,N)} ] \right)_{\mu_1 \cdots \mu_{j-k}},
 \label{eq:block-dcmp-q=not0-tilde-app}  
\end{equation}
where $a^{(k,s,N)}$ are totally symmetric 4D-traceless 4D-transverse 
rank-$(j-k-s)$ tensor fields of $\SO(3,1)$. 
An operation $a \mapsto \tilde{E}[a]$ on a symmetric 
$\SO(3,1)$ tensor $a$ is given by (\ref{eq:def-E-tilde}).

In order to see that the parametrization of 
$A_{z^k \mu_1 \cdots \mu_{j-k}}$ by 
$(a^{(k,s,N)})_{\mu_1 \cdots \mu_{j-k-s}}$'s above is indeed a
decomposition, one needs to see that $a^{(k,s,N)}$'s can be retrieved 
from $A_{z^k \mu_1 \cdots \mu_{j-k}}$, so that the degrees of freedom 
$a^{(k,s,N)}$ are independent.
For this purpose, it is convenient to derive some relations analogous to 
(\ref{eq:recursion-E-by-eta}). First of all, note that 
$E [D[a]] = D[E[a]]$ and\footnote{%
\begin{equation}
 E^t D^{s-2t}[ a ] = \sum
  \eta_{\mu_{p_1} \mu_{p_2}} \cdots \eta_{\mu_{p_{2t-1}} \mu_{p_{2t}} }
  \partial_{\mu_{p_{2t+1}} } \cdots \partial_{\mu_{p_s}} 
  \left[ a \right]_{\mu_1 \cdots \check{} \check{} \cdots
  \mu_{r+s}},
\label{eq:def-DandE}
\end{equation}
where the sum is taken over all possible ordered choices of 
$p_1, p_2, \cdots, p_s \in \{ 1, \cdots, j \}$ such that 
$p_i \neq p_j$ for $i \neq j$. 
} $\tilde{E}[ D[a]] = D[\tilde{E}[a]]$ for a totally symmetric 
$\SO(3,1)$ tensor $a$. 
If the rank-$r$ tensor $a$ is also 4D-transverse and 4D-traceless, then 
one can derive the following relations:
\begin{eqnarray}
 \partial^{\hat{\rho}} 
  \left( E^t D^{s-2t} \left[ a \right] \right)_{\rho \mu_2 \cdots
  \mu_{r+s}} & = & 
  -\Delta^2 (s-2t) E^t D^{s-2t-1}\left[a \right]
  + (2t) E^{t-1} D^{s-2t+1} \left[ a \right], 
    \label{eq:recursion-ED-by-partial}\\
 \eta^{\hat{\rho} \hat{\sigma}}
  \left(E^t D^{s-2t} \left[ a \right] \right)_{\rho \sigma \mu_3
  \cdots \mu_{r+s}} & = &
   - \Delta^2 (s-2t)(s-2t-1) E^t D^{s-2t-2} \left[ a \right] 
   \nonumber \\
  & & + 4t(r+s-t+1) E^{t-1} D^{s-2t} \left[ a \right]. 
    \label{eq:recursion-ED-by-eta}
\end{eqnarray}
\begin{eqnarray}
\partial^{\hat{\rho}}
 \left( \tilde{E}^N D^{s-2N} [ a ]\right)_{\rho \mu_2 \cdots \mu_{r+s}} & = &  
-(s-2N) \Delta^2 \tilde{E}^N D^{s-2N-1} [ a ],  \\
 \left( \eta^{\hat{\rho}\hat{\sigma}} -
        \frac{ \partial^{\hat{\rho}} \partial^{\hat{\sigma}} }
             {\partial^2} 
 \right)
 \left(\tilde{E}^N D^{s-2N}[ a ]
 \right)_{\rho \sigma \mu_3 \cdots \mu_{r+s}} & = & 
 4N(r+N+1/2) \tilde{E}^{N-1} D^{s-2N}[ a ]. 
\end{eqnarray}
With the relations above, it is now possible to compute 
\begin{eqnarray}
&&
 \left(\eta^{\hat{\mu}_1 \hat{\mu}_2}
      - \frac{\partial^{\hat{\mu}_1} \partial^{\hat{\mu}_2}}
             {\partial^2} \right)
   \cdots 
 \left(\eta^{\hat{\mu}_{2p-1} \hat{\mu}_{2p}}
    - \frac{\partial^{\hat{\mu}_{2p-1}} \partial^{\hat{\mu}_{2p}}}
           {\partial^2} \right)
 \; 
 \frac{\partial^{\hat{\mu}_{2p+1}}}{\partial^2} \cdots
 \frac{\partial^{\hat{\mu}_{2p+q}}}{\partial^2}  
 \; 
 \left( \tilde{E}^N D^{s-2N}[a] \right)_{\mu_1 \cdots \mu_{r+s}}
   \nonumber  \\
& = & \left\{ \begin{array}{ll}
   \frac{b^{(r)}_{s-2p-q, N-p}}{b^{(r)}_{s,N}}
   \left(\tilde{E}^{N-p} D^{s-2N-q}[a] \right)_{\mu_{2p+q+1} \cdots \mu_{r+s}}
      & {\rm if~} p \leq N {\rm ~and~} q \leq s-2N,\\
0 & {\rm otherwise}
      \end{array}\right. 
\label{eq:extract-SO(3,1)-tensor}
\end{eqnarray}
where we assume that $a$ is a totally symmetric 4D-traceless
4D-transverse rank-$r$ tensor of $\SO(3,1)$.
In the last line, 
\begin{equation}
 b^{(r)}_{s,N} := \frac{1}{4^N N! (s-2N)!}
  \frac{\Gamma\left(r+3/2\right)}{\Gamma \left( r+N+3/2 \right)}.
\label{eq:def-of-b}
\end{equation}
%
%
%

It is now clear how to retrieve $a^{(k,s,N)}$ 
from $A_{z^k \mu_1 \cdots \mu_{j-k}}$ given by
(\ref{eq:block-dcmp-q=not0-tilde}=\ref{eq:block-dcmp-q=not0-tilde-app}). 
First, one has to multiply 
$\eta^{\hat{\rho}\hat{\sigma}} - \partial^{\hat{\rho}} 
\partial^{\hat{\sigma}}/\partial^2$ and
$\partial^{\hat{\sigma}}/\partial^2$ to $A_{z^k \mu_1 \cdots \mu_{j-k}}$
as many times as possible in order to obtain $a^{(k,s,N)}$ with larger 
$N$ and $(s-2N)$. Then $a^{(k,s,N)}$'s with smaller $N$ or $(s-2N)$ can 
be determined by multiplying 
$\eta^{\hat{\rho}\hat{\sigma}} - \partial^{\hat{\rho}} 
\partial^{\hat{\sigma}}/\partial^2$ and
$\partial^{\hat{\sigma}}/\partial^2$ fewer times. 

Let us now return to the eigenmode equation for the cases with 
$\Delta^\mu \neq 0$. Following precisely the same argument as in 
section \ref{ssec:appendix-eigen-q=0}, one can see that the eigenmode 
equation can be separated into the following independent equations 
labeled by $k, s$ and $N$: 
\begin{eqnarray}
& &
 \left[R^2 \Delta_j - \left[(2k+1)j-2k^2+3k \right]+ {\cal E}\right] \; 
      a^{(k,s,N)}     \nonumber \\
& + &  2z k(s+1-2N) (\partial^2) \; a^{(k-1,s+1,N)}  \nonumber \\
& + & k(k-1)(s+2-2N)(s+1-2N) (\partial^2) \; a^{(k-2,s+2,N)}  \\
&& + 4k(k-1)(N+1) (j-m+N+3/2) \; a^{(k-2,s+2,N+1)}  \nonumber \\
& - & 2z \; a^{(k+1,s-1,N)}
+ a^{(k+2,s-2,N-1)}
+ (\partial^2)^{-1} \; a^{(k+2,s-2,N)} = 0 
\qquad {\rm for~}{}^\forall k,s,N.   \nonumber 
\label{eq:EGeq-q=not0-matrix-m}
\end{eqnarray}
The relations (\ref{eq:recursion-ED-by-partial},
\ref{eq:recursion-ED-by-eta}) was used to evaluate the 2nd--4th terms 
of (\ref{eq:Eigenequation-4+1-app}).
Onc can see that $a^{(k,s,N)}$'s with a common value of $m := k+s$
form a coupled eigenmode equations, but those with different $m$'s 
do not. Thus, $a^{(k,s,N)}(z; \Delta^\mu)$'s with $k+s = m$ form 
the $m$-th subspace of $A_{m_1 \cdots m_j}(z; \Delta^\mu)$, and the 
eigenmode equation becomes block diagonal in the decomposition 
into the subspaces labeled by $m =0, \cdots, j$.

The eigenmode equation on the $m$-th subspace is given by 
the equation above with $0 \leq k =(m-s) \leq m$, and 
$0 \leq N \leq [s/2]$. Thus, the total number of equations is 
\begin{equation}
 \sum_{s=0}^m \left( [s/2] + 1 \right), 
\label{eq:nmb-eqn-m-sector-app}
\end{equation} 
and the same number of eigenvalues should be obtained from the $m$-th sector.

\subsubsection{Examples}

{\bf The sector} $m=0$: There is only one field $a^{(0,0,0)}$ in this
sector, and the eigenmode equation is 
\begin{equation}
 \left[\Delta_j - \frac{j}{R^2} \right] a^{(0,0,0)}(z; \Delta^\mu) 
 = - \frac{{\cal E}}{R^2} a^{(0,0,0)}(z; \Delta^\mu).
\end{equation}
Assuming a power series expansion for the solution to this equation, 
beginning with some power $z^{2-j-i\nu}$, the eigenvalue is 
determined as a function of $(i\nu)$:
%
\begin{equation}
 {\cal E}_{0,0} = (j+4+\nu^2), \nonumber 
\end{equation}
and the wavefunction can be chosen as  
\begin{eqnarray}
 a^{(0,0,0)}(z; \Delta^\mu)_{\mu_1 \cdots \mu_j} & = &  
 \epsilon^{(0,0,0)}_{\mu_1 \cdots \mu_j} \;
 \Psi^{(j)}_{i\nu}(- \Delta^2, z), \\
\Psi^{(j)}_{i\nu}(\Delta^2,z) & := & 
  \frac{2}{\pi}\sqrt{\frac{\nu \sinh (\pi \nu)}{2R}} \; 
  e^{(j-2)A} K_{i\nu}(\Delta z).
\end{eqnarray}

{\bf The sector} $m=1$: The eigenmode equation in this sector becomes 
\begin{equation}
 \left[ \begin{array}{cc}
  R^2 \Delta_j - j & -2z \\ -2z \Delta^2 & R^2 \Delta_j - (3j+1) 
	\end{array}\right]
\left(  \begin{array}{c}
 a^{(0,1,0)} \\ a^{(1,0,0)}
	\end{array}\right) = 
 - {\cal E}
\left(  \begin{array}{c}
 a^{(0,1,0)} \\ a^{(1,0,0)}
	\end{array}\right). 
\label{eq:EGeq-q=not0-m=1}
\end{equation}
Assuming the power series expansion in $z$, beginning with 
$z^{2-j-i\nu}$ terms, we obtain two eigenvalues depending on $i\nu$. 
They are given by evaluating $R^2 \Delta_j - j$ and 
$R^2 \Delta_j - (3j+1)$ on $z^{2-j-i\nu}$: 
\begin{equation}
 {\cal E}_{0,0} = (j+4+\nu^2), \quad {\rm and} \quad 
 {\cal E}_{1,0} = (3j+5+\nu^2).
\end{equation}

{\bf The sector} $m=2$: The eigenmode equation becomes 
%
%
\begin{equation}
 \left(
(R^2 \Delta_j + {\cal E}) {\bf 1}_{4\times 4} + 
 \left[ \begin{array}{cccc}
  -j & & -2z & 1/\partial^2 \\
  & -j & & 1 \\
  4z \partial^2 &  & -(3j+1) & -2z \\
  4\partial^2 & 8j-4 & 4z \partial^2 & -(5j-2)
	\end{array}\right]
\right)
\left( \begin{array}{c}
 a^{(0,2,0)} \\
 a^{(0,2,1)}  \\
 a^{(1,1,0)}  \\
 a^{(2,0,0)}
       \end{array}\right) = 0.
\end{equation}
The indicial equation relating the exponent $(2-j-i\nu)$ at $z = 0$ and 
the eigenvalues split into two parts; three eigenvalues of this matrix 
\begin{equation}
 \left(\begin{array}{ccc}
  -j & & 1 \\ & -j & 1 \\ 4 & (8j-4) & -(5j-2)
       \end{array}\right),
\end{equation}
determine $-{\cal E}-(4+\nu^2)$ for the three eigenmodes, 
and $-({\cal E}-(4+\nu^2)) = - (3j+1)$ for the last eigenmode.
Therefore, the four eigenvalues in the $m=2$ sector are 
\begin{equation}
 {\cal E}_{0,0}=(j+4+\nu^2), \quad 
 {\cal E}_{1,0}=(3j+5+\nu^2), \quad 
 {\cal E}_{2,0}=(5j+4+\nu^2), \quad 
 {\cal E}_{2,1}=(j+2+\nu^2).
\end{equation}
%


In all the examples above, the $m$-th sector consists of 
eigenmodes with eigenvalues ${\cal E}_{n,l}$ for 
$0 \leq n \leq m$, $0 \leq l \leq [n/2]$. The number of eigenmodes 
is, of course, the same as (\ref{eq:nmb-eqn-m-sector-app}).

\subsection{Wavefunctions of 5D-Traceless 5D-Transverse Modes}
\label{ssec:appendix-5D-trltrv}

As we discussed toward the end of section \ref{ssec:eigen-q=not0}, 
it is possible to require for a rank-$j$ totally symmetric tensor 
field configuration $A_{m_1 \cdots m_j}(z; \Delta^\mu)$ to be 
an eigenmode and to be 5D-traceless 5D transverse
(\ref{eq:cond-5D-traceless}, \ref{eq:cond-5D-transverse}) at the same
time. We will see in the following that the these two extra conditions 
(\ref{eq:cond-5D-traceless}, \ref{eq:cond-5D-transverse}) leave
precisely one eigenmode in each one of the block-diagonal sector labeled
by $m = 0, \cdots, j$. We will further determined the wavefunction
profile of such eigenmodes. 

Let us first rewrite the 5D-traceless condition 
(\ref{eq:cond-5D-traceless}) in a more convinient form. 
\begin{equation}
  \eta^{\hat{\rho} \hat{\sigma}} 
  A_{z^{k-2}\rho\sigma \mu_1 \cdots \mu_{j-k}}
   + A_{z^k \mu_1 \cdots \mu_{j-k}} = 0,   
\label{eq:5D-trac-cond}
\end{equation}
which, in the $m$-th sector, means 
\begin{equation}
 a^{(k,s,N)} = (s+2-2N)(s+1-2N) \Delta^2 a^{(k-2,s+2,N)}
    + 4 (N+1)(j-m+N+3/2) a^{(k-2,s+2,N+1)}  
\label{eq:5D-trac-cond-EZ}
\end{equation}
for $N=0, \cdots , [s/2]$; $k+s = m$ is understood. 
Under the 5D-traceless condition, the 5D-transverse condition 
\begin{equation}
 (k-1) \eta^{\hat{\rho} \hat{\sigma}}
     A_{z^{k-2}\rho\sigma \mu_1 \cdots \mu_{j-k}} + 
  z \partial^{\hat{\rho}} 
     A_{z^{k-1}\rho \mu_1 \cdots \mu_{j-k}} + 
  \left(z \partial_z + (k-4) \right)
     A_{z^k \mu_1 \cdots \mu_{j-k}} = 0, 
\end{equation}
becomes
\begin{equation}
   z \partial^{\hat{\rho}} 
       A_{z^{k-1}\rho \mu_1 \cdots \mu_{j-k}} + 
   \left(z \partial_z -3 \right) A_{z^k \mu_1 \cdots \mu_{j-k}} = 0.
  \label{eq:5D-transv-cond-splfd}
\end{equation}
In the $m$-th sector ($k+s = m$), therefore, 
\begin{equation}
 (s+1-2N) \Delta^2 a^{(k-1,s+1,N)} = z^3 \partial_z z^{-3} a^{(k,s,N)}
\label{eq:5D-transv-cond-splfd-EZ}
\end{equation}
for $N=0,\cdots, [s/2]$. Hereafter, we use a simplified notation 
${\cal D} := z^3 \partial_z z^{-3}$.
One can see that all of $a^{(k,s,N)}$'s with $k+s = m$ and $N \leq
[s/2]$ can be determined from $a^{(m,0,0)}$ by using the 
relations (\ref{eq:5D-trac-cond-EZ}, \ref{eq:5D-transv-cond-splfd-EZ}).
This observation already implies that there can be at most one 
eigemode in a given $m$-th sector that satisfies both the 
5D-traceless and 5D-transverse conditions. 

For now, let us assume that there is one, and proceed to determine 
the wavefunction. 
The wavefunction---$z$-dependence---of $a^{(m.0.0)}(z; \Delta^\mu)$ 
can be determined from the eigenmode equation
(\ref{eq:EGeq-q=not0-matrix-m}) with $k = m$, $s=N=0$.
Using (\ref{eq:5D-trac-cond}) and (\ref{eq:5D-transv-cond-splfd}), 
we can rewrite the equation as 
\begin{equation}
 \left[ R^2 \Delta_j - \{ (2m+1)j -m^2 + 2m \} 
       - 2 m \left(z \partial_z - 3 \right)
       +{\cal E} \right] 
   a^{(m,0,0)}(z;\Delta)
 = 0. 
\end{equation}
For this equation, 
\begin{equation}
 \left(a^{(m,0,0)}(z;\Delta)\right)_{\mu_1 \cdots \mu_{j-m}} = 
 \epsilon_{\mu_1 \cdots \mu_{j-m}} 
  \left(\frac{z}{R}\right)^{2-j} (\Delta z)^m K_{i\nu}(\Delta z), 
    \qquad 
{\cal E} = (j+4+\nu^2), 
\end{equation}
is a solution, where $\epsilon_{\mu_1 \cdots \mu_{j-m}}$ is a
$z$-independent 4D-traceless 4D-transverse totally symmetric 
rank-$(j-m)$ tensor of $\SO(3,1)$. From the value of the eigenvalue, 
it turns out that the 5D-traceless 5D-transverse mode in the $m$-th
sector corresponds to the $(n,l,m) = (0,0,m)$ node. 
The $z$-dependence we determined above implies that 
\begin{equation}
 \Psi^{(j);0,0}_{i\nu; 0,0,m} (-\Delta^2,z) 
  \propto   (\Delta z)^{m} \Psi^{(j);0,0}_{i\nu; 0,0,0} (-\Delta^2, z).
\label{eq:result-Psi-00}
\end{equation}
This result corresponds to the $(s,N) = (0,0)$ case of (\ref{eq:00m-mode-fcn}).
The normalization constant $N_{j,m}$ is determined later in this 
section.

Let us now proceed to determine other $\Psi^{(j);s,N}_{i\nu;0,0,m}$, 
not just for $(s,N)=(0,0)$. 
Using the 5D-transverse condition, (\ref{eq:5D-transv-cond-splfd-EZ}), 
$a^{(m-1,1,0)}(z; \Delta)$ can be determined from $a^{(m,0,0)}(z;\Delta)$. 
\begin{equation}
 a^{(m-1,1,0)} = \frac{{\cal D}}{\Delta^2} a^{(m,0,0)}, \qquad 
 \Psi^{(j);1,0}_{i\nu;0,0,m} = 
   \frac{{\cal D}}{\Delta} \Psi^{(j);0,0}_{i\nu;0,0,m}.
\end{equation}

In order to determine the $s=2$ components $a^{(m-2,2,N)}$ 
($N=0,1$) of the $(n,l)=(0,0)$ mode in the $m$-th sector, 
one has to use both the 5D-transverse condition and 5D-traceless condition: 
\begin{eqnarray}
 2 \Delta^2 a^{(m-2,2,0)} & = & {\cal D} \; a^{(m-1,1,0)}, \\ 
 2 \Delta^2 a^{(m-2,2,0)}
 -  4(j-m+3/2) a^{(m-2,2,1)} & = & a^{(m,0,0)} .
\end{eqnarray}
Therefore, 
\begin{equation}
 a^{(m-2,2,0)} = \frac{1}{2\Delta^2} 
   \left(\frac{{\cal D}}{\Delta}\right)^2 a^{(m,0,0)},  \qquad 
 a^{(m-2,2,1)} = \frac{1}{4(j-m+3/2)}
   \left\{ \left(\frac{{\cal D}}{\Delta}\right)^2 - 1 \right\}
   a^{(m,0,0)}, 
%
\end{equation}
After factoring out a normalization factor
$b^{(j-m)}_{s,N}/\Delta^{s-2N}$ and the common 4D-tensor 
$\epsilon^{(0,0,m)}$, we obtain 
\begin{equation}
 \Psi^{(j);2,0}_{i\nu;0,0,m} = \left(\frac{{\cal D}}{\Delta}\right)^2
    \Psi^{(j);0,0}_{i\nu;0,0,m}, \qquad 
 \Psi^{(j);2,1}_{i\nu;0,0,m} = 
   \left\{ \left(\frac{{\cal D}}{\Delta}\right)^2 - 1 \right\}
   \Psi^{(j);0,0}_{i\nu;0,0,m}. 
\end{equation}

The 5D-transverse conditions (\ref{eq:5D-transv-cond-splfd-EZ}) 
determine the $s=3$ components $a^{(m-3,3,N)}(z;\Delta)$ ($N=0,1$) 
from the $s=2$ components.
\begin{equation}
 a^{(m-3,3,0)} = \frac{1}{6\Delta^3} 
   \left(\frac{{\cal D}}{\Delta}\right)^3 a^{(m,0,0)},  \quad 
 a^{(m-3,3,1)} = \frac{1}{4(j-m+3/2)\Delta}
   \left\{ \left(\frac{{\cal D}}{\Delta}\right)^3
         -  \left(\frac{{\cal D}}{\Delta}\right) \right\}
   a^{(m,0,0)}, 
\end{equation}
and after factoring out the normalization factor 
$b^{(j-m)}_{s,N}/\Delta^{s-2N}$ and $\epsilon^{(0,0,m)}$ as before, 
we obtain 
\begin{equation}
 \Psi^{(j);3,0}_{i\nu;0,0,m} =
    \left(\frac{{\cal D}}{\Delta}\right)^3 \Psi^{(j);0,0}_{i\nu;0,0,m},
   \qquad 
\Psi^{(j);3,1}_{i\nu;0,0,m} = 
   \left\{  \left(\frac{{\cal D}}{\Delta}\right)^3 
          -  \left(\frac{{\cal D}}{\Delta}\right) \right\}
     \Psi^{(j);0,0}_{i\nu;0,0,m}.
\end{equation}
The $s=3$ components determined purely by the conditions
(\ref{eq:5D-transv-cond-splfd-EZ}) satisfy the 5D-traceless condition 
(\ref{eq:5D-trac-cond-EZ}) with the $s=1$ component:
\begin{equation}
 6\Delta^2 a^{(m-3,3,0)} - 4(j-m+3/2)a^{(m-3,3,1)} =
   \frac{{\cal D}}{\Delta^2} a^{(m,0,0)} = a^{(m-1,1,0)}.
\end{equation}

In this way, the wavefunctions all the 
$\Psi^{(j);s,N}_{i\nu;0,0,m}(-\Delta^2, z)$ are determined, and
the result is 
\begin{equation}
  \Psi^{(j);s,N}_{i\nu;0,0,m}(-\Delta^2,z) =
    \sum_{a=0}^N (-)^a {}_N C_a 
      \left( \frac{{\cal D}}{\Delta}\right)^{s-2a} 
        \left[(z \Delta)^m \Psi^{(j);0,0}_{i\nu;0,0,0}(-\Delta^2, z)
	\right] \times N_{j,m}.
\label{eq:00m-mode-fcn-app}
\end{equation}
(\ref{eq:00m-mode-fcn}). The only remaining concern 
was that the there are more conditions from (\ref{eq:5D-trac-cond-EZ}, 
\ref{eq:5D-transv-cond-splfd-EZ}) than the number of components 
$a^{(k,s,N)}$ in the $m$-th sector; there can be at most one eigenmodes 
satisfying these 5D-traceless 5D-transverse conditions, as we stated
earlier, but there may be no eigenomde left, if the conditions are
overdetermining. We have confirmed, however, that the wavefunctions 
(\ref{eq:00m-mode-fcn}) satisfy all of the relations given by 
(\ref{eq:5D-trac-cond-EZ}, \ref{eq:5D-transv-cond-splfd-EZ}).

\subsubsection{Normalization}

We have yet to determine the normalization factor $N_{j,m}$; 
as in the main text, we choose (\ref{eq:wvfc-normalization-cond}) 
to be the normalization condition. Orthogonal nature among the
eigenmodes is guaranteed because of the Hermitian nature of the 
operator $\alpha' \left( \nabla^2 - M^2 \right)$.
It is thus sufficient to focus only on the divergent part of the 
integral in the normalization condition in order to determine $N_{j,m}$.

The divergent part of the integral in (\ref{eq:wvfc-normalization-cond})
comes only from terms with $s = m$, $k=0$, $(0 \leq N \leq [m/2])$ 
and $a = 0$. For a given $m$, 
\begin{eqnarray}
&& [ \epsilon \cdot \epsilon'] \; \delta(\nu-\nu') 
   \label{eq:normalization-temp-app}\\
& \sim & 
 N_{j,m}^2 \int_0 dz \sqrt{-g(z)} e^{-2jA} \nonumber \\
& & \quad 
  \left(\sum_{N=0}^{[m/2]} \tilde{E}^N D^{m-2N}[\epsilon^{(0,0,m)}]
     \frac{b^{(j-m)}_{m,N}}{\Delta^{m-2N}} 
     \frac{z^3 \partial_z^{m} z^{-3}}{\Delta^m}
     (z\Delta)^m \Psi^{(j);0,0}_{i\nu;0,0,m}(- \Delta^2,z)
  \right)_{\mu_1 \cdots \mu_j}
   \nonumber  \\
 & & \quad 
  \left(\sum_{M=0}^{[m/2]} \tilde{E}^M D^{m-2M}[\epsilon^{(0,0,m)}]
     \frac{b^{(j-m)}_{m,M}}{\Delta^{m-2M}} 
     \frac{z^3 \partial_z^{m} z^{-3}}{\Delta^m}
     (z\Delta)^m \Psi^{(j);0,0}_{i\nu;0,0,m}(- \Delta^2,z)  
  \right)^{\hat{\mu}_1 \cdots \hat{\mu}_j}.
   \nonumber 
\end{eqnarray}
Divegent part of the integral in this expression comes from 
\begin{eqnarray}
 & &
 \left(\frac{2}{\pi}\right)^2 \frac{\nu \sinh(\pi\nu)}{2}
  \int dx x^{2j-5}
    \left[x^3 \partial_x^m x^{-1-j+m} K_{i\nu}(x) \right]
    \left[x^3 \partial_x^m x^{-1-j+m} K_{i\nu'}(x) \right] \nonumber \\
& \simeq & 
   \prod_{p=1}^m \left[ (j-p+1)^2 + \nu^2 \right] \delta(\nu-\nu') 
 = \frac{ \Gamma(j+1-i\nu) \Gamma(j+1+i\nu) }
        { \Gamma(j+1-m-i\nu) \Gamma(j+1-m+i\nu) } \delta(\nu- \nu').
   \nonumber 
\end{eqnarray}
Noting that 
\begin{eqnarray}
&&
   \left(\sum_{N=0}^{[m/2]} \tilde{E}^N D^{m-2N}[\epsilon^{(0,0,m)}]
     \frac{b^{(j-m)}_{m,N}}{\Delta^{m-2N}} \right)
  \left(\sum_{M=0}^{[m/2]} \tilde{E}^M D^{m-2M}[\epsilon^{'(0,0,m)}]
     \frac{b^{(j-m)}_{m,M}}{\Delta^{m-2M}} \right), \nonumber  \\
&=& 
 \frac{j!}{(j-m)!} \left( \sum_{N=0}^{[m/2]} b^{(j-m)}_{m,N} \right) 
 \; \epsilon^{(0,0,m)}_{\mu_1 \cdots \mu_{j-m}} \cdot 
    \epsilon^{'(0,0,m) \; \hat{\mu}_1 \cdots \hat{\mu}_{j-m}}, \nonumber  
\end{eqnarray}
we find that (\ref{eq:normalization-temp-app}) implies 
\begin{eqnarray}
 N_{j,m}^{-2} & = &
 \frac{ \Gamma(j+1-i\nu) \Gamma(j+1+i\nu) }
        { \Gamma(j+1-m-i\nu) \Gamma(j+1-m+i\nu) } 
 \frac{j!}{(j-m)!} \left( \sum_{N=0}^{[m/2]} b^{(j-m)}_{m,N} \right), 
   \nonumber \\ 
& = &  
  \frac{\Gamma(j+1-i\nu)}{\Gamma(j+1-m-i\nu)}
  \frac{\Gamma(j+1+i\nu)}{\Gamma(j+1-m+i\nu)} \; 
  {}_jC_m 
  \frac{\Gamma(3/2+j-m)}{2^m\Gamma(3/2+j)}
  \frac{\Gamma(2+2j)}{\Gamma(2+2j-m)}. 
   \nonumber 
\end{eqnarray}
%

\subsection{A Note on Wavefunction of Massless Vector Field}
\label{ssec:vector}

For a rank-1 tensor (vector) field on AdS$_5$, we can determine 
the wavefunction of the $(n,l,m) = (1,0,1)$ eigenmode, 
not just for the $(n,l,m) = (0,0,m)$ modes with $m = 0,1$.
With the eigenvalue ${\cal E}_{1,0} = (3j+5+\nu^2)|_{j=1}$, 
\begin{equation}
 a^{(0,1,0)} = \epsilon^{(1,0,0)} z^2 K_{i\nu}(\Delta z), \qquad 
 a^{(1,0,0)} = \epsilon^{(1,0,0)}
  \partial_z \left(z^2 K_{i\nu}(\Delta z) \right). 
\end{equation}
is the eigenvector solution to (\ref{eq:EGeq-q=not0-m=1}).

The $(n,l,m) = (0,0,1)$ mode and $(n,l,m)=(1,0,1)$ mode are
independent, even after the mass-shell condition (\ref{eq:mass-shell}) 
for generic vector fields in the bosonic string theory. However, 
for the massless vector field $A_m$ obtained by simple dimensional 
reduction of the massless vector field $A^{(Y)}_M$ with 
$Y = \left\{1,0,0 \right\}$, those two modes become degenerate. 
To see this, note that $c_y = -4$ for this mode, so that the 
mass-shell condition (\ref{eq:mass-shell}) implies, 
\begin{equation}
 (j+4+\nu^2 + c_y)|_{j=1} = 0 \quad (0,0,1){\rm ~mode}, \qquad 
 (3j+5+\nu^2+c_y)|_{j=1} = 0 \quad (1,0,1){~\rm mode},
\end{equation}
or equivalently, $i\nu = 1$ and $i\nu = 2$, respectively, for these two
modes. It is now obvious that the terms proprotional to 
$(\epsilon \cdot \epsilon)$ in (\ref{eq:photon-wvfc}) are in the 
form of this $(n,l,m) = (1,0,1)$ mode. With the relations 
$x^3 \partial_x \left[x^{-3+2} K_1(x) \right] =
  - x^3 \left[x^{-1} K_2(x)\right]$ and  $
\partial_x \left[x^2 K_2(x) \right] = - x^2 K_1(x)$, one can also see 
that the wavefunction for the $(n,l,m)=(0,0,1)$ mode is also
proportional to the form given in (\ref{eq:photon-wvfc}).


\section*{Acknowledgment}

We thank Wen Yin for discussion, with whom we worked together at earlier 
stage of this project, and Teruhiko Kawano for useful comments.
The work is supported in part by JSPS Research Fellowships for 
Young Scientists (RN), World Premier International Research Center Initiative
(WPI Initiative), MEXT, Japan (RN, TW) and a Grant-in-Aid for 
Scientific Research on Innovative Areas 2303 (TW).


\begin{thebibliography}{99}
%
%
\bibitem{PS-01-PRL}
%
  J.~Polchinski and M.~J.~Strassler,
  ``Hard scattering and gauge / string duality,''  Phys.\ Rev.\ Lett.\
	{\bf 88}, 031601 (2002)  [hep-th/0109174]. 
%
\bibitem{PS-02-DIS}
%
  J.~Polchinski and M.~J.~Strassler,
  ``Deep inelastic scattering and gauge / string duality,''  JHEP {\bf
	0305}, 012 (2003)  [hep-th/0209211].
%
\bibitem{BPST-06}
%
  R.~C.~Brower, J.~Polchinski, M.~J.~Strassler and C.~-ITan,
  ``The Pomeron and gauge/string duality,''  JHEP {\bf 0712}, 005 (2007)
	[hep-th/0603115].
%


%
 \bibitem{NW-first}
%
%
  R.~Nishio and T.~Watari,
  ``Investigating Generalized Parton Distribution in Gravity Dual,''
	 Phys.\ Lett.\ B {\bf 707}, 362 (2012)  [arXiv:1105.2907
	 [hep-ph]]; \\ 
%
%
  R.~Nishio and T.~Watari,
  ``High--Energy Photon--Hadron Scattering in Holographic QCD,''  Phys.\ Rev.\ D {\bf 84}, 075025 (2011)  [arXiv:1105.2999 [hep-ph]].  
%


%
%
\bibitem{NW-more}
%
R.~Nishio and T.~Watari, work in progress.
%



%
\bibitem{Hatta-07}
  Y.~Hatta, E.~Iancu and A.~H.~Mueller,
  ``Deep inelastic scattering at strong coupling from gauge/string duality: The Saturation line,''  JHEP {\bf 0801}, 026 (2008)  [arXiv:0710.2148 [hep-th]].  
%




%
\bibitem{CC-DIS}
%
  L.~Cornalba, M.~S.~Costa and J.~Penedones,
  ``Deep Inelastic Scattering in Conformal QCD,''  JHEP {\bf 1003}, 133
	(2010)  [arXiv:0911.0043 [hep-th]].
%



%
\bibitem{BDST-10}
%
  R.~C.~Brower, M.~Djuric, I.~Sarcevic and C.~-ITan,
  ``String-Gauge Dual Description of Deep Inelastic Scattering at Small-$x$,''  JHEP {\bf 1011}, 051 (2010)  [arXiv:1007.2259 [hep-ph]].  
%










%
\bibitem{GPD-review} 
%
  M.~Diehl,
  ``Generalized parton distributions,''  Phys.\ Rept.\  {\bf 388}, 41
 (2003)  [hep-ph/0307382]; \\ 
%
  A.~V.~Belitsky and A.~V.~Radyushkin,
  ``Unraveling hadron structure with generalized parton distributions,''
	Phys.\ Rept.\  {\bf 418}, 1 (2005)  [hep-ph/0504030].
%


%
\bibitem{GPD-transv-profile}
%
  M.~Burkardt,
  ``Impact parameter dependent parton distributions and off forward
	parton distributions for zeta ---> 0,''  Phys.\ Rev.\ D {\bf
	62}, 071503 (2000)  [Erratum-ibid.\ D {\bf 66}, 119903 (2002)]
	[hep-ph/0005108]; \\ 
%
%
  J.~P.~Ralston and B.~Pire,
  ``Femtophotography of protons to nuclei with deeply virtual Compton
	scattering,''  Phys.\ Rev.\ D {\bf 66}, 111501 (2002)
	[hep-ph/0110075]; \\
%
  M.~Diehl,
  ``Generalized parton distributions in impact parameter space,''  Eur.\
	Phys.\ J.\ C {\bf 25}, 223 (2002)  [Erratum-ibid.\ C {\bf 31},
	277 (2003)]  [hep-ph/0205208]; \\
  M.~Burkardt,
  ``Impact parameter space interpretation for generalized parton distributions,''  Int.\ J.\ Mod.\ Phys.\ A {\bf 18}, 173 (2003)  [hep-ph/0207047].  


%
%




%



%
\bibitem{Ji}
  X.~-D.~Ji,
  ``Deeply virtual Compton scattering,''  Phys.\ Rev.\ D {\bf 55}, 7114 (1997)  [hep-ph/9609381].  
%


%
\bibitem{dual-para}
%
  A.~V.~Belitsky, B.~Geyer, D.~Mueller and A.~Schafer,
  ``On the leading logarithmic evolution of the off forward
	distributions,''  Phys.\ Lett.\ B {\bf 421}, 312 (1998)
	[hep-ph/9710427]; \\
%
%
%
  M.~V.~Polyakov,
  ``Hard exclusive electroproduction of two pions and their
	resonances,''  Nucl.\ Phys.\ B {\bf 555}, 231 (1999)
	[hep-ph/9809483]; \\ 
%
%
  M.~V.~Polyakov and A.~G.~Shuvaev,
  ``On'dual' parametrizations of generalized parton distributions,''
	hep-ph/0207153. 
%



%
%



\bibitem{ER}

  A.~V.~Efremov and A.~V.~Radyushkin,
  ``Asymptotical Behavior of Pion Electromagnetic Form-Factor in QCD,''  Theor.\ Math.\ Phys.\  {\bf 42}, 97 (1980)  [Teor.\ Mat.\ Fiz.\  {\bf 42}, 147 (1980)].  
%



%
%
%


%
\bibitem{BM-conf-viol}
%
  A.~V.~Belitsky and D.~Mueller,
  ``Next-to-leading order evolution of twist-2 conformal operators: The
	Abelian case,''  Nucl.\ Phys.\ B {\bf 527}, 207 (1998)
	[hep-ph/9802411].  
%




\bibitem{Klebanov-W}
%
  I.~R.~Klebanov and E.~Witten,
  ``Superconformal field theory on three-branes at a Calabi-Yau
	singularity,''  Nucl.\ Phys.\ B {\bf 536}, 199 (1998)
	[hep-th/9807080].
%



%
\bibitem{FGG-confl-HG}
  S.~Ferrara, R.~Gatto and A.~F.~Grillo,
  ``Conformal invariance on the light cone and canonical dimensions,''
	Nucl.\ Phys.\ B {\bf 34}, 349 (1971).
%

%




%
\bibitem{Mueller-PRD98}
%
  D.~Mueller,
  ``Restricted conformal invariance in QCD and its predictive power for
	virtual two photon processes,''  Phys.\ Rev.\ D {\bf 58}, 054005
	(1998)  [hep-ph/9704406].
%



\bibitem{Mueller-S-05}
  D.~Mueller and A.~Schafer,
  ``Complex conformal spin partial wave expansion of generalized parton distributions and distribution amplitudes,''  Nucl.\ Phys.\ B {\bf 739}, 1 (2006)  [hep-ph/0509204].  
%

%
\bibitem{KK-Mueller-07}
%
  K.~Kumericki, D.~Mueller and K.~Passek-Kumericki,
  ``Towards a fitting procedure for deeply virtual Compton scattering at
	next-to-leading order and beyond,''  Nucl.\ Phys.\ B {\bf 794},
	244 (2008)  [hep-ph/0703179 [HEP-PH]].
%










%
\bibitem{GKP-02}
%
  S.~S.~Gubser, I.~R.~Klebanov and A.~M.~Polyakov,
  ``A Semiclassical limit of the gauge / string correspondence,''
	Nucl.\ Phys.\ B {\bf 636} (2002) 99  [hep-th/0204051].
%
	
%





\bibitem{Klebanov-Strassler}
%
  I.~R.~Klebanov and M.~J.~Strassler,
  ``Supergravity and a confining gauge theory: Duality cascades and chi
	SB resolution of naked singularities,''  JHEP {\bf 0008}, 052
	(2000)  [hep-th/0007191]. 
%








%
\bibitem{SFT-Witten-cubic}
%
  E.~Witten,
  ``Noncommutative Geometry and String Field Theory,''  Nucl.\ Phys.\ B
	{\bf 268}, 253 (1986).
%



\bibitem{SFT-interaction}
%
%
  E.~Cremmer, A.~Schwimmer and C.~B.~Thorn,
  ``The Vertex Function in Witten's Formulation of String Field
	Theory,''  Phys.\ Lett.\ B {\bf 179}, 57 (1986); \\
%
%
%
  D.~J.~Gross and A.~Jevicki,
  ``Operator Formulation of Interacting String Field Theory,''  Nucl.\
	Phys.\ B {\bf 283}, 1 (1987); 
%
%
%
  D.~J.~Gross and A.~Jevicki,
  ``Operator Formulation of Interacting String Field Theory. 2.,''
	Nucl.\ Phys.\ B {\bf 287}, 225 (1987); \\
%
%
%
%
%
  W.~Taylor,
  ``D-brane effective field theory from string field theory,''  Nucl.\
	Phys.\ B {\bf 585}, 171 (2000)  [hep-th/0001201].



%
\bibitem{SFT-Hata-lec.note}
%
H.~Hata, ``String theory and string field theory,'' 
a lecture note (in Japanese) for the 49th summer school for 
young generations, 
edt. by graduate students at Kyoto University.
%


%
\bibitem{SFT-Giddings-Veneziano}
%
  S.~B.~Giddings,
  ``The Veneziano Amplitude from Interacting String Field Theory,''
	Nucl.\ Phys.\ B {\bf 278}, 242 (1986).
%
%


%
\bibitem{SFT-Thorn}
%
  C.~B.~Thorn,
  ``Perturbation Theory for Quantized String Fields,''  Nucl.\ Phys.\ B {\bf 287}, 61 (1987).  
%





%
\bibitem{AdSCFT-review}
%
  O.~Aharony, S.~S.~Gubser, J.~M.~Maldacena, H.~Ooguri and Y.~Oz,
  ``Large N field theories, string theory and gravity,''  Phys.\ Rept.\
	{\bf 323}, 183 (2000)  [hep-th/9905111].
%


\end{thebibliography}
\end{document}